\documentclass{aa}
\usepackage{latexsym,graphicx,amssymb,verbatim,natbib,float,amsmath}
\usepackage{pdflscape}
\usepackage{adjustbox}
\usepackage{hyperref}
\usepackage{color}
\begin{document}
\title{Cosmic-ray energy spectrum and composition up to the ankle $-$ the case for a second Galactic component }
\author{S.~Thoudam\inst{1,2}$^,$\thanks{E-mail: satyendra.thoudam@lnu.se}, J.P.~Rachen\inst{1}, A.~van~Vliet\inst{1}, A.~Achterberg\inst{1}, S.~Buitink\inst{3}, H.~Falcke\inst{1,4,5}, J.R.~H\"{o}randel\inst{1,4}}
\institute{Department of Astrophysics/IMAPP, Radboud University, P.O. Box 9010, 6500 GL Nijmegen, The Netherlands
\and
Now at: Department of Physics and Electrical Engineering, Linn\'{e}universitetet, 35195 V\"{a}xj\"{o}, Sweden
\and
Astronomical Institute, Vrije Universiteit Brussel, Pleinlaan 2, 1050 Brussels, Belgium
\and
NIKHEF, Science Park Amsterdam, 1098 XG Amsterdam, The Netherlands
\and
ASTRON, Postbus 2, 7990 AA Dwingeloo, The Netherlands
}

\date{\today}

\abstract{Motivated by the recent high-precision measurements of cosmic rays by several new-generation experiments, we have carried out a detailed study to understand the observed energy spectrum and composition of cosmic rays with energies up to about  $10^{18}$~eV. Our study shows that a single Galactic component with subsequent energy cut-offs in the individual spectra of different elements, optimised to explain the observed elemental spectra below ${\sim}\,10^{14}$~eV and the `knee' in the all-particle spectrum, cannot explain the observed all-particle spectrum above ${\sim}\,2\times\,10^{16}$~eV. We discuss two approaches for a second component of Galactic cosmic rays -- re-acceleration at a Galactic wind termination shock, and supernova explosions of Wolf-Rayet stars, and show that the latter scenario can explain almost all observed features in the all-particle spectrum and the composition up to ${\sim}\,10^{18}$~eV, when combined with a canonical extra-galactic spectrum expected from strong radio galaxies or a source population with similar cosmological evolution. In this two-component Galactic model, the knee at ${\sim}\,3\times 10^{15}\,$~eV and the `second knee' at ${\sim}\,10^{17}\,$~eV in the all-particle spectrum are due to the cut-offs in the first and second components, respectively. We also discuss several variations of the extra-galactic component, from a minimal contribution to scenarios with a significant component below the `ankle' (at ${\sim}\,4\times10^{18}$~eV), and find that extra-galactic contributions in excess of regular source evolution are neither indicated nor in conflict with the existing data. We also provide arguments that an extra-galactic contribution is unlikely to dominate at or below the second knee. Our main result is that the second Galactic component predicts a composition of Galactic cosmic rays at and above the second knee that largely consists of helium or a mixture of helium and CNO nuclei, with a weak or essentially vanishing iron fraction, in contrast to most common assumptions. This prediction is in agreement with new measurements from LOFAR and the Pierre Auger Observatory which indicate a strong light component and a rather low iron fraction between ${\sim}\,10^{17}$ and $10^{18}$~eV.}

\keywords{Galaxy --- cosmic rays --- diffusion --- ISM: supernova remnants --- Stars: winds --- Stars: Wolf-Rayet}

\authorrunning{Thoudam et al.}
\titlerunning{Cosmic-ray energy spectrum and composition up to the ankle}
\maketitle

\section{Introduction}
Until a decade ago, the cosmic ray spectrum from ${\sim}\,10\,$~GeV to ${\sim}\,10^{11}\,$~GeV was seen as a power law with two main  features: a steepening from a spectral index $\gamma \approx -2.7$ to $\gamma\approx -3.1$ at about $3\times 10^6\,$~GeV, commonly called the `knee', and a flattening back to $\gamma \approx -2.7$ at about $4\times 10^{9}\,$~GeV, consequently denoted as the `ankle'. Phenomenological explanations for the knee have been given due to propagation effects in the Galaxy \citep{Ptuskin1993}, progressive cutoffs in the spectra of nuclear components from hydrogen to lead \citep{Hoerandel2003a}, or re-acceleration at shocks in a Galactic wind \citep{Voelk2004}, but left open the question of the primary Galactic accelerators producing these particles. Explanations based on source physics have been mostly built on the assumption that supernova remnants, on grounds of energetics known as one of the most promising sources for cosmic rays \citep{BaadeZwicky1934}, accelerate cosmic rays at shocks ploughing into the interstellar medium to energies up to about $10^{5{-}6}\,$~GeV \citep{Lagage1983,  Axford1994}. This may extend to ${\sim}\,10^{8}\,$~GeV if they are propagating in fast and highly magnetised stellar winds \citep{VoelkBiermann1988, Biermann1993}, or if non-linear effects in the acceleration process are considered \citep{BellLucek2001}. The combination of such components could eventually explain cosmic rays below and above the knee as a superposition of components of different nuclei, as shown, for example by \citet{Stanev1993}. At energies above $10^9$\,~GeV this steep component was assumed to merge into a flatter extra-galactic component \citep{Rachen1993, Berezinsky2004}, explaining the ankle in the spectrum. For this extra-galactic component, sources on all scales have been proposed: From clusters of galaxies \citep{KangRyuJones1996} through radio galaxies \citep{RachenBiermann1993}, compact AGN jets \citep{Mannheim2001} to gamma-ray bursts \citep{Waxman1995}. It was commonly assumed to be dominated by protons. Eventually, at ${\sim}\,10^{11}\,$~GeV the cosmic ray spectrum was believed to terminate in the so-called GZK cutoff \citep{Greisen1966, ZatsepinKuzmin1966} due to interaction with cosmic microwave background (CMB) photons.

Recent measurements of cosmic rays by several new generation experiments have severely challenged this simple view. At low energies, below ${\sim}\,10^{6}$~GeV, satellite and balloon-borne experiments such as ATIC-2 \citep{Panov2007}, CREAM \citep{Yoon2011}, TRACER \citep{Obermeier2011}, PAMELA \citep{Adriani2014}, AMS-02 \citep{Aguilar2014, Aguilar2015a, Aguilar2015b}, and \textit{Fermi-}LAT \citep{Abdo2009} have measured the energy spectra of various elements of cosmic rays ranging from protons to heavier nuclei such as iron as well as the leptonic component of cosmic rays, and anti-particles such as positrons and anti-protons. Some of their results, for example the rise of the positron fraction above ${\sim}\,10$~GeV \citep{Aguilar2013}, the harder energy spectrum of helium nuclei with respect to the proton spectrum \citep{Adriani2011}, and the spectral hardening of both the proton and helium nuclei at TeV energies \citep{Yoon2011}, are difficult to explain using standard models of cosmic-ray acceleration in supernova remnants and their subsequent propagation in the Galaxy. At high energies, that is above ${\sim}\,10^{6}$~GeV, ground-based experiments such as KASCADE-Grande \citep{Apel2013}, the Tibet III array \citep{Amenomori2008}, IceTop \citep{Aartsen2013}, the Pierre Auger Observatory \citep{Ghia2015} and the Telescope Array \citep{Abu-Zayyad2013} have carried out detailed measurements of the all-particle energy spectrum and the composition of cosmic rays. First, they confirm a third major break in the spectrum, a steepening to $\gamma \approx -3.3$ above about $10^{8}\,$~GeV, which has been suggested before both by the Fly's Eye stereo energy spectrum \citep{BirdEtAl1994} and theoretical arguments about the structure of the ankle \citep{Berezinsky1988, Rachen1993}. It has anatomically been named the `second knee' \citep{Hoerandel2006}. While this still fits with the original view, the cosmic-ray composition measurements at these energies pose a severe challenge: Instead of gradually becoming heavier as expected, the data show that the composition reaches a maximum mean mass at energies around $6\times 10^{7}$~GeV, and then becomes gradually lighter again up to the ankle. Finally, above the ankle the composition becomes heavier again. It has been shown that the observed spectrum and composition at the highest energies can be explained by a mixed-composition extra-galactic source spectrum with progressive cutoffs at ${\sim}\,Z\times 5{\times}10^{9}$~GeV, where $Z$ is the nuclear charge \citep{Aloisio2014}. This would imply that there is no significant impact of the GZK effect in cosmic ray propagation except through photo-disintegration of nuclei. In addition, the measurement of an ankle-like feature in the light component of cosmic rays at $\sim 10^8$~GeV by the KASCADE-Grande experiment \citep{Apel2013}, and the new revelation of a strong light component and a very small iron component by the LOFAR measurements between $\sim (1-4)\times10^8$~GeV \citep{Buitink2016}, and by the Pierre Auger Observatory above ${\sim}\,7\times10^8$~GeV \citep{Aab2014} add further challenges to the standard model.

The new data have led to a number of theoretical modifications of the standard model. The spectral hardening at TeV energies has been explained as due to the hardening in the source spectrum of cosmic rays \citep{Biermann2010a, Ohira2011, Yuan2011, Ptuskin2013}, as a propagation effect \citep{Tomassetti2012, Blasi2012}, the effect of re-acceleration by weak shocks \citep{Thoudam2014} or the effect of nearby sources \citep{Thoudam2012, Thoudam2013, Erlykin2012}. At high energies, the increasing mean mass around the knee still fits well the idea of progressive cut-offs \citep{Hoerandel2003a}, if the nuclear species are constrained to masses up to iron and thus limited to energies below about $3{\times}10^{7}\,$~GeV. The light composition around the ankle revived interest in the so-called `proton dip model', which explains the ankle feature as due to an extra-galactic propagation effect of protons producing electron-positron pairs at the CMB \citep{Berezinsky1988, Berezinsky2006}. This would imply that the cosmic ray spectrum below the ankle is, at least in part, of extra-galactic origin. While the recent measurement of ${\sim}\,40\%$ proton fraction at the ankle by the Pierre Auger Collaboration \citep{Aab2014} has raised problems with this approach, as the model is compatible only with more than $80\%$ protons \citep{Berezinsky2006}, a number of new models have been suggested, involving compact sources with significant photo-disintegration of nuclei during acceleration \citep{Globus2015, Unger2015}, or as a component with primordial element composition accelerated at clusters of galaxies and limited by pair production losses in the CMB \citep{Rachen2016}. However,  with all these new ideas, big questions  remain open: How does the cosmic ray component at the knee connect to the one at the second knee to ankle regime, and where is the transition from Galactic to extra-galactic cosmic rays?
 
In this work, we revisit the basic models of Galactic cosmic-ray production in view of the currently available data. We start by  developing a detailed model description for low-energy cosmic rays assuming them to be primarily produced inside supernova remnants (SNRs) present in the interstellar medium (hereafter, these cosmic rays will be referred to as the `SNR-CRs'). This model, described in Section \ref{sec-SNR-CR}, has been demonstrated to explain the observed spectral hardening of protons and helium nuclei in the TeV region and, at the same time, explains the observed composition of cosmic rays at low energies \citep{Thoudam2014}. The model prediction will be extended to high energies, and compared with the observed all-particle energy spectrum. It will be shown that SNR-CRs cannot explain the observed energy spectrum above ${\sim}\,10^{7}$~GeV. We then revisit two possibilities for a second Galactic component in Section \ref{sec-additional-component}: (a) The re-acceleration of SNR-CRs escaped into the Galactic halo by the Galactic wind termination shocks  \citep{Jokipii1987, Zirakashvili2006}, and (b) the contribution of cosmic rays from the explosions of Wolf-Rayet stars in the Galaxy  \citep{Biermann1993}. The possibility of a second Galactic component has also been discussed in \cite{Hillas2005} who  considered Type II SNRs expanding into a dense wind of their precursor stars. For both the scenarios considered in the present work, we assume the extra-galactic proton component used by \cite{Rachen1993} to obtain proper results for total spectrum and composition at energies just below the ankle in Section \ref{sec-total-spectrum}. In Section \ref{section-EG-models} we then check the effect of other hypotheses for the extra-galactic component, using (1) a phenomenological `minimal model' derived from composition results measured at the Pierre Auger Observatory \citep{Matteo2015}, (2) the minimal model plus the `primordial cluster component' introduced by \cite{Rachen2016}, and (3) the `extra-galactic ankle' model by \cite{Unger2015}. In Section \ref{sec-discussion}, we present a discussion of our results and their implications, and other views on the cosmic rays below $10^9$~GeV, followed by our conclusions in Section \ref{sec-conclusion}.  

\section{Cosmic rays from supernova remnants (SNR-CRs)}
\label{sec-SNR-CR}
Although the exact nature of cosmic-ray sources in the Galaxy is not yet firmly established, supernova remnants are considered to be the most plausible candidates both from the theoretical and the observational points of view. It has been theoretically established that shock waves associated with supernova remnants can accelerate particles from the thermal pool to a non-thermal distribution of  energetic particles. The underlying acceleration process, commonly referred to as the diffusive shock acceleration process, has been studied quite extensively, and it produces a power-law spectrum of particles with a spectral index close to $2$ \citep{Krymskii1977, Bell1978, Blandford1978, Drury1983, Ptuskin2010, Caprioli2011}, which is in good agreement with the values inferred from radio observation of supernova remnants \citep{Green2009}. Moreover, the total power of ${\sim}\,10^{42}$ ergs s$^{-1}$ injected by supernova explosions into the Galaxy, considering a supernova explosion energy of ${\sim}\,10^{51}$ ergs and an explosion frequency of ${\sim}\,1/30$ yr$^{-1}$, is more than sufficient to maintain the cosmic-ray energy content of the Galaxy. In addition to the radio measurements, observational evidence for the presence of high-energy particles inside supernova remnants is provided by the detection of non-thermal X-rays \citep{Vink2003, Parizot2006} and TeV gamma rays from a number of supernova remnants \citep{Aharonian2006, Aharonian2008, Albert2007}. For instance, the detection of TeV gamma rays up to energies close to $100$~TeV from the supernova remnant RX J1713.7-3946 by the H.E.S.S. Cherenkov telescope array indicates that particles with energies up to ${\sim}\,1$ PeV can be accelerated inside supernova remnants \citep{Aharonian2007}.

\subsection{Transport of SNR-CRs in the Galaxy}
After acceleration by strong supernova remnant shock waves, cosmic rays escape from the remnants and undergo diffusive propagation through the Galaxy. During the propagation, some fraction of cosmic rays may further get re-accelerated due to repeated encounters with expanding supernova remnant shock waves in the interstellar medium \citep{Wandel1988, Berezhko2003}. This re-acceleration is expected to be produced mainly by older remnants, with weaker shocks, because of their bigger sizes. Therefore, the re-acceleration is expected to generate a particle spectrum which is steeper than the initial source spectrum of cosmic rays produced by strong shocks. This model has been described in detail in \cite{Thoudam2014}, and it has been shown that the re-accelerated cosmic rays can dominate the GeV energy region while the non-re-accelerated cosmic rays dominate at TeV energies, thereby explaining the observed spectral hardening in the TeV region. Below, we briefly summarise some key features of the model which are important for the present study. 

\begin{figure*}
\centering
\includegraphics*[width=\columnwidth,height=0.6\columnwidth,angle=0,clip]{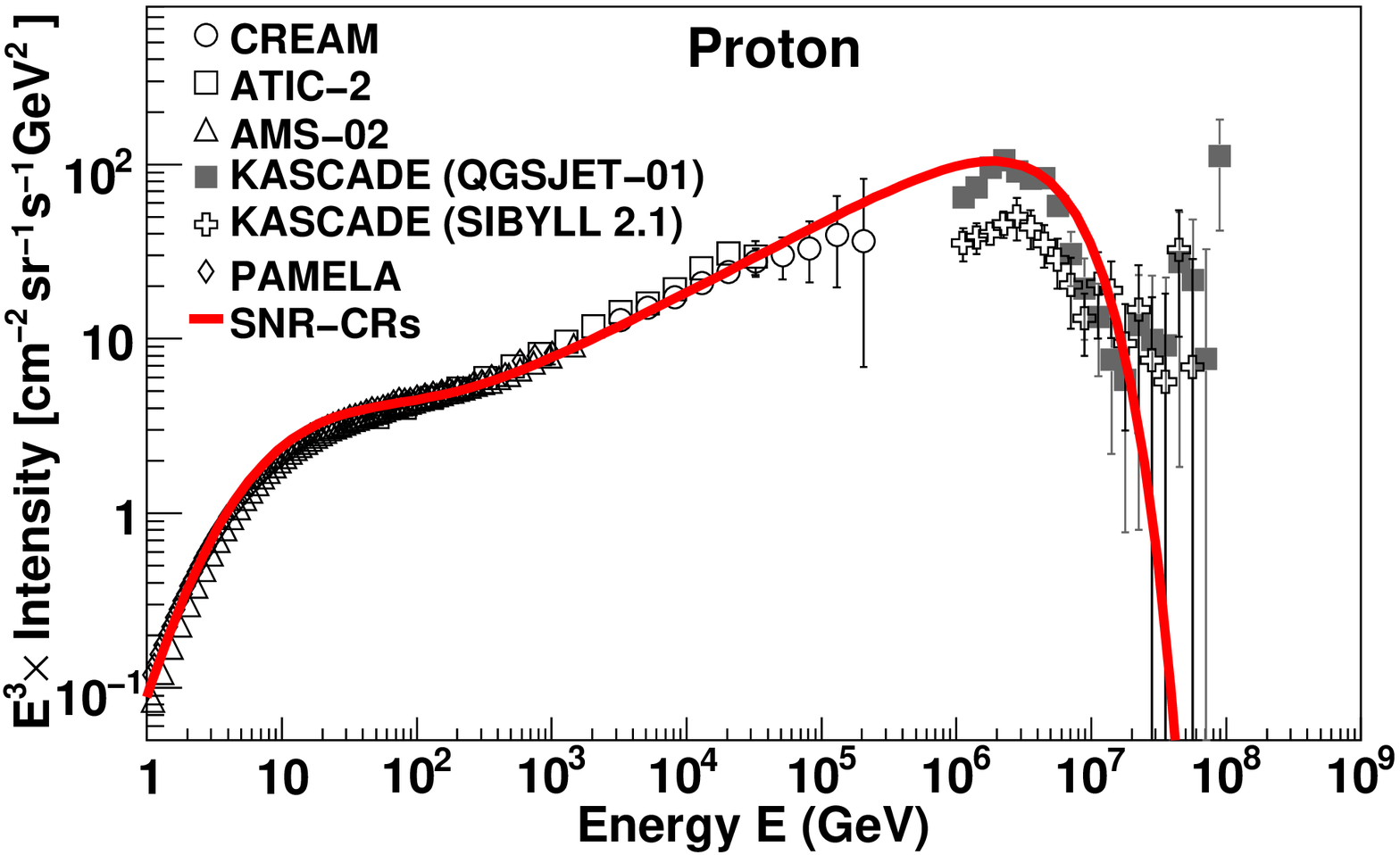}
\includegraphics*[width=\columnwidth,height=0.6\columnwidth,angle=0,clip]{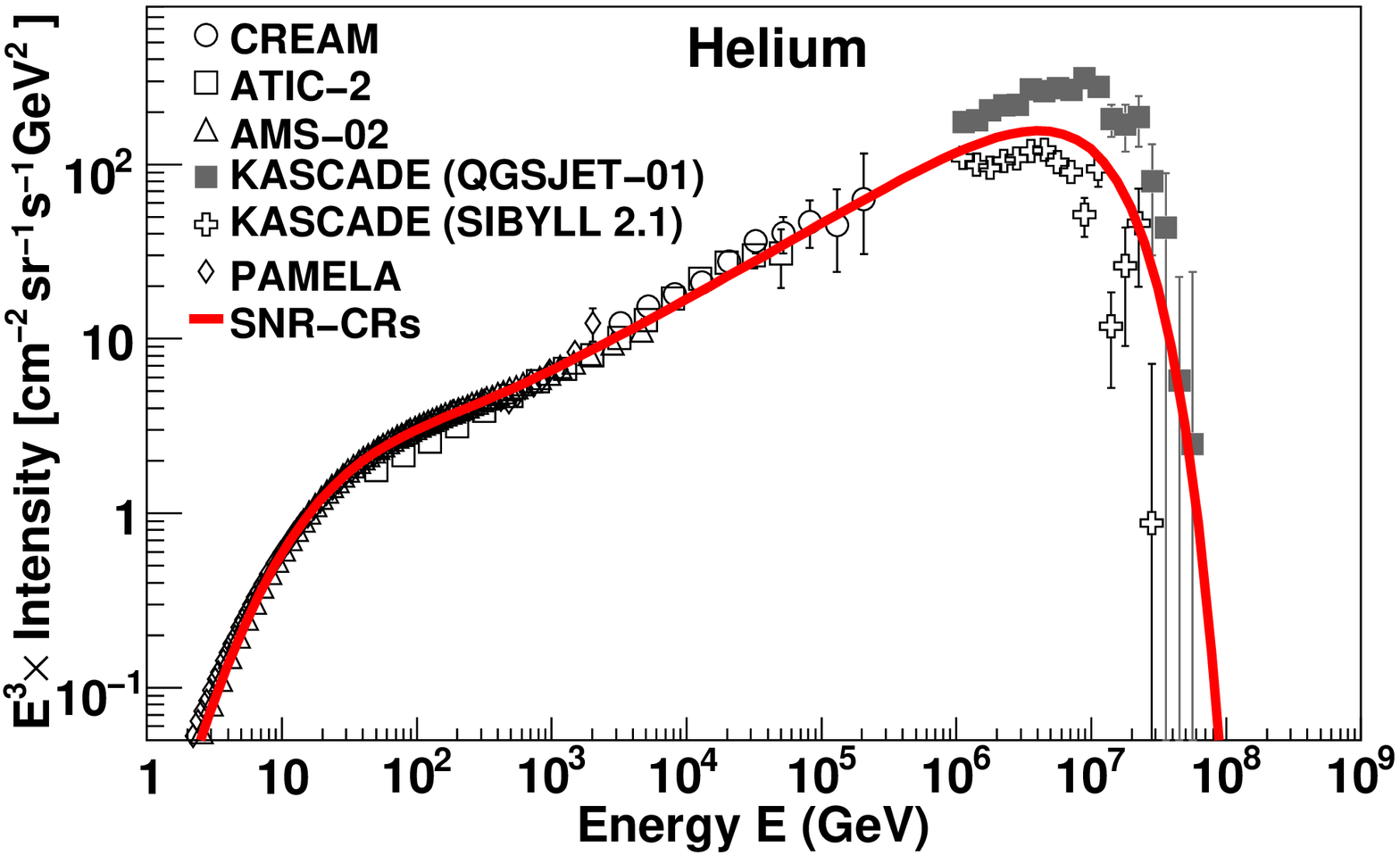}\\
\includegraphics*[width=\columnwidth,height=0.6\columnwidth,angle=0,clip]{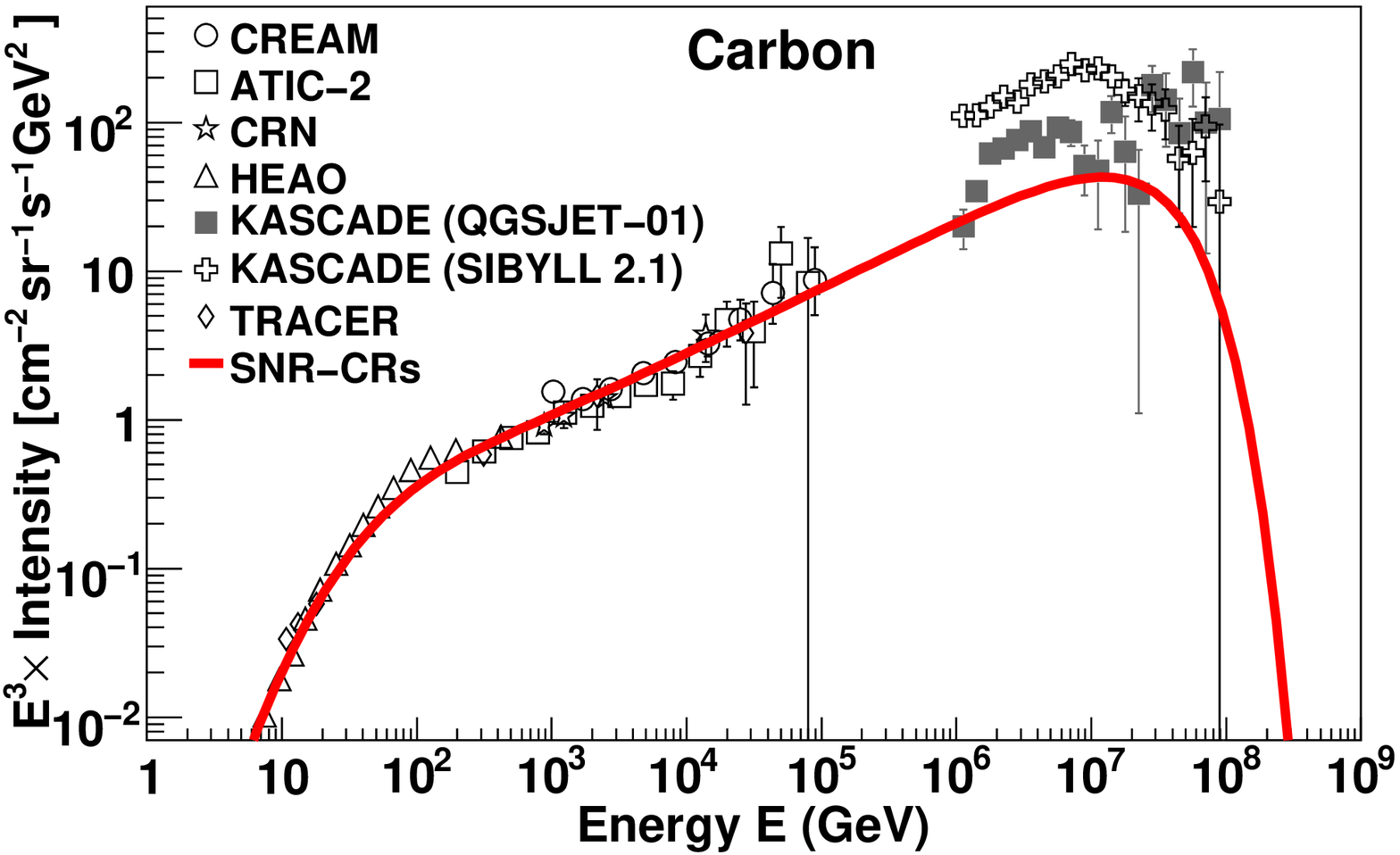}
\includegraphics*[width=\columnwidth,height=0.6\columnwidth,angle=0,clip]{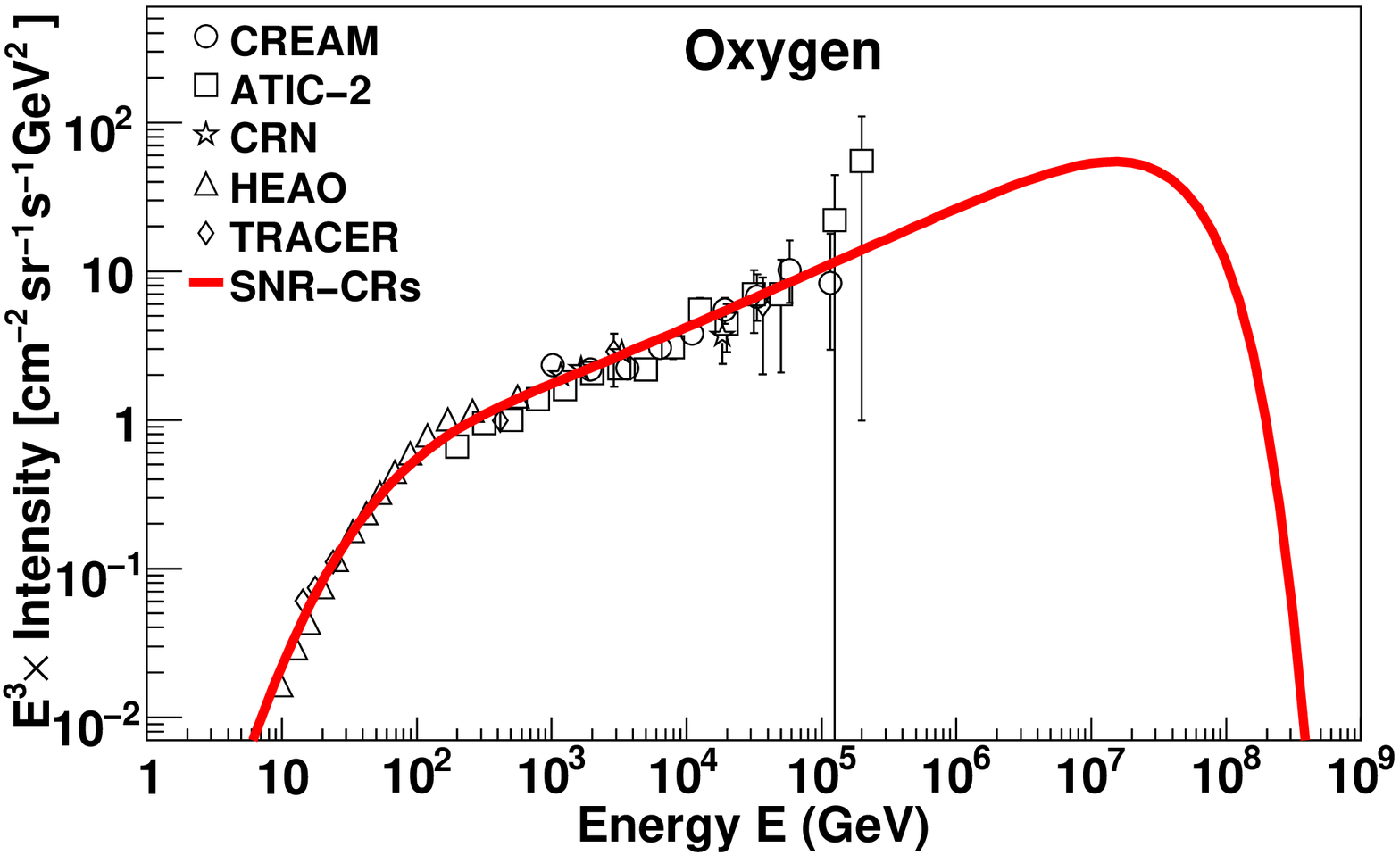}\\
\includegraphics*[width=\columnwidth,height=0.6\columnwidth,angle=0,clip]{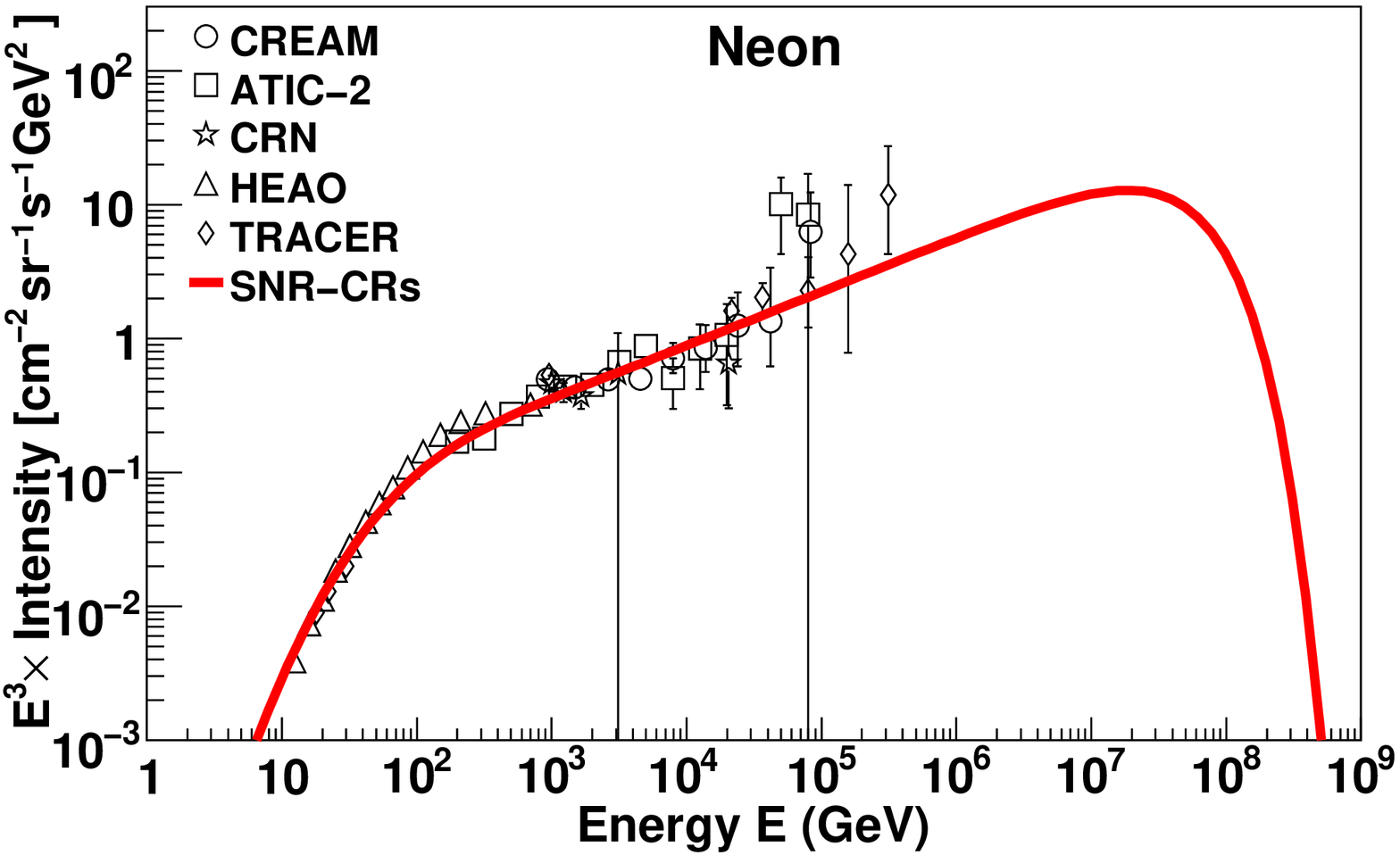}
\includegraphics*[width=\columnwidth,height=0.6\columnwidth,angle=0,clip]{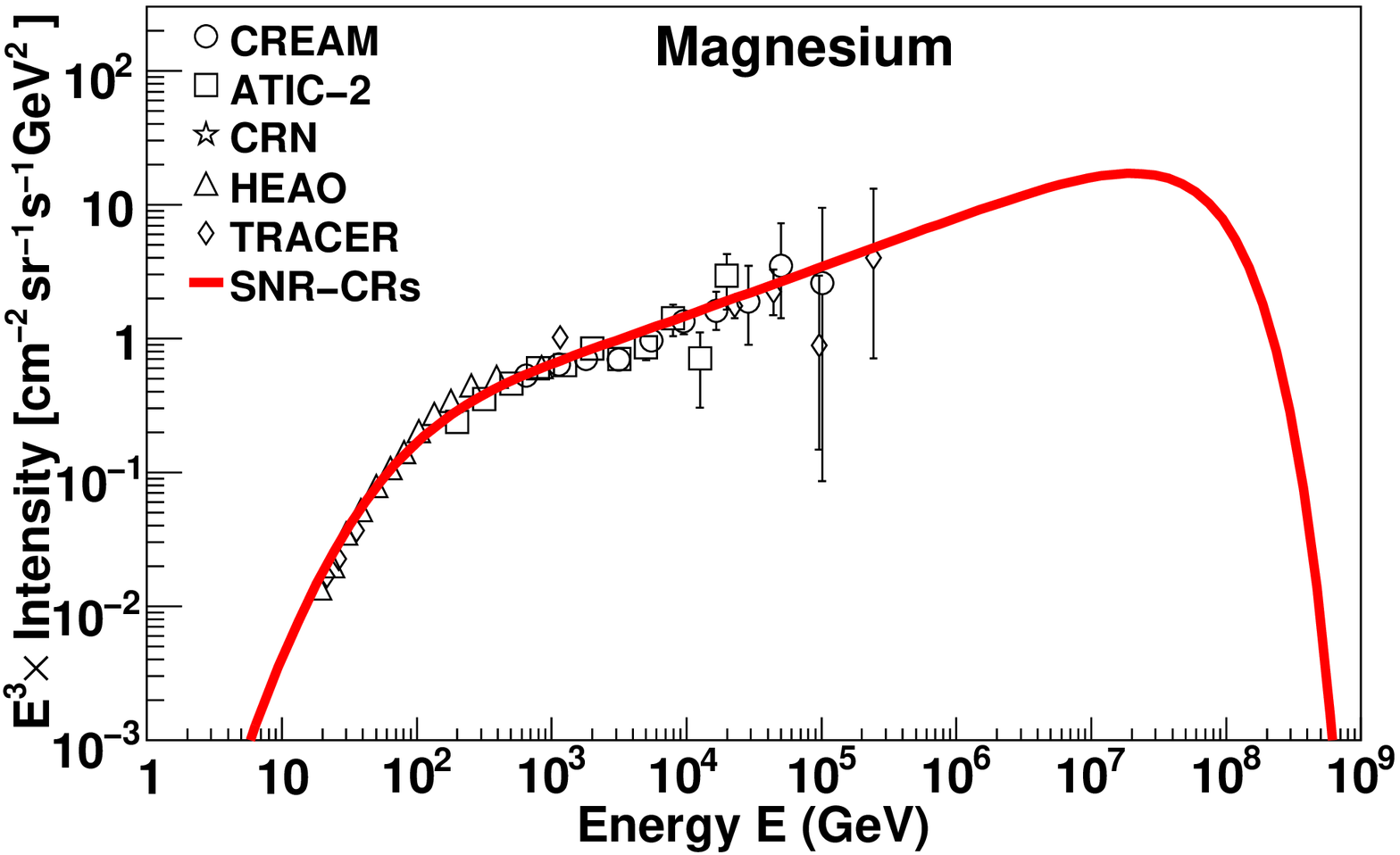}\\
\includegraphics*[width=\columnwidth,height=0.6\columnwidth,angle=0,clip]{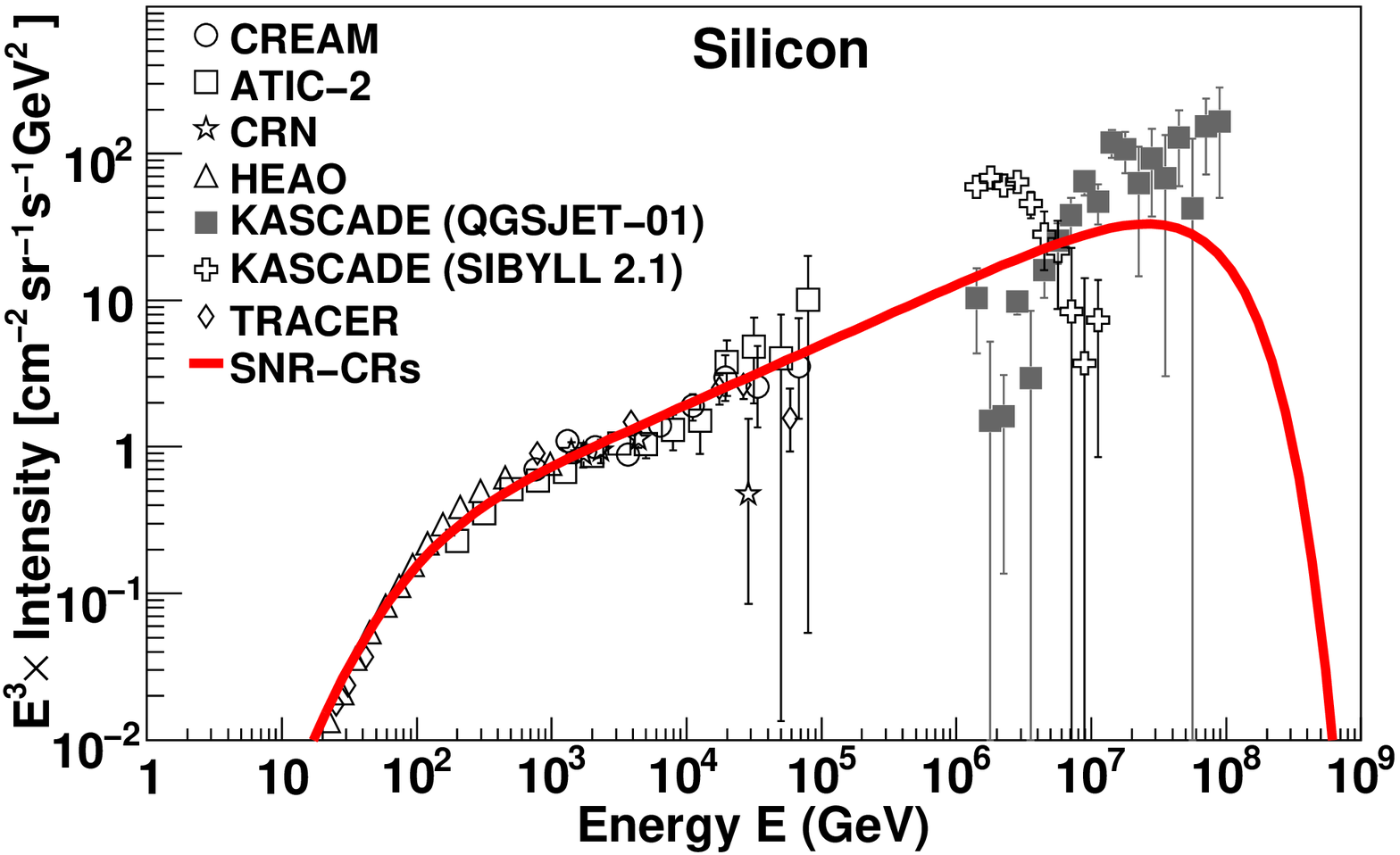}
\includegraphics*[width=\columnwidth,height=0.6\columnwidth,angle=0,clip]{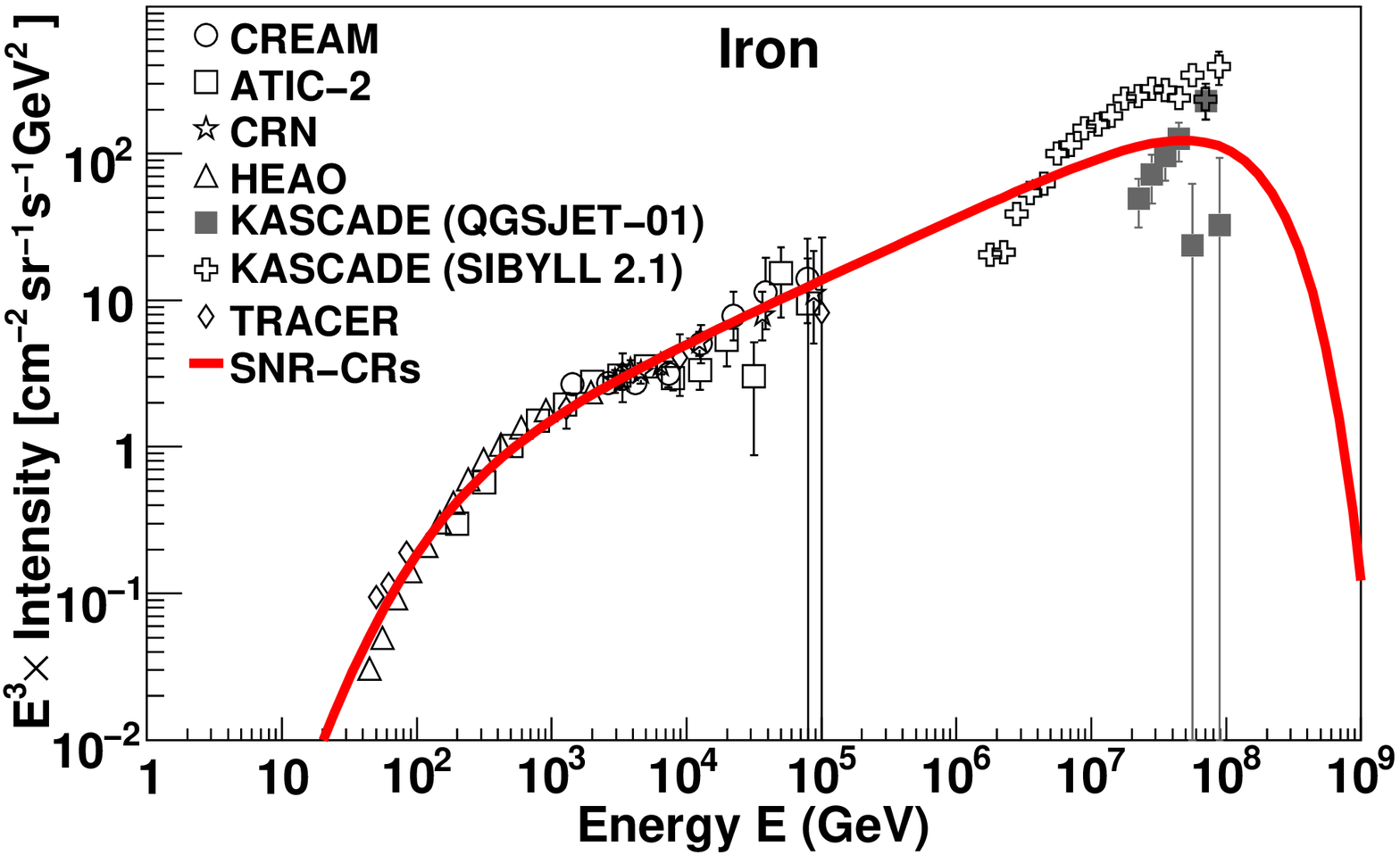}\\
\caption{\label {fig-low-energy} Energy spectra for different cosmic-ray elements. {\it Solid line}: Model prediction for the SNR-CRs. {\it Data}: CREAM \citep{Ahn2009, Yoon2011}, ATIC-2 \citep{Panov2007}, AMS-02 \citep{Aguilar2015a, Aguilar2015b}, PAMELA \citep{Adriani2011}, CRN \citep{Mueller1991, Swordy1990}, HEAO \citep{Engelmann1990}, TRACER \citep{Obermeier2011}, and KASCADE \citep{Antoni2005}. Cosmic-ray source parameters ($q,f$) used in the calculation are given in Table \ref{table-SNR-CRs}. For the other model parameters $(D_0, a, \eta, s)$, see text for details.}
\end{figure*}

The steady-state transport equation for cosmic-ray nuclei in the Galaxy in the re-acceleration model is described by,
\begin{align}
\label{eq-transport}
\nabla\cdot(D&\nabla N)-\left[\bar{n} v\sigma+\xi\right]\delta(z)N\nonumber\\
&+\left[\xi sp^{-s}\int^p_{p_0}du\;N(u)u^{s-1}\right]\delta(z)=-Q\delta(z),
\end{align}
where we have adopted a cylindrical geometry for the propagation region described by the radial $r$ and vertical $z$ coordinates with $z=0$ representing the Galactic plane. We assume the region to have a constant halo boundary at $z=\pm L$, and no boundary in the radial direction. This is a reasonable assumption for cosmic rays at the galacto-centric radius of the Sun as the majority of them are produced within a radial distance ${\sim}\,L$ from the Sun \citep{Thoudam2008}. Choosing a different (smaller) halo height for the Galactic centre region, as indicated by the observed \textit{WMAP} haze \citep{Biermann2010b}, will not produce significant effects in our present study. $N(r,z,p)$ represents the differential number density of the cosmic-ray nuclei with momentum/nucleon $p$, and $Q(r,p)\delta(z)$ is the injection rate of cosmic rays per unit volume by supernova remnants in the Galaxy. The diffusive nature of the propagation is represented by the first term in Equation \ref{eq-transport}. The diffusion coefficient $D(\rho)$ is assumed to be a function of the particle rigidity $\rho$ as, $D(\rho)=D_0\beta(\rho/\rho_0)^a$, where $D_0$ is the diffusion constant, $\beta=v/c$ with $v(p)$ and $c$ representing the velocity of the particle and the velocity of light respectively, $\rho_0=3$~GV is a constant, and $a$ is the diffusion index. The rigidity is defined as $\rho=Apc/Ze$, where $A$ and $Z$ represent the mass number and the charge number of the nuclei respectively, and $e$ is the charge of an electron. The second term in Equation \ref{eq-transport} represents the loss of particles during the propagation due to inelastic interaction with the interstellar matter, and also due to re-acceleration to higher energies, where $\bar{n}$ represents the surface density of matter in the Galactic disk, $\sigma(p)$ is the inelastic interaction cross-section, and $\xi$ corresponds to the rate of re-acceleration. We take $\xi=\eta V\bar{\nu}$, where $V=4\pi \Re^3/3$ is the volume occupied by a supernova remnant of radius $\Re$ re-accelerating the cosmic rays, $\eta$ is a correction factor that is introduced to account for the actual unknown size of the remnants, and $\bar{\nu}$ is the frequency of supernova explosions  per unit surface area in the Galactic disk. The term containing  the integral in Equation \ref{eq-transport} represents the gain in the number of particles due to re-acceleration from lower energies. The effect of Galactic wind and ionisation losses which are important mostly at low energies, below $\sim 1$~GeV/nucleon, are not included explicitly in the transport equation. Instead, we introduce a low-momentum cut-off, $p_0{\sim}\,100$ MeV/nucleon, in the particle distribution to account for the effect on the number of low-energy particles available for re-acceleration in the presence of these processes \citep{Wandel1987}. We assume that re-acceleration instantaneously produces a power-law spectrum of particles with spectral index $s$. The source term $Q(r,p)$ can be expressed as $Q(r,p)=\bar{\nu} \mathrm{H}[R-r]\mathrm{H}[p-p_0]Q(p)$, where $\mathrm{H}(m)=1 (0)$ for $m>0 (<0)$ represents a Heaviside step function, and the source spectrum $Q(p)$ is assumed to follow a power-law in total momentum with an exponential cut-off which, in terms of momentum/nucleon, can be written as 
\begin{equation}
\label{eq-source}
Q(p)=AQ_0 (Ap)^{-q}\exp\left(-\frac{Ap}{Zp_\mathrm{c}}\right),
\end{equation}
where $Q_0$ is a normalisation constant which is proportional to the amount of energy $f$ channelled into cosmic rays by a single supernova event, $q$ is the spectral index, and $p_\mathrm{c}$ is the cut-off momentum for protons. The exponential cut-off in Equation \ref{eq-source} represents a good approximation for particles at the shock produced by the diffusive shock acceleration mechanism (see e.g.  \citealp{Malkov2001}). We assume that the maximum energy for cosmic-ray nuclei produced by the supernova shock is $Z$ times the maximum energy for protons. Based on the observed high concentration of supernova remnants and atomic and molecular hydrogen near the Galactic disk, in Equation \ref{eq-transport}, we assume that both cosmic-ray sources and interstellar matter are distributed in the disk (i.e. at $z=0$). The distributions are assumed to be uniform, and extended up to a radius $R$. 
  
Recalling the analytical solution of Equation \ref{eq-transport} derived in \cite{Thoudam2014}, the cosmic-ray density at the position $r=0$ for $p>p_0$ follows,
\begin{align}
\label{eq-solution}
N&(z,p)=\bar{\nu} R\int^{\infty}_0 dk\; \frac{\sinh\left[k(L-z)\right]}{\sinh(kL)}\times \frac{\mathrm{J_1}(kR)}{B(p)}\Biggl\{Q(p)\nonumber\\
&+\xi sp^{-s}\int^p_{p_0}dp^\prime{p^\prime}^s Q(p^\prime)\mathcal{A}(p^\prime)\exp\left(\xi s\int^p_{p^\prime} \mathcal{A}(u)du\right)\Biggr\},
\end{align}
where $\mathrm{J_1}$ is a Bessel function of order 1, and the functions $B$ and $\mathcal{A}$ are given by,
\begin{align}
\label{eq-l-and-A}
&B(p)=2D(p)k\coth(kL)+\bar{n}v(p)\sigma(p)+\xi\nonumber\\
&\mathcal{A}(u)=\frac{1}{uB(u)}.
\end{align}
From Equation \ref{eq-solution}, the cosmic-ray density at the Earth can be obtained by taking $z=0$ considering that our Solar system lies close to the Galactic plane.

\begin{table}
\centering
\caption{\label{table-SNR-CRs}Source spectral indices, $q$, and energy injected per supernova, $f$, for the different species  of cosmic rays used in the calculation of the SNR-CRs spectra shown in Figures \ref{fig-low-energy} and \ref{fig-all-particle-SNR}.}
\bigskip
\begin{tabular}{c|c|c}
\hline
Particle type	&	$q$		&	$f$ ($\times 10^{49}$ ergs) \\         
\hline
Proton		&	$2.24$	&	$6.95$\\
Helium		&	$2.21$	&	$0.79$\\
Carbon		&	$2.21$	&	$2.42\times 10^{-2}$\\
Oxygen		&	$2.25$	&	$2.52\times 10^{-2}$\\
Neon		&	$2.25$	&	$3.78\times 10^{-3}$\\
Magnesium	&	$2.29$	&	$5.17\times 10^{-3}$\\
Silicon		&	$2.25$	&	$5.01\times 10^{-3}$\\
Iron			&	$2.25$	&	$4.95\times 10^{-3}$\\
\hline
\end{tabular}
\end{table}
\begin{figure*}
\centering
\includegraphics*[width=0.8\textwidth,angle=0,clip]{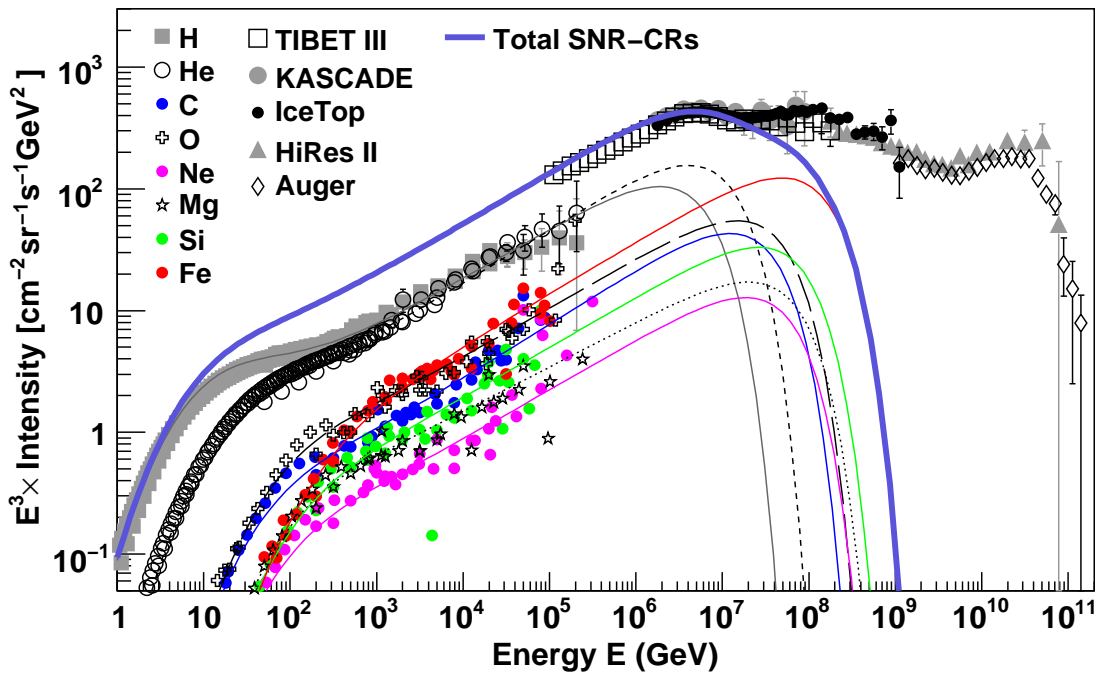}
\caption{\label {fig-all-particle-SNR} Contribution of SNR-CRs to the all-particle cosmic-ray spectrum. The thin lines represent spectra for the individual elements, and the thick-solid line represents the total contribution. The calculation assumes an exponential cut-off energy for protons at $E_\mathrm{c}=4.5\times 10^6$ GeV. Other model parameters, and the low-energy data are the same as in Figure \ref{fig-low-energy}. Error bars are shown only for the proton and helium data. High-energy data: KASCADE \citep{Antoni2005}, IceTop \citep{Aartsen2013}, Tibet III \citep{Amenomori2008}, the Pierre Auger Observatory \citep{Schulz2013}, and HiRes II \citep{Abbasi2009}.}
\end{figure*}

\subsection{Model prediction for the low-energy measurements}
\label{subsection-low-energy}
By comparing the abundance ratio of boron-to-carbon nuclei predicted by the model with the measurements, the cosmic-ray propagation parameters ($D_0,a$) and the re-acceleration parameters ($\eta,s$) have been obtained to be, $D_0=9\times 10^{28}$~cm$^2$~s$^{-1}$, $a=0.33$, $\eta=1.02$, and $s=4.5$ \citep{Thoudam2014}. We adopt these values in our present study. The supernova remnant radius is taken to be $\Re=100$~pc. The inelastic interaction cross-section for protons is taken from  \cite{Kelner2006}, and for heavier nuclei, the cross-sections are taken  from \cite{Letaw1983}. The surface matter density is taken as the averaged density in the Galactic disk within a radius equal to the size of the diffusion boundary $L$. We choose $L=5$~kpc, which gives an averaged surface density of atomic hydrogen of $\bar{n}=7.24\times 10^{20}$~atoms~cm$^{-2}$ \citep{Thoudam2013}. An extra $10\%$  is further added to $\bar{n}$ to account for the helium abundance in the interstellar medium. The radial extent of the source distribution is taken as $R=20$~kpc. Each supernova explosion is assumed to release a total kinetic energy of $10^{51}$~ergs, and the supernova explosion frequency is taken as $\bar{\nu}=25$~SNe~Myr$^{-1}$~kpc$^{-2}$. The latter corresponds to a rate of ${\sim}\,3$ supernova explosions per century in the Galaxy.

Using the values of various parameters mentioned above, the energy spectra of SNR-CRs for different elements are calculated. In Figure \ref{fig-low-energy}, results for eight elements (proton, helium, carbon, oxygen, neon, magnesium, silicon and iron, which represent the dominant species at low energies) are compared with the measured data at low energies. The source parameters $(q,f)$ for the individual elements are kept free in the calculation, and they are optimised based on the observed individual spectra at low energies. The parameter values that best reproduce the measured data are listed in Table \ref{table-SNR-CRs}. The source spectral indices are in the range of $2.21-2.29$, and out of the total of $8\%$ of the supernova explosion energy channelled into SNR-CRs, the largest fraction goes into protons at the level of $6.95\%$, followed by helium nuclei with $0.79\%$. The calculated spectra reproduce the measured data quite well including the behaviour of spectral hardening at TeV energies observed for protons and helium nuclei. In our model, the absence of such a spectral hardening for heavier nuclei is explained as due to the increasing effect of inelastic collision over re-acceleration with the increase in mass \citep{Thoudam2014}.

\subsection{Extrapolation of the SNR-CR spectrum to high energies}
\label{section-low-to-high-energy}
In Figure \ref{fig-low-energy}, we also show an extrapolation of the model prediction to high energies. For protons, helium, carbon, silicon and iron nuclei, the predictions are compared with the available measurements from the KASCADE experiment above ${\sim}\,10^6$~GeV. The calculation assumes an exponential cut-off for the proton source spectrum at $E_\mathrm{c}=4.5\times 10^{6}$~GeV, and for the heavier nuclei at $ZE_\mathrm{c}$. This value of $E_\mathrm{c}$, which is obtained by comparing the predicted all-particle spectrum with the observed all-particle spectrum as shown in Figure \ref{fig-all-particle-SNR}, represents the maximum $E_\mathrm{c}$ value permitted by the measurements. While obtaining the all-particle spectrum shown in Figure \ref{fig-all-particle-SNR}, we also include contributions from the sub-dominant primary cosmic-ray elements $(Z<26)$, calculated using elemental abundances at $10^3$~GeV given in  \cite{Hoerandel2003a} and a source index of 2.25. Their total contribution amounts up to $\sim 8\%$ of the all-particle  spectrum. The predicted all-particle spectrum agrees with the data up to ${\sim}\,2\times 10^{7}$~GeV, and reproduces the observed knee at the right position. Choosing $E_\mathrm{c}$ values larger than $4.5\times 10^{6}$~GeV will produce an all-particle spectrum which is inconsistent both with the observed knee position and the intensity above the knee. Although our estimate for the best-fit $E_\mathrm{c}$ value does not rely on the proton measurements at high energies, it can be noticed from Figure \ref{fig-low-energy} that both the predicted proton and helium spectra are in good agreement (within systematic uncertainties) with the KASCADE data. For carbon, silicon and iron nuclei, the agreement with the data  is less convincing, which may be related to the larger systematic uncertainties in the shapes of the measured spectra.

From Figure \ref{fig-all-particle-SNR}, it can be observed that, at energies around the knee, the all-particle spectrum is predicted to be dominated by helium nuclei, not by protons. The CREAM measurements have shown that helium nuclei become more abundant than protons at energies ${\sim}\,10^5$~GeV. Such a trend is also consistent with the KASCADE measurements above ${\sim}\,10^6$~GeV (see Figure \ref{fig-low-energy}). Based on our prediction, helium nuclei dominate the all-particle spectrum up to ${\sim}\,1.5\times\,10^7$ GeV, while above, iron nuclei dominate. The maximum energy of SNR-CRs, which corresponds to the fall-off energy of iron nuclei, is $26\times E_\mathrm{c}=1.2\times\,10^8$ GeV. Although this energy is close to the position of the second knee, the predicted intensity is not  enough to explain the observed intensity around the second knee. Our result shows that SNR-CRs alone cannot account for the observed cosmic rays above ${\sim}\,2\times\,10^7$~GeV. At $10^8$ GeV, they contribute only ${\sim}\,30\%$ of the observed data.

\section{Additional component of Galactic cosmic rays}
\label{sec-additional-component}
Despite numerous studies, it is not clearly understood at what energy the transition from Galactic to extra-galactic cosmic rays (EG-CRs) occurs. Although it was pointed out soon after the discovery of the CMB and the related GZK effect that it is possible to construct an all-extra-galactic spectrum of cosmic rays containing both the knee and the ankle as features of cosmological propagation \citep{Hillas1967}, the most natural explanation was assumed to be that the transition occurs at the ankle, where a steep Galactic component is taken over by a flatter extra-galactic one. To obtain a sharp feature like the ankle in such a construction, it is necessary to assume a cut-off in the Galactic component to occur immediately below it \citep{Rachen1993, Axford1994}, thus this scenario is naturally expecting a second knee feature. For a typical Galactic magnetic field strength of $3$~$\mu$G, the Larmor radii for cosmic rays of energy $Z\times 10^8$~GeV is $36$~pc, much smaller than the size of the diffusion halo of the Galaxy, which is typically considered to be a few kpc in cosmic-ray propagation studies, keeping comic rays around the second knee well confined in the Galaxy. This suggests that the Galactic cut-off at this energy must be intrinsic to a source population or acceleration mechanism different from the standard supernova remnants we have discussed above. In an earlier work, \cite{Hillas2005} considered an additional Galactic component resulting from Type II supernova remnants in the Galaxy expanding into a dense slow wind of the precursor stars. In the following, we discuss two other possible scenarios. The first is the re-acceleration of SNR-CRs by Galactic wind termination shocks in the Galactic halo \citep{Jokipii1987, Zirakashvili2006}, and the second is the contribution of cosmic rays from the explosions of Wolf-Rayet stars in the Galaxy \citep{Biermann1993}. Both these ideas have been explored in the past when detailed measurements of the cosmic-ray spectrum and composition at low and high energies were not available. Using new measurements of cosmic rays and astronomical data (like the Wolf-Rayet wind composition), our study can provide a more realistic estimate of the cosmic-ray contribution from these two possible mechanisms. In the following, the re-accelerated cosmic rays from Galactic wind termination shocks will be referred to as `GW-CRs', and cosmic rays from Wolf-Rayet stars as `WR-CRs'. Some ramifications of these basic scenarios will be discussed in Section \ref{sec-discussion}, after investigating the effect of different extra-galactic contributions below the ankle in Section \ref{section-EG-models}.

\subsection{Re-acceleration of SNR-CRs by Galactic wind termination shocks (GW-CRs)}
\label{section-galactic-wind}
The effect of Galactic winds on the transport of cosmic rays in the Galaxy has been discussed quite extensively \citep{Lerche1982a, Bloemen1993, Strong1998, Jones2001, Breit2002}. For cosmic rays produced by sources in the Galactic disk such as the SNR-CRs, the effect of winds on their transport is expected to be negligible above a few GeV as the transport is expected to be dominated mainly by the diffusion process. However, Galactic winds can lead to the production of an additional component of cosmic rays which can dominate at high energies. Galactic winds, which start at a typical velocity of about few km/s near the disk, reach supersonic speeds at distances of a few tens of kpc away from the disk. At about a hundred kpc distance or so, the wind flow terminates resulting into the formation of termination shocks. These shocks can catch the SNR-CRs escaping from the disk into the Galactic halo, and re-accelerate them via the diffusive shock acceleration process. The reaccelerated cosmic rays can return to the disk through diffusive propagation against the Galactic wind outflow. For an energy dependent diffusion process, only the high-energy particles may be effectively able to reach the disk. 
\begin{figure}
\centering
\includegraphics*[width=\columnwidth,height=0.65\columnwidth,angle=0,clip]{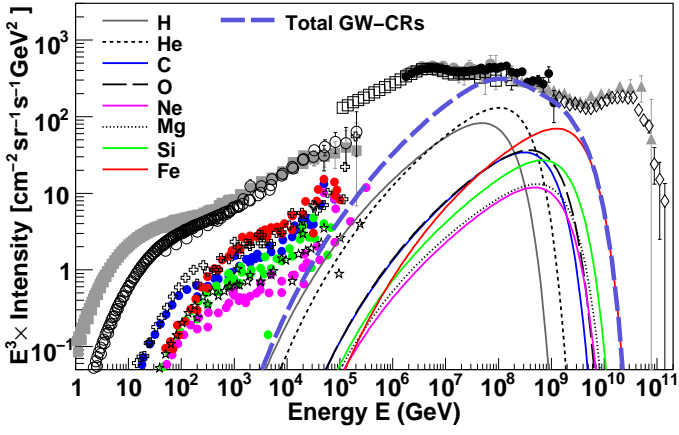}
\caption{\label {fig-spectrum-only-wind} Contribution of GW-CRs to the all-particle cosmic-ray spectrum. The thin lines represent spectra for the individual elements, and the thick dashed line represents the total contribution. The injection fraction, $k_\mathrm{w}=14.5\%$, and the exponential cut-off energy for protons, $E_\mathrm{sh}=9.5\times 10^7$ GeV. See text for the other model parameters. Data are the same as in Figure \ref{fig-all-particle-SNR}.}
\end{figure}

To obtain the contribution of GW-CRs, we will first calculate the escape rate of SNR-CRs from the {\textit{inner}} diffusion boundary, then propagate the escaped cosmic rays through the Galactic wind region, and calculate the cosmic-ray flux injected into the Galactic wind termination shocks. The escaped flux of SNR-CRs from the diffusion boundary, $F_\mathrm{esc}$, can be calculated as,
\begin{equation}
\label{eq-fesc}
F_\mathrm{esc}=\left[D\nabla N\right]_{z=\pm L}=\left[D\frac{dN}{dz}\right]_{z=\pm L},
\end{equation}
where $N(z,p)$ is given by Equation \ref{eq-solution}. Equation \ref{eq-fesc} assumes that cosmic rays escape only through the diffusion boundaries located at $z=\pm L$. Under this assumption, the total escape rate of SNR-CRs is given by,
\begin{equation}
\label{eq-qesc}
Q_\mathrm{esc}=F_\mathrm{esc}\times 2A_\mathrm{esc},
\end{equation}
where $A_\mathrm{esc}=\pi R^2$ is the surface area of one side of the cylindrical diffusion boundary which is assumed to have the same radius as the Galactic disk, and the factor $2$ is to account for the two boundaries at $z=\pm L$. The propagation of the escaped SNR-CRs in the Galactic wind region is governed by the following transport equation:
\begin{equation}
\label{eq-transport-wind}
\nabla.(D_\mathrm{w}\nabla N_\mathrm{w}-\textbf{V}N_\mathrm{w})+\frac{\partial}{\partial p}\left\lbrace\frac{\nabla.\textbf{V}}{3}pN_\mathrm{w}\right\rbrace=-Q_\mathrm{esc}\delta(\textbf{r}),
\end{equation}
where we have assumed a spherically symmetric geometry characterised by the radial variable $r$, $D_\mathrm{w}$ represents the diffusion coefficient of cosmic rays in the wind region which is taken to be spatially constant, $N_\mathrm{w}(r,p)$ is the cosmic-ray number density, $\textbf{V}=\tilde{V}r\hat{r}$ is the wind velocity which is assumed to increase linearly with $r$ and directed radially outwards, $\tilde{V}$ is a constant that denotes the velocity gradient, and $Q_\mathrm{esc}(p)$ is given by Equation \ref{eq-qesc}. The exact nature of the Galactic wind is not known. The spatial dependence of the wind velocity considered here is based on the model of magnetohydrodynamic wind driven by cosmic rays, which shows that the wind velocity increases linearly with distance from the Galactic disk until it reaches an asymptotic value at a distance of around $100$~kpc \citep{Zirakashvili1996}. The second term on the left-hand side of Equation \ref{eq-transport-wind} represents the loss of particles due to advection by the Galactic wind, and the third term represents momentum loss due to the adiabatic expansion of the wind flow which is assumed to be spherically symmetric. In writing Equation \ref{eq-transport-wind}, considering that the size of the wind region is much larger than the size of the escaping region of the SNR-CRs, we neglect the size of the escaping region and consider $Q_\mathrm{esc}$ to be a point source located at $r=0$. By solving Equation \ref{eq-transport-wind} analytically, the density of cosmic rays at distance $r$ is given by (see Appendix \ref{appendix-A}),
\begin{align}
\label{eq-solution-wind}
N_\mathrm{w}(r,p)=\frac{\sqrt{\tilde{V}}p^2}{8\pi^{3/2}}\int^\infty_0 &dp^\prime \frac{Q_\mathrm{esc}(p^\prime)}{\left[{\int^{p^\prime}_p u D_\mathrm{w}(u)du}\right]^{3/2}}\nonumber\\
&\times \exp\left({-\frac{r^2\tilde{V} p^2}{4\int^{p^\prime}_p u D_\mathrm{w}(u)du}}\right).
\end{align}
From Equation \ref{eq-solution-wind}, the cosmic-ray flux with momentum/nucleon $p$ at the termination shock is obtained as,
\begin{equation}
\label{eq-finj}
F_\mathrm{w}(p)=\left [-D_\mathrm{w}\frac{\partial N_\mathrm{w}}{\partial r}+\textbf{V}N_\mathrm{w}\right ]_{r=R_\mathrm{sh}},
\end{equation}
where $R_\mathrm{sh}$ represents the radius of the termination shock. The total rate of cosmic rays injected into the termination shock is given by,
\begin{equation}
Q_\mathrm{inj}(p)=F_\mathrm{w}(p)\times A_\mathrm{sh},
\end{equation}
where $A_\mathrm{sh}=4\pi R_\mathrm{sh}^2$ is the surface area of the termination shock. Assuming that only a certain fraction, $k_\mathrm{sh}$, participates in the re-acceleration process, the cosmic-ray spectrum produced by the termination shock under the test particle approximation can be written as \citep{Drury1983},
\begin{equation}
Q_\mathrm{sh}(p)=\gamma p^{-\gamma} \exp\left(-\frac{Ap}{Zp_\mathrm{sh}}\right)\int^p_{p_0} k_\mathrm{sh} Q_\mathrm{inj}(u)u^{\gamma-1} du,
\end{equation}
where we have introduced an exponential cut-off in the spectrum at momentum $Z p_\mathrm{sh}$ with $p_\mathrm{sh}$ representing the maximum momentum for protons, and $\gamma$ is the spectral index. In our calculation, $p_\mathrm{sh}$ and $k_\mathrm{sh}$ will be kept as model parameters, and their values will be determined based on the measured all-particle spectrum.
\begin{table}
\centering
\caption{\label{table-C-to-He}Relative abundances of different cosmic-ray species with respect to helium for two different Wolf-Rayet wind compositions used in our model \citep{Pollock2005}.}
\bigskip
\begin{tabular}{c|c|c}
\hline
Particle type	&	$\mathrm{C/He}=0.1$		&	$\mathrm{C/He}=0.4$ \\         
\hline
Proton		&	$0$	&	$0$\\
Helium		&	$1.0$	&	$1.0$\\
Carbon		&	$0.1$	&	$0.4$\\
Oxygen		&	$3.19\times 10^{-2}$	&	$7.18\times 10^{-2}$\\
Neon		&	$0.42\times 10^{-2}$	&	$1.03\times 10^{-2}$\\
Magnesium		&	$2.63\times 10^{-4}$	&	$6.54\times 10^{-4}$\\
Silicon		&	$2.34\times 10^{-4}$	&	$5.85\times 10^{-4}$\\
Iron		&	$0.68\times 10^{-4}$	&	$1.69\times 10^{-4}$\\
\hline
\end{tabular}
\end{table}

After re-acceleration, the transport of cosmic-rays from the termination shock towards the Galactic disk also follows Equation \ref{eq-transport-wind}. In the absence of adiabatic losses, the density of re-accelerated cosmic rays at the  Earth (taken to be at $r=0$) is given by,
\begin{equation}
\label{eq-sol-GW-CRs}
N_\mathrm{GW-CRs}(p)=\frac{Q_\mathrm{sh}}{4\pi D_\mathrm{w}R_\mathrm{sh}} \exp\left[-\frac{\tilde{V}R^2_\mathrm{sh}}{2D_\mathrm{w}}\right]
\end{equation}

The diffusion in the wind region is assumed to be much faster than near the Galactic disk as the level of magnetic turbulence responsible for particle scattering is expected to decrease with the distance away from the Galactic disk. We assume $D_\mathrm{w}$ to follow the same rigidity dependence as $D$, and take $D_\mathrm{w}=10 D$. For the wind velocity, we take the velocity gradient $\tilde{V}=15$~km/s/kpc. This value of $\tilde{V}$ is within the range predicted in an earlier study using an advection-diffusion propagation model \citep{Bloemen1993}, but slightly larger than the constraint given in \cite{Strong1998}. It may be noted that as long as both $D_\mathrm{w}$ and $\tilde{V}$ are within a reasonable range, it is not their individual values that is important in determining the flux of GW-CRs, but their ratio $\tilde{V}/D_\mathrm{w}$, as can be seen from Equation \ref{eq-sol-GW-CRs}. The larger this ratio, the more the flux will be suppressed, and vice-versa.

The distance to the termination shock can be estimated by balancing the Galactic wind ram pressure, $P_\mathrm{w}=\rho V^2_\mathrm{t}$, against the intergalactic pressure, $P_\mathrm{IGM}$, at the position of the termination shock, where $\rho$ is the mass density of the wind and $V_\mathrm{t}=\tilde{V} R_\mathrm{sh}$ represents the terminal velocity of the wind. The ram pressure is related to the total mechanical luminosity of the wind at the termination shock as, $L_\mathrm{w}=2\pi R^2_\mathrm{sh} P_\mathrm{w} V_\mathrm{t}$. Using this, we obtain,
\begin{equation}
\label{eq-shock-radius}
R_\mathrm{sh}=\left(\frac{L_\mathrm{w}}{2\pi P_\mathrm{IGM} \tilde{V}}\right)^{1/3}.
\end{equation}
For Galactic wind driven by cosmic rays \citep{Zirakashvili1996}, the total mechanical luminosity of the wind cannot be larger than the total power of the cosmic rays. From Section \ref{subsection-low-energy}, the total power invested in SNR-CRs (which dominates the overall cosmic-ray energy density in our model) is ${\sim}\,8\%$ of the mechanical power injected by supernova explosions in the Galaxy. This corresponds to a total power of ${\sim}\,8\times 10^{40}$ ergs s$^{-1}$ injected into SNR-CRs. Using this, and taking an intergalactic pressure of $P_\mathrm{IGM}=10^{-15}$ ergs cm$^{-3}$ \citep{Breit1991}, we obtain $R_\mathrm{sh}=96$~kpc from Equation \ref{eq-shock-radius}. The spectral indices $\gamma$ are taken to be the same as the source indices of the SNR-CRs listed in Table \ref{table-SNR-CRs}. Having fixed these parameter values, the spectra of the GW-CRs calculated using Equation \ref{eq-sol-GW-CRs} are shown in Figure \ref{fig-spectrum-only-wind}. Spectra for the individual elements and also the total contribution are shown. The same particle injection fraction of $k_\mathrm{sh}=14.5\%$ is applied to all the elements, and the maximum proton energy corresponding to $p_\mathrm{sh}$ is taken as $E_\mathrm{sh}=9.5\times 10^7$ GeV. These values are chosen so that the total GW-CR spectrum reasonably agrees with the observed all-particle spectrum between ${\sim}\,10^8$ and $10^9$~GeV. 

\begin{figure}
\centering
\includegraphics*[width=\columnwidth,height=0.65\columnwidth,angle=0,clip]{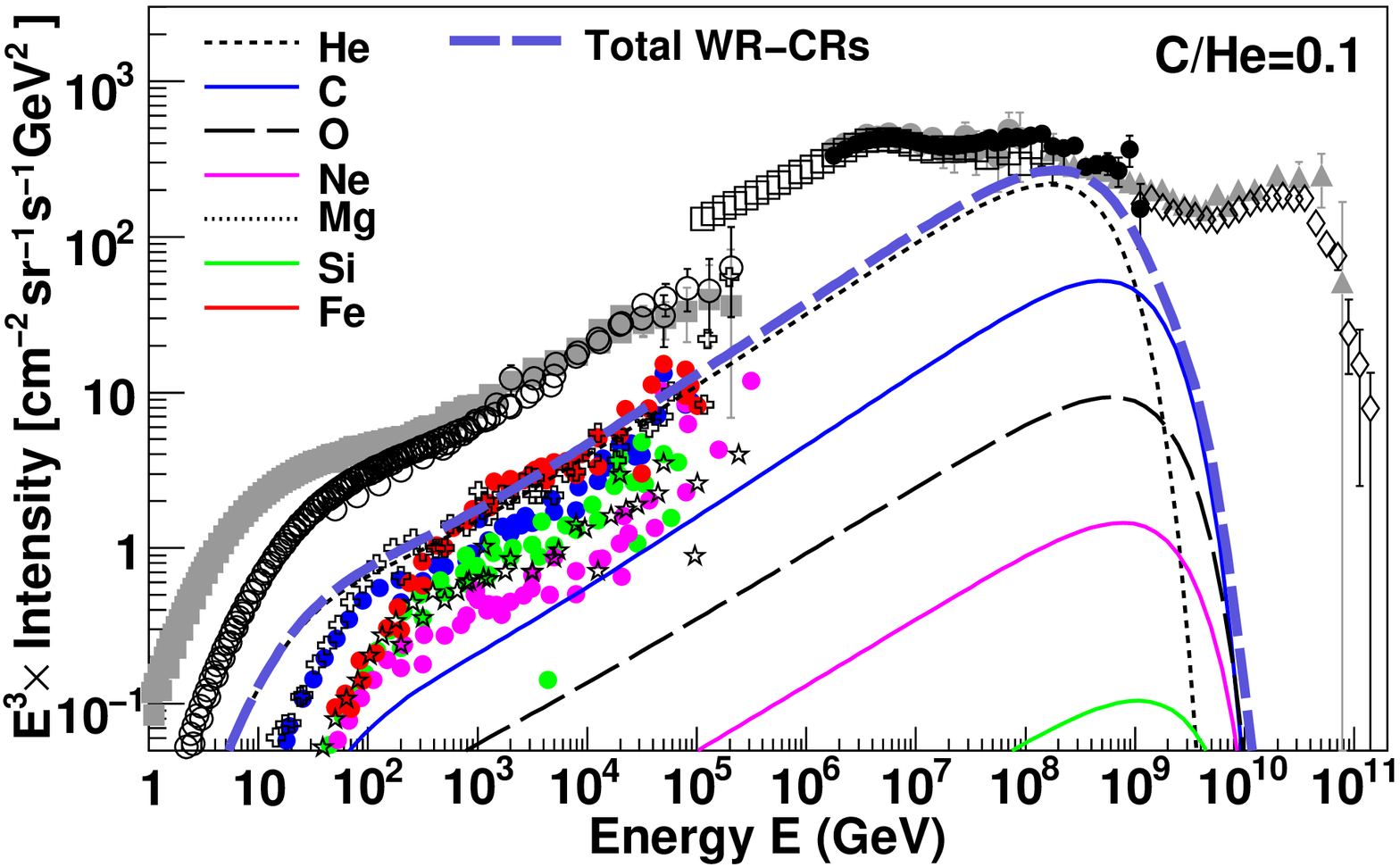}\\
\includegraphics*[width=\columnwidth,height=0.65\columnwidth,angle=0,clip]{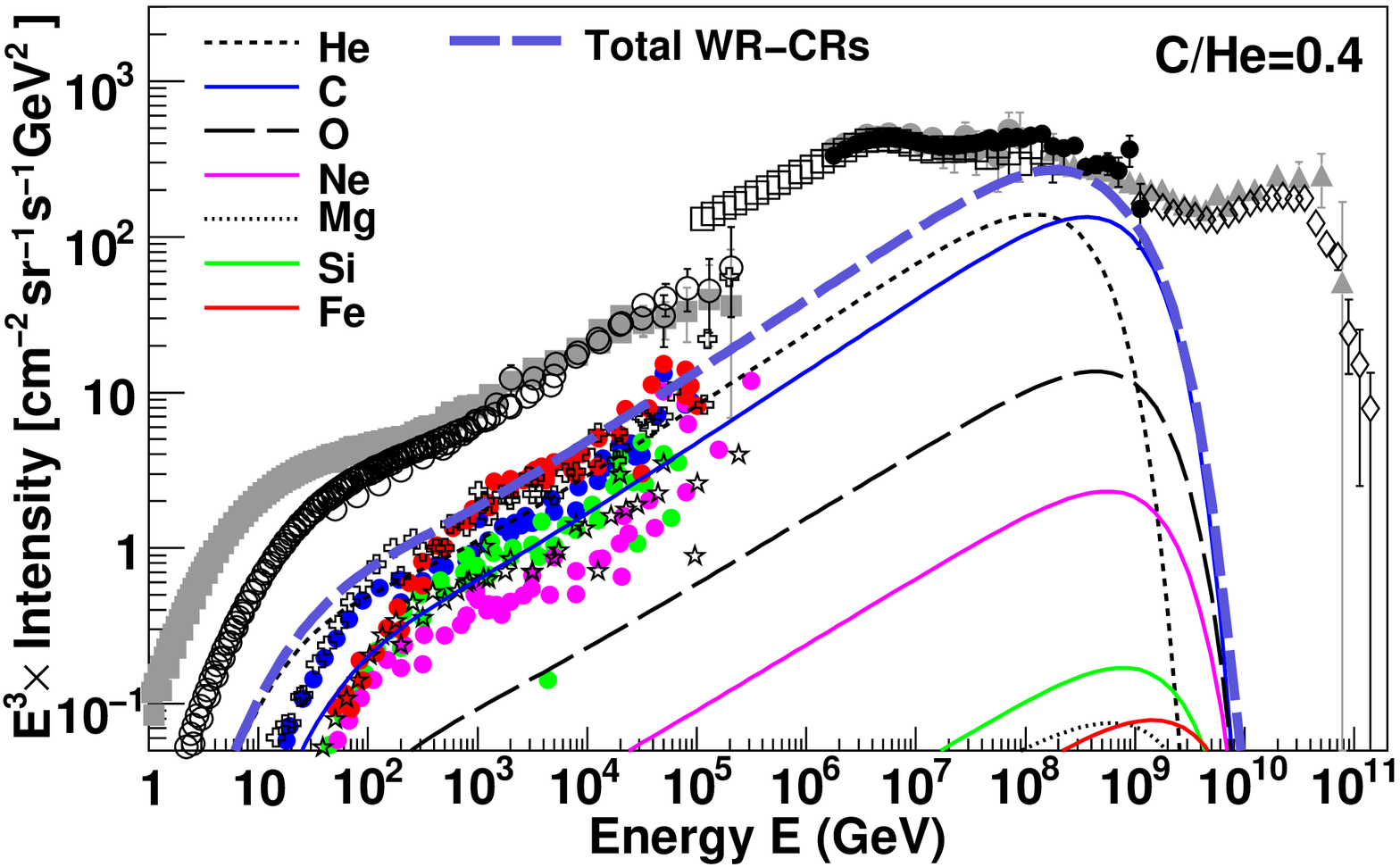}
\caption{\label {fig-spectrum-only-WR} Contribution of WR-CRs to the all-particle spectrum. {\it Top}: $\mathrm{C/He=0.1}$. {\it Bottom}: $\mathrm{C/He}=0.4$. The thin lines represent spectra for the individual elements, and the thick dashed line represents the total contribution. The calculation assumes an exponential energy cut-off for protons at $E_\mathrm{c}=1.8\times 10^8$~GeV for $\mathrm{C/He=0.1}$, and $E_\mathrm{c}=1.3\times 10^8$~GeV for $\mathrm{C/He=0.4}$. See text for the other model parameters. Data: same as in Figure \ref{fig-all-particle-SNR}.}
\end{figure}

The GW-CRs produce a negligible contribution at low energies. This is due to the increasing effect of advection over diffusion at these energies, preventing particles from reaching the Galactic disk. Higher energy particles, which diffuse relatively faster, can overcome the advection and reach the disk more effectively. The flux suppression at low energies is more significant for heavier nuclei like iron which is due to their slower diffusion relative to lighter nuclei at the same total energy. Adding adiabatic losses to Equation \ref{eq-sol-GW-CRs} will lead to further suppression of the flux at low energies. But, at energies of our interest, that is above ${\sim}\,10^7$~GeV, the result will not be significantly affected as the particle diffusion time, $t_\mathrm{dif}= R^2_\mathrm{sh}/(6D_\mathrm{w}$), is significantly less than the adiabatic energy loss time, $t_\mathrm{ad}=1/\tilde{V}=6.52\times 10^7$~yr. The steep spectral cut-offs at high energies are due to the exponential cut-offs introduced in the source spectra.
\begin{table}
\centering
\caption{\label{table-SNR-CRs-WR}Injection energy of SNR-CRs used in the calculation of all-particle spectrum in the WR-CR  model (Figure \ref{fig-total-spectrum-WR}).}
\bigskip
\begin{tabular}{c|c|c}
\hline
Particle type	&	$\mathrm{C/He}=0.1$		&	$\mathrm{C/He}=0.4$ \\
		&	$f(\times 10^{49}$ ergs)		&       $f(\times 10^{49}$ ergs)\\
\hline
Proton		&	$8.11$				&	$8.11$\\
Helium		&	$0.67$				&	$0.78$\\
Carbon		&	$2.11\times 10^{-2}$		&	$0.73\times 10^{-2}$\\
Oxygen		&	$2.94\times 10^{-2}$		&	$2.94\times 10^{-2}$\\
Neon		&	$4.41\times 10^{-3}$		&	$4.41\times 10^{-3}$\\
Magnesium	&	$6.03\times 10^{-3}$		&	$6.03\times 10^{-3}$\\
Silicon		&	$5.84\times 10^{-3}$		&	$5.84\times 10^{-3}$\\
Iron			&	$5.77\times 10^{-3}$		&	$5.77\times 10^{-3}$\\
\hline
\end{tabular}
\end{table}
\begin{figure*}
\centering
\includegraphics*[width=0.8\textwidth,angle=0,clip]{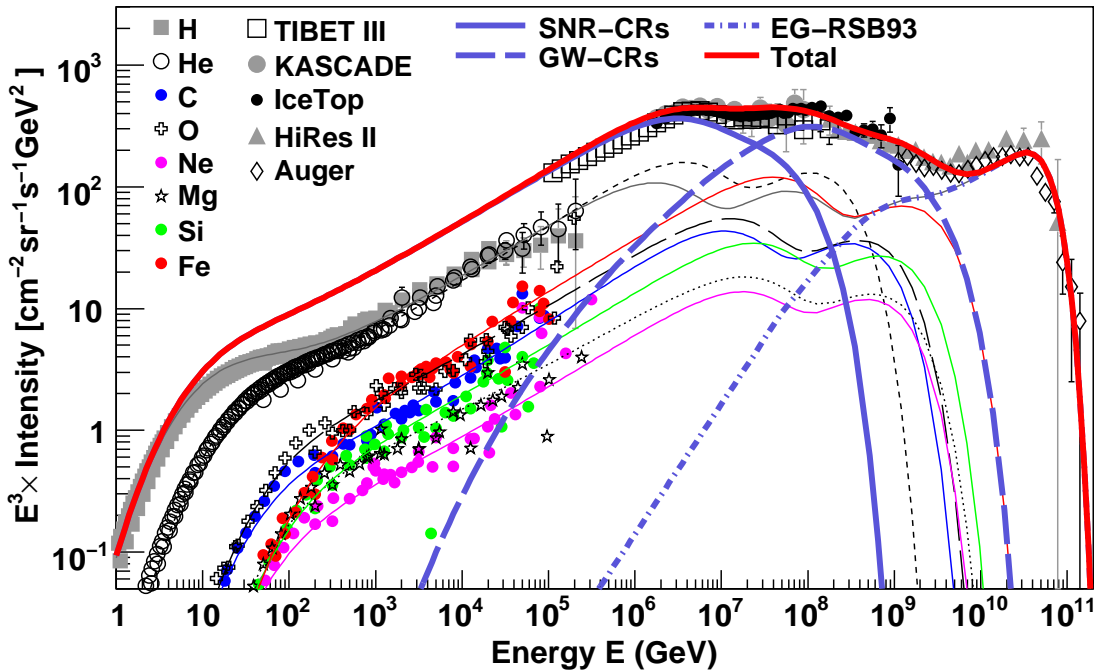}
\caption{\label {fig-total-spectrum-wind} Model prediction for the all-particle spectrum using the Galactic wind re-acceleration model. The thick solid blue line represents the total SNR-CRs, the thick dashed line represents GW-CRs, the thick dotted-dashed line represents extra-galactic cosmic rays (EG-RSB93) taken from \cite{Rachen1993}, and the thick solid red line represents the total all-particle spectrum. The thin lines represent total spectra for the individual elements. For the SNR-CRs, an exponential energy cut-off for protons at $E_\mathrm{c}=3\times 10^6$ GeV is assumed. See text for the other model parameters. Data are the same as in Figure \ref{fig-all-particle-SNR}.}
\end{figure*}

\subsection{Cosmic rays from Wolf-Rayet star explosions (WR-CRs)}
\label{subsection-WR}
While the majority of the supernova explosions in the Galaxy occur in the interstellar medium, a small fraction is expected to occur in the winds of massive progenitors like Wolf-Rayet stars \citep{Gal-Yam2014}. Magnetic fields in the winds of Wolf-Rayet stars can reach of the order of $100$~G, and it has been argued that a strong supernova shock in such a field can lead to particle acceleration of energies up to ${\sim}\,3\times 10^{9}$~GeV \citep{Biermann1993, Stanev1993}.

Since the distribution of Wolf-Rayet stars in the Galaxy is concentrated close to the Galactic disk (see e.g. \cite{Rosslowe2015}), the propagation of WR-CRs can also be described by Equation \ref{eq-transport} with the source term replaced by $Q(r,p)=\bar{\nu}_\mathrm{0} \mathrm{H}[R-r]\mathrm{H}[p-p_0]Q(p)$, where $\bar{\nu}_\mathrm{0}$ represents the frequency of Wolf-Rayet supernova explosions per unit surface area in the Galactic disk, and the source spectrum $Q(p)$ follows Equation \ref{eq-source}. We assume that each Wolf-Rayet supernova explosion releases a kinetic energy of $10^{51}$~ergs, same as the normal supernova explosion in the interstellar medium. From the estimated total number of Wolf-Rayet stars of ${\sim}\,1200$ in the Galaxy and an average lifetime of ${\sim}\,0.25$~Myr for these stars \citep{Rosslowe2015}, we estimate a frequency of ${\sim}\,1$ Wolf-Rayet explosion in every $210$ years. This corresponds to ${\sim}\,1$ Wolf-Rayet explosion in every $7$ supernova explosions occurring in the Galaxy. The source indices of the different cosmic-ray species and the propagation parameters for the WR-CRs are taken to be the same as for the SNR-CRs.

The contribution of WR-CRs to the all-particle spectrum is shown in Figure \ref{fig-spectrum-only-WR}. The results are for two different compositions of the Wolf-Rayet winds available in the literatures: Carbon-to-helium (C/He) ratio of $0.1$ (top panel) and $0.4$ (bottom panel), given in \cite{Pollock2005}. The abundance ratios of different elements with respect to helium for the two different wind compositions are listed in Table \ref{table-C-to-He}. In our calculation, these ratios are assumed to be proportional to the relative amount of supernova explosion energy injected into different elements. The overall normalisation of the total WR-CR spectrum and the maximum energy of the proton source spectrum are taken as free parameters. Their values are determined based on the observed all-particle spectrum between ${\sim}\,10^8$ and $10^9$~GeV. For $\mathrm{C/He}=0.1$, we obtain an injection energy of $1.3\times 10^{49}$~ergs into helium nuclei from a single supernova explosion and a proton source spectrum cut-off of $1.8\times 10^8$~GeV, while for $\mathrm{C/He}=0.4$, we obtain $9.4\times 10^{48}$~ergs and $1.3\times 10^8$~GeV respectively. For both the progenitor wind compositions, the total amount of energy injected into cosmic rays by a single supernova explosion is approximately $5$ times less than the total energy injected into SNR-CRs by  a supernova explosion in the Galaxy. The total WR-CR spectrum for the $\mathrm{C/He}=0.1$ case is dominated by helium nuclei up to ${\sim}\,10^9$~GeV, while for the $\mathrm{C/He}=0.4$ case, helium nuclei dominate up to ${\sim}\,2\times 10^8$~GeV. At higher energies, carbon nuclei dominate. One major difference of the WR-CR spectra from the GW-CR spectrum (Figure \ref{fig-spectrum-only-wind}) is the absence of the proton component, and a very small contribution of the heavy elements like magnesium, silicon and iron. Another major difference is the much larger flux of WR-CRs than the GW-CRs below ${\sim}\,10^5$~GeV. Below the knee, the total WR-CR spectrum is an order of magnitude less than the total SNR-CRs spectrum (Figure \ref{fig-all-particle-SNR}).
\begin{figure*}
\centering
\includegraphics*[width=0.8\textwidth,angle=0,clip]{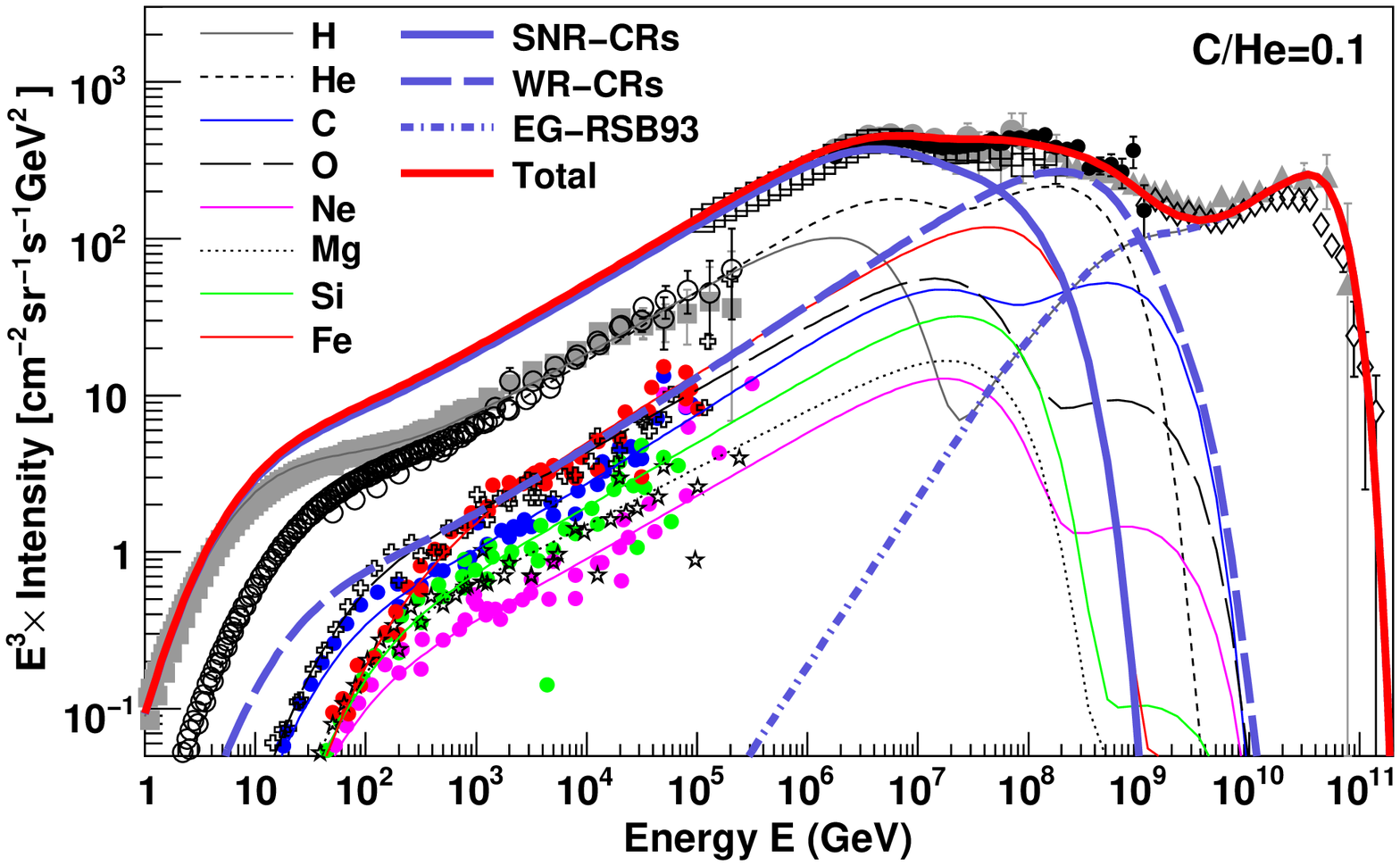}\\
\includegraphics*[width=0.8\textwidth,angle=0,clip]{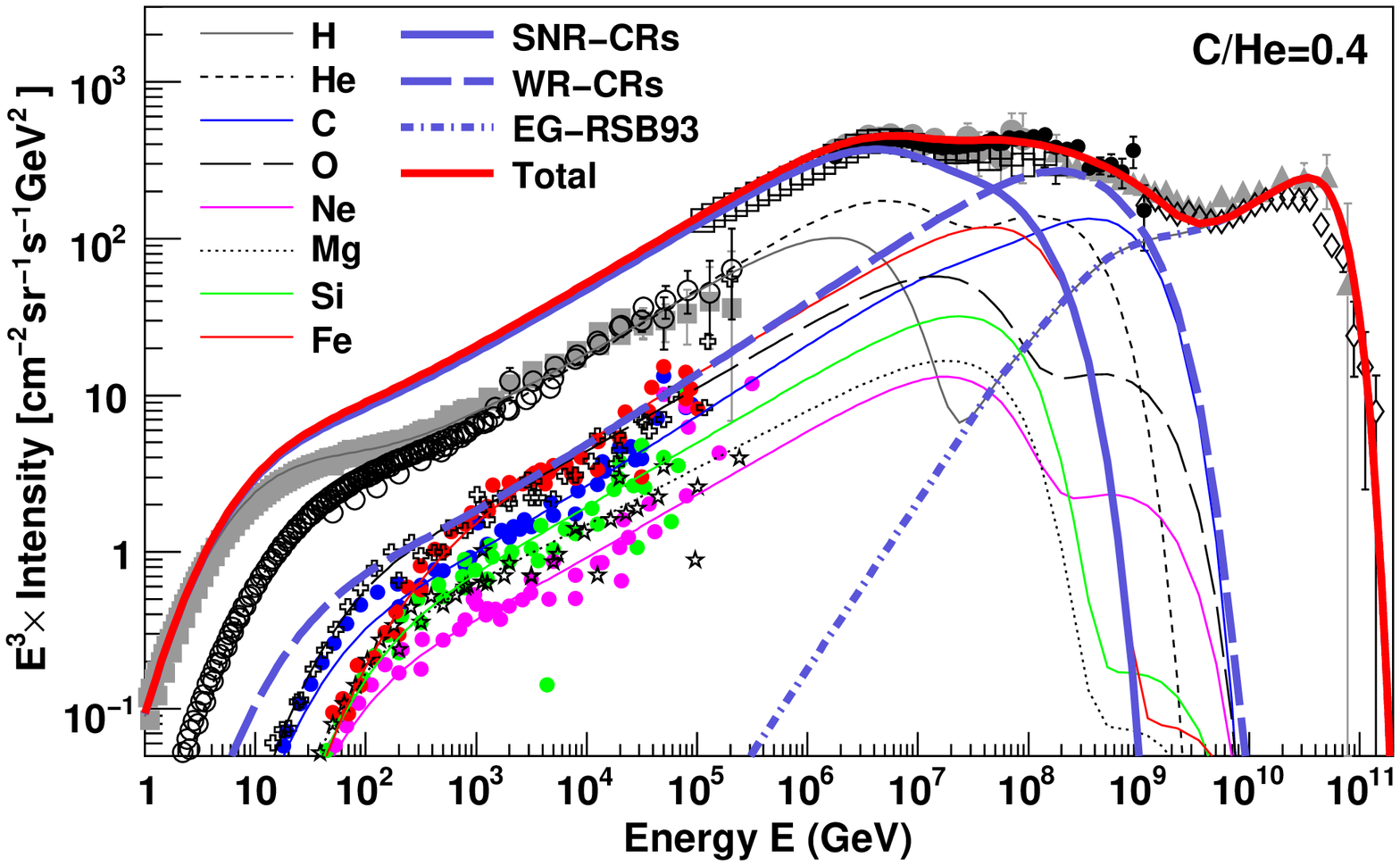}
\caption{\label {fig-total-spectrum-WR} Model prediction for the all-particle spectrum using the Wolf-Rayet stars model. {\it Top}: $\mathrm{C/He}=0.1$. {\it Bottom}: $\mathrm{C/He}=0.4$. The thick solid blue line represents the total SNR-CRs, the thick dashed line represents WR-CRs, the thick dotted-dashed line represents extra-galactic cosmic rays (EG-RSB93) taken from \cite{Rachen1993}, and the thick solid red line represents the total all-particle spectrum. The thin lines represent total spectra for the individual elements. For the SNR-CRs, an exponential energy cut-off for protons at $E_\mathrm{c}=4.1\times 10^6$ GeV is assumed. See text for the other model parameters. Data are the same as in Figure \ref{fig-all-particle-SNR}.}
\end{figure*}

\section{All-particle spectrum and composition of cosmic rays at high energies}
\label{sec-total-spectrum}
The all-particle spectrum obtained by combining the contributions of SNR-CRs, GW-CRs and EG-CRs is compared with the measured data in Figure \ref{fig-total-spectrum-wind}. For the SNR-CRs shown in the figure, we have slightly reduced the value of $E_\mathrm{c}$ from $4.5\times10^6$~GeV (as used in Figure \ref{fig-all-particle-SNR}) to $3\times10^6$ GeV in order to reproduce the measurements better around the knee. The extra-galactic contribution, denoted by EG-RSB93 in the figure, is taken from \cite{Rachen1993}, which represents a pure proton population with a source spectrum of $E^{-2}$ and an exponential cut-off at $10^{11}$~GeV as expected from strong radio galaxies or sources with a similar cosmological evolution. Also shown in the figure are the spectra of the individual elements. The model prediction reproduces the observed elemental spectra as well as the observed features in the all-particle spectrum.

The total spectra for the two WR-CR scenarios are shown in Figure \ref{fig-total-spectrum-WR}. For the SNR-CRs, here we take $E_\mathrm{c}=4.1\times10^6$~GeV, and a slightly lower value of $\nu$ which corresponds to $6$ out of every $7$ supernova explosions in the Galaxy (assuming a fraction $1/7$ going into Wolf-Rayet supernova explosions as deduced in the previous section). The injection energy $f$ for the different elements of the SNR-CRs has been re-adjusted accordingly, so that the sum of SNR-CRs and WR-CRs for the individual elements agree with the measured elemental spectra at low energies. The $f$ values are listed in Table \ref{table-SNR-CRs-WR}. The cosmic-ray propagation parameters are the same as in Figure \ref{fig-all-particle-SNR}. The predicted all-particle spectra are in good agreement with the measurements. The WR-CR scenarios are found to reproduce the second knee and the ankle better than the GW-CR model.
\begin{figure}
\centering
\includegraphics*[width=\columnwidth,angle=0,clip]{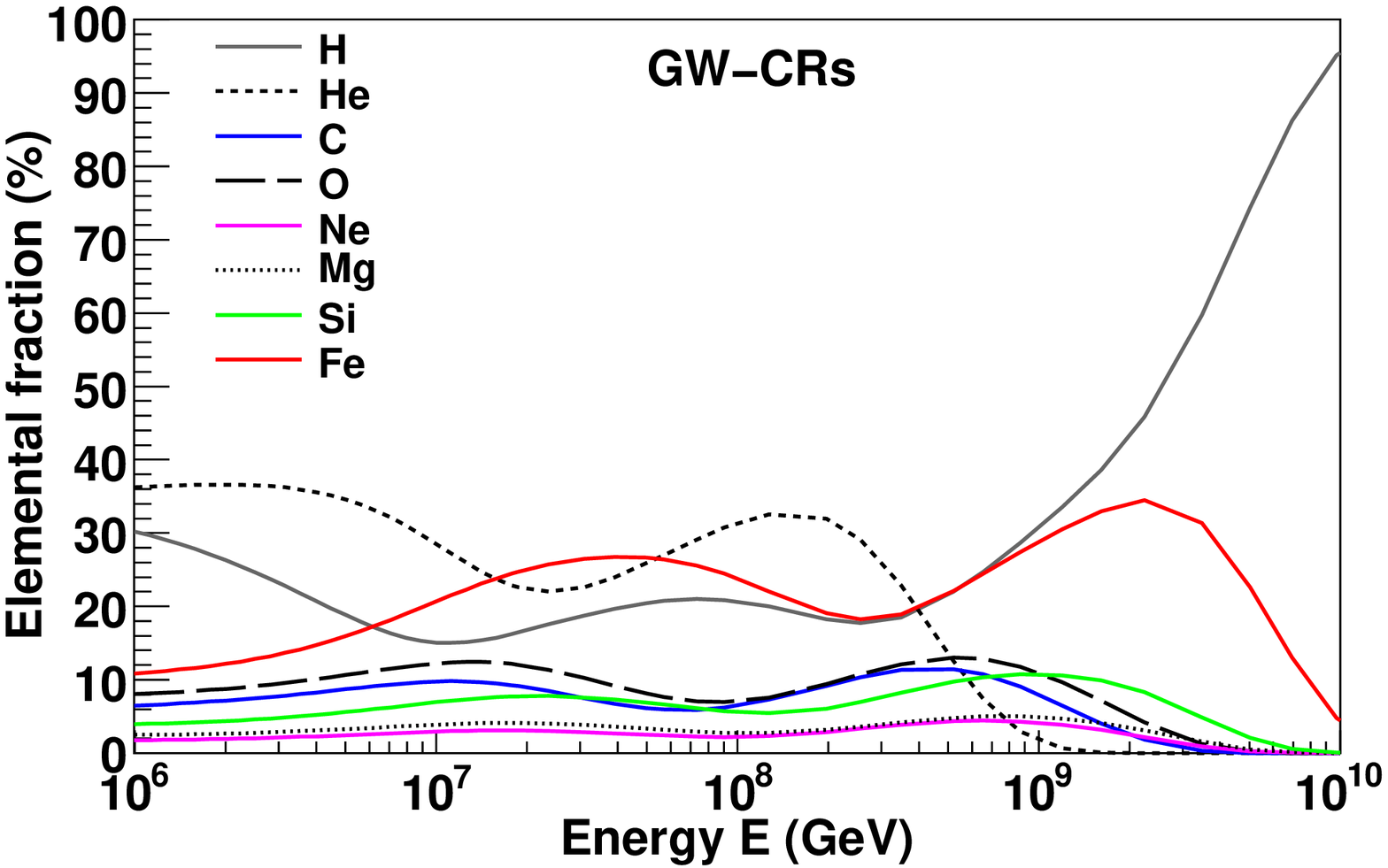}\\
\includegraphics*[width=\columnwidth,angle=0,clip]{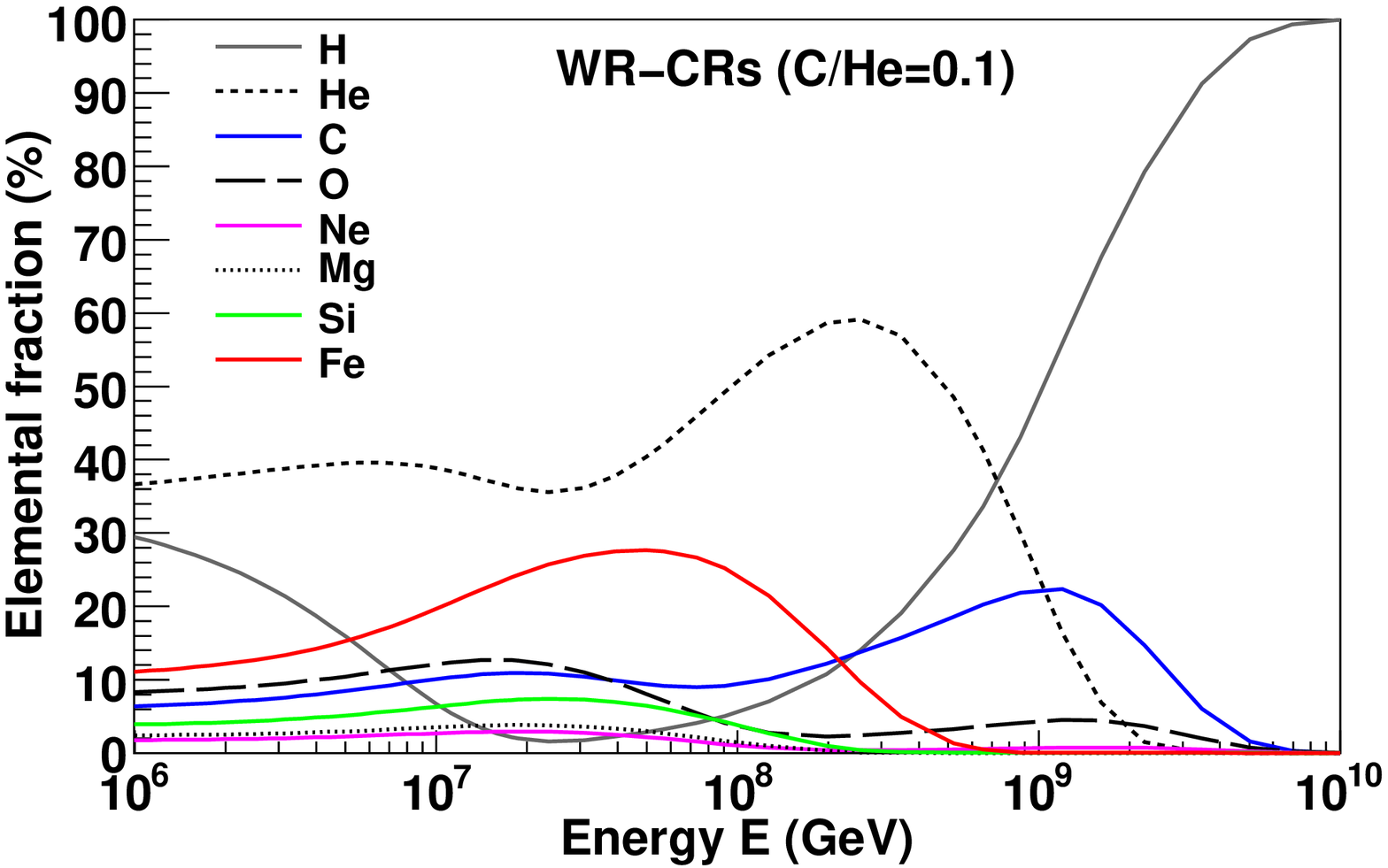}\\
\includegraphics*[width=\columnwidth,angle=0,clip]{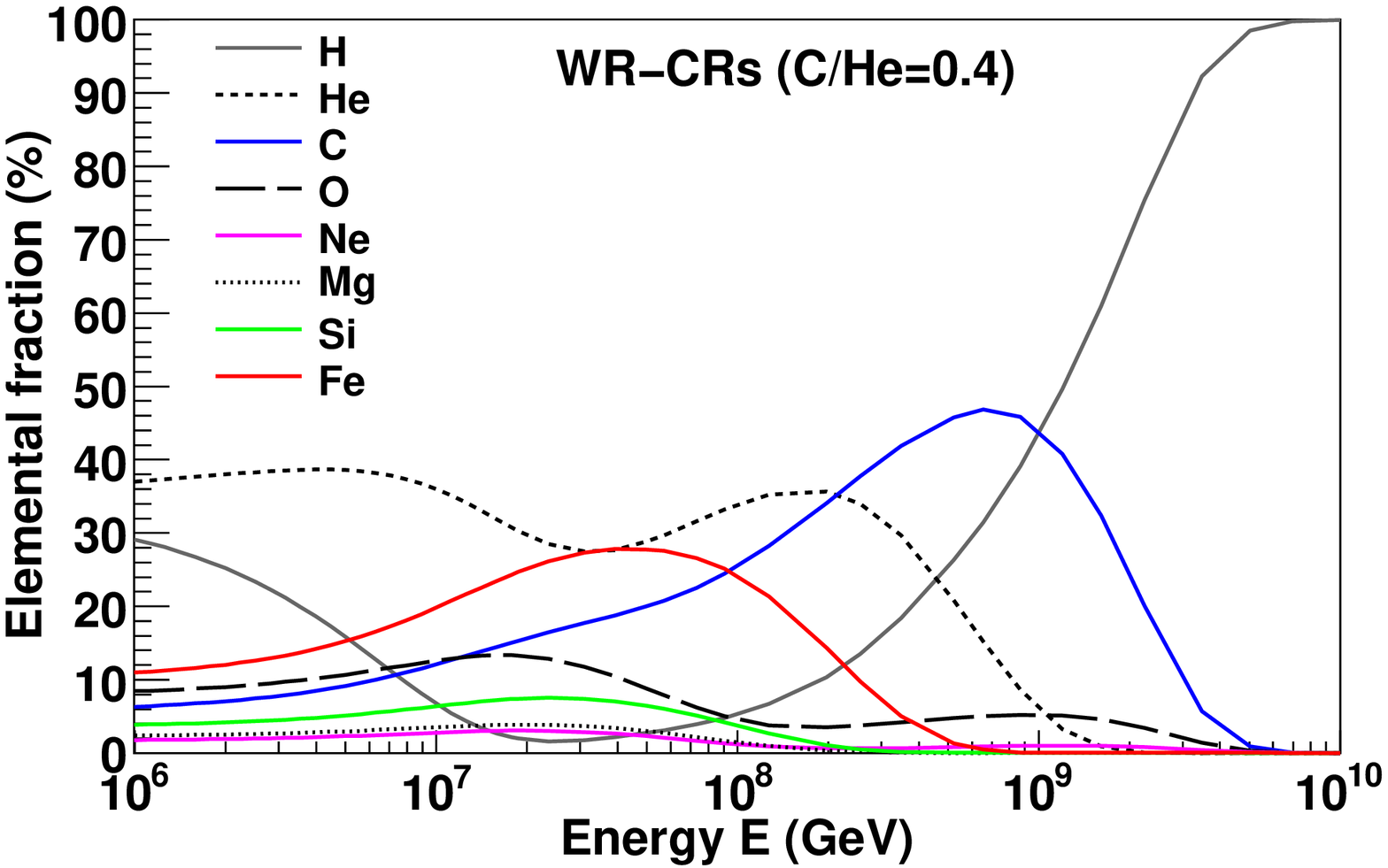}
\caption{\label {fig-fraction} Elemental fraction predicted by the different models of the additional Galactic component. {\it Top}: GW-CRs, {\it middle}: WR-CRs ($\mathrm{C/He}=0.1$), and {\it bottom}: WR-CRs ($\mathrm{C/He}=0.4$).}
\end{figure}

In Figure \ref{fig-fraction}, we show the elemental fraction at high energies predicted by the GW-CR and WR-CR models. In all the models, the composition consists of a large fraction of helium nuclei over a wide energy range. The maximum helium fraction is found in the case of WR-CR ($\mathrm{C/He}$=0.1) scenario, where the fraction reaches up to ${\sim}\,63\%$ at energy ${\sim}\,2\times 10^8$ GeV. In contrast to common perceptions, the WR-CR scenarios predict a composition of Galactic cosmic rays dominated mainly by helium (in the $\mathrm{C/He}=0.1$ case) or carbon nuclei (in the $\mathrm{C/He}=0.4$) near the transition energy region from Galactic to extra-galactic cosmic rays. The GW-CR model predicts an almost equal contribution of helium and iron nuclei at the transition region.

The cosmic-ray composition at energies above ${\sim}\,3\times 10^5$~GeV is not quite as well-measured as at lower energies. Above ${\sim}\,10^6$~GeV, KASCADE has provided spectral measurements for groups of elements by measuring the electron and muon numbers of extensive air showers induced by cosmic rays in the Earth's atmosphere. Several other experiments such as LOFAR, TUNKA, and the Pierre Auger Observatory have also provide composition measurements at high energies by measuring the depth of the shower maximum $(X_\mathrm{max})$. Heavier nuclei interact higher in the atmosphere, resulting in smaller values of $X_\mathrm{max}$ as compared to lighter nuclei. For comparison with theoretical predictions, we often use the mean logarithmic mass, $\langle\mathrm{lnA}\rangle$, of the measured cosmic rays which can be obtained from the measured $X_\mathrm{max}$ values using the relation \citep{Hoerandel2003b},
\begin{equation}
\label{eq-xmax}
\langle\mathrm{lnA}\rangle=\left(\frac{X_\mathrm{max}-X^\mathrm{p}_\mathrm{max}}{X^\mathrm{Fe}_\mathrm{max}-X^\mathrm{p}_\mathrm{max}}\right)\times \mathrm{ln} A_\mathrm{Fe},
\end{equation}
where $X^\mathrm{p}_\mathrm{max}$ and $X^\mathrm{Fe}_\mathrm{max}$ represent the average depths of the shower maximum for protons and iron nuclei respectively given by Monte-Carlo simulations, and $A_\mathrm{Fe}$ is the mass number of iron nuclei.

\begin{figure*}
\centering
\includegraphics*[width=0.75\textwidth,angle=0,clip]{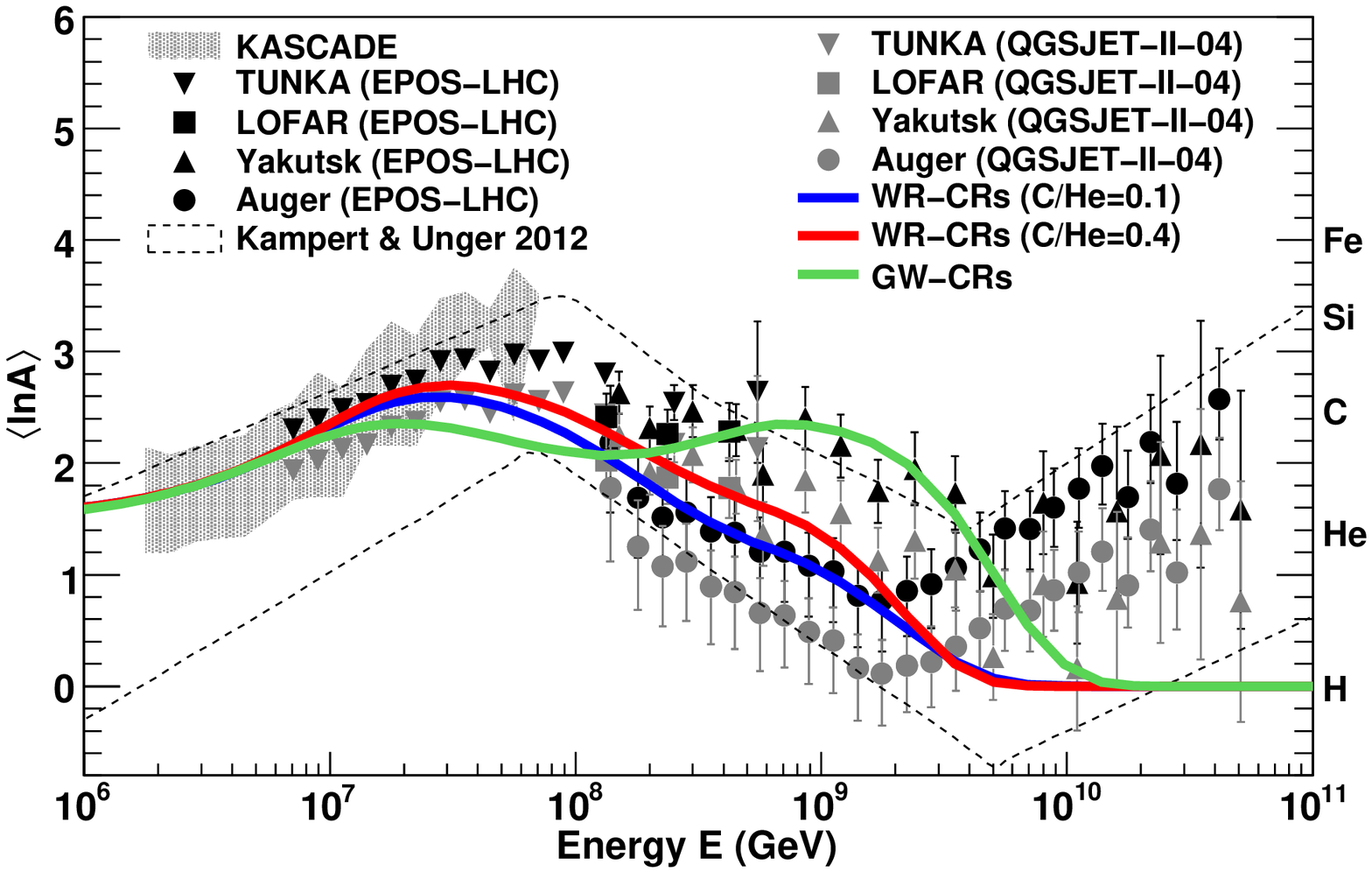}
\caption{\label {fig-lnA} Mean logarithmic mass, $\langle\mathrm{lnA}\rangle$, of cosmic rays predicted using the three different models of the additional Galactic component: WR-CRs ($\mathrm{C/He}=0.1$), WR-CRs ($\mathrm{C/He}=0.4$), and GW-CRs. {\it Data}: KASCADE \citep{Antoni2005}, TUNKA \citep{Berezhnev2013}, LOFAR \citep{Buitink2016}, Yakutsk \citep{Knurenko2010}, the Pierre Auger Observatory \citep{Porcelli2015}, and the different optical measurements compiled in \cite{KampertUnger2012}. The two sets of data points correspond to two different hadronic interaction models (EPOS-LHC and QGSJET-II-04) used to convert $X_\mathrm{max}$ values to $\langle\mathrm{lnA}\rangle$.}
\end{figure*}

In Figure \ref{fig-lnA}, the $\langle\mathrm{lnA}\rangle$ values predicted by the different models are compared with the measurements from different experiments. Although all our model predictions are within the large systematic uncertainties of the measurements, at energies above ${\sim}\,10^7$~GeV, the GW-CR model deviates from the general trend of the observed composition which reaches a maximum mean mass at ${\sim}\,6\times\,10^7$~GeV, and becomes gradually lighter up to the ankle. However, in the narrow energy range of ${\sim}\,(1-5)\times\,10^8$~GeV, the behaviour of the GW-CR model is in good agreement with the measurements from TUNKA, LOFAR and Yakutsk experiments which show a nearly constant composition that is different from the behaviour observed by the Pierre Auger Observatory at these energies. Understanding the systematic differences between the different measurements at these energies will be important for further testing of the GW-CR model. Up to around the ankle, the WR-CR models show an overall better agreement with the measurements than the GW-CR model. At around $(3-5)\times 10^7$~GeV, the WR-CR models seem to slightly under predict the KASCADE measurements, and they are more in agreement with the TUNKA measurements. Cosmic-ray composition measured by experiments like KASCADE, which measures the particle content of air showers on the ground, is known to have a large systematic difference from the composition measured with fluorescence and Cherenkov light detectors using $X_\mathrm{max}$ measurements \citep{Hoerandel2003b}. The large discrepancy between the model predictions and the data above the ankle is due to the absence of heavy elements in the EG-CR model considered in our calculation. The effect of choosing other models of EG-CRs will be discussed in the next section.

\section{Test with different models of extra-galactic cosmic rays}
\label{section-EG-models}
Despite of the dominance of the ankle-transition model in the general discussion, it has often been pointed out that the essential high-energy features of the cosmic ray spectrum, that is the ankle and, in part, even the second knee, can be explained by propagation effects of extra-galactic protons in the cosmologically evolving microwave background \citep{Hillas1967, Berezinsky1988, Berezinsky2006, Hillas2005, Aloisio2012, Aloisio2014}. While the most elegant and also most radical formulation of this hypothesis, the so-called `proton dip model', is meanwhile considered disfavoured by the proton fraction at the ankle measured by the Pierre Auger Observatory \citep{Aab2014}, the light composition below the ankle recently reported by the LOFAR measurement \citep{Buitink2016} and a potential `light ankle' at about $10^{8}\,$~GeV found by the KASCADE-Grande experiment \citep{Apel2013} have reinstated the interest in such models, and led to a number of ramifications, all predicting a more or less significant contribution of extra-galactic cosmic rays below the ankle. As such a component can greatly modify the model parameters, in particular the maximum energy, for the additional Galactic component -- if not removing its necessity altogether -- we study this effect using the WR-CR models, which show an overall best agreement with the data below the ankle, as a Galactic paradigm.

Before, however, discussing a stronger extra-galactic component below the ankle, we want to think about the \textit{minimal} extra-galactic contribution we can have, if we assume the largely heavy spectrum above the ankle is all extra-galactic and consider their propagation over extra-galactic distances.  To construct this `minimal model', we follow \cite{Matteo2015} and use the Monte-Carlo simulation code CRPropa 3.0 \citep{Batista2016}, which takes into account all important interaction processes undergone by EG-CRs while propagating through the CMB and the extra-galactic background light, and also the energy loss associated with the cosmological expansion. The effects of uncertainties in the simulations are discussed in \cite{Batista2015}. We assume the sources to be uniformly distributed in a comoving volume, and they produce cosmic rays with a spectrum given by \citep{Matteo2015},
\begin{align}
\label{eq-spectrum-Xgal}
Q_\mathrm{EG}&=K_0 F_\mathrm{j}\left(\frac{E}{E_0}\right)^{-\gamma},	& \frac{E}{Z}< R_\mathrm{c}\nonumber\\
&=K_0 F_\mathrm{j}\left(\frac{E}{E_0}\right)^{-\gamma}\exp\left(1-\frac{E}{ZR_\mathrm{c}}\right), & \frac{E}{Z}>R_\mathrm{c}
\end{align}
where $K_0$ is a normalisation constant, $F_\mathrm{j}$ is the injection fraction which depends on the type of the nuclei $j$, $E_0=10^9$~GeV, $\gamma$ is the source spectral index which is assumed to be the same for the different nuclei, and $R_\mathrm{c}$ is the rigidity at which the spectrum deviates from a power law. The model parameters are determined by simultaneously fitting the cosmic-ray energy spectrum, $X_\mathrm{max}$ and variance of $X_\mathrm{max}$ above the ankle  observed at the Pierre Auger Observatory. We adopt the CTG\footnote{CRPropa with the default TALYS photo-disintegration cross sections and the EBL model of \cite{Gilmore2012}} model for our calculation \citep{Matteo2015}, and consider that the sources inject protons, helium, nitrogen and iron nuclei. The best-fit model parameters values are $\gamma=0.73$, $R_\mathrm{c}=3.8\times 10^9$~GV, $F_\mathrm{H}=0\%$, $F_\mathrm{He}=0\%$, $F_\mathrm{N}=98.69\%$ and $F_\mathrm{Fe}=1.31\%$. In this model, the EG-CR spectrum below ${\sim}\,10^{10}$~GeV is dominated by protons and helium nuclei which  are secondary products  from the photo-disintegration of heavier nuclei during the propagation. At higher energies up to ${\sim}\,6\times 10^{10}$~GeV, the spectrum is dominated by the CNO group. Above ${\sim}\,3\times 10^{10}$~GeV, the spectrum exhibits a steep cut-off which is mostly due to the intrinsic cut-off in the injection spectrum, and not due to the GZK absorption during the propagation. This gives an overall best agreement with the measured data \citep{Matteo2015}. 

The first assumption we consider for an additional component of light particles below the ankle is based on the same physics, that is photo-disintegration of energetic nuclei in photon backgrounds, but considering this effect already in potentially densely photon loaded sources during acceleration. The physical motivation for this scenario is the acceleration of heavy nuclei at external/internal shocks in gamma ray bursts \citep{Murase2008, GlobusGRB2015}, or in tidal disruption events \citep{FarrarGruzinov2009}. Two variants of this assumptions have been recently suggested: the first, by \citet{Globus2015}, assumes that diffusion losses in the source are faster than the photo-disintegration time scale over a large range of energies, leading to a significantly steeper spectrum of the secondary protons than for the escaping residual nuclei, while in the second model by \citet{Unger2015} only the highest energy particles have an escape time which is smaller than the photo-disintegration time. While the predictions of the former model for secondary protons below the ankle are phenomenologically quite similar to the extra-galactic component of \citet{Rachen1993} at these energies, that is an approximate $E^{-2}$ source spectrum with a cosmological evolution $\propto (1+z)^{3.5}$, the second model \citet[hereafter the `UFA model']{Unger2015} predicts a strong pure-proton component concentrated only about one order of magnitude in energy below the ankle. Within their fiducial model, they consider a mix with a pure iron Galactic cosmic-ray component in \citet{Unger2015}. For our study, we use results which are optimised for a pure nitrogen Galactic composition\footnote{Michael Unger, private communication.}, which is closer to our predicted composition for the WR-CR model ($\mathrm{C/He}=0.4$) around the second knee.

A second assumption for an additional extra-galactic component is based on a universal scaling argument, which links the energetics of extra-galactic cosmic-ray sources on various scales and predicts that a dominant contribution to extra-galactic cosmic rays is expected from clusters of galaxies, accelerating a primordial proton-helium mix at their accretion shocks during cosmological structure formation \citep{Rachen2016}. As it has been shown already by \citet{Kang1997} that, for canonical assumptions on the diffusion coefficient around shocks (e.g. Bohm diffusion), the particle acceleration in this scenario is limited by pair-production losses in the CMB, this extra-galactic component is rather expected not to reach ultra-high energies, except for very optimistic assumptions on the acceleration process, but to be confined to energies \textit{below} the ankle. As so far no detailed Monte-Carlo propagation for this model has been calculated, we use here the analytical approximation developed in \citet{Rachen2016}. Assuming that both injection and acceleration of primordial protons and helium nuclei are only dependent on particle rigidity, the model predicts a succession of a proton and helium component with increasing energy, which are fixed in relative normalisation by the know primordial abundances. The more energetic helium component sharply cuts off at the ankle, merging into the cosmic-ray spectrum produced by extra-galactic sources at smaller scales, for which acceleration even in the conservative case is not limited by CMB or other photon interactions, and thus reaches the so-called Hillas limit, $E=Z e B R$, if $B$ is the typical magnetic field, and $R$ the typical size of the accelerator \citep{Hillas1984}. In our treatment, we hereby keep the exact cut-off energy and the total normalisation of this primordial cluster shock component as free parameters and determine them from fitting the all-particle spectrum, where we use the minimal model derived above as the second extra-galactic component extending into ultra-high energies. This model is henceforth denoted as `PCS model'.

\begin{figure}[H]
\centering
\includegraphics*[width=\columnwidth,angle=0,clip]{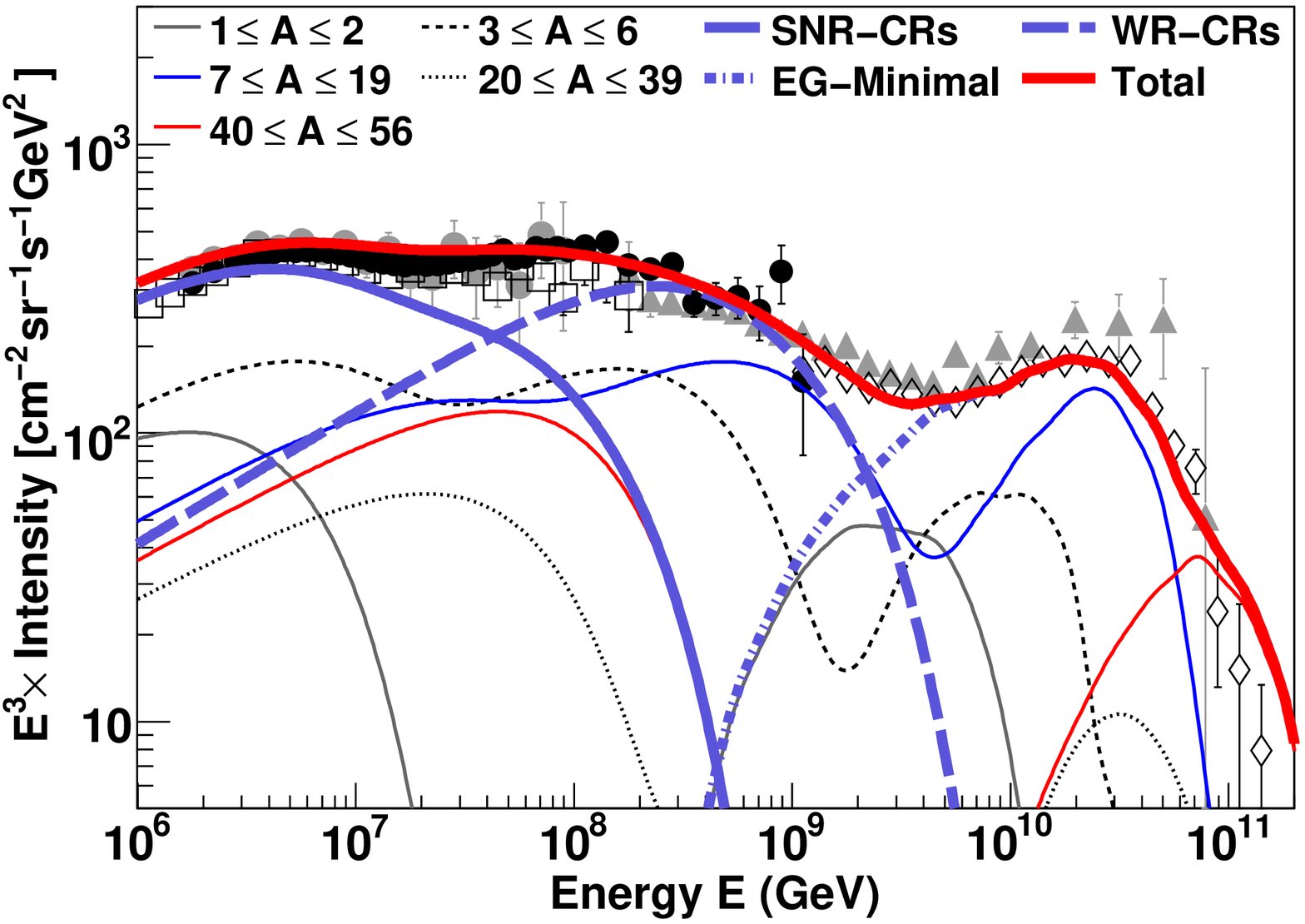}\\
\includegraphics*[width=\columnwidth,angle=0,clip]{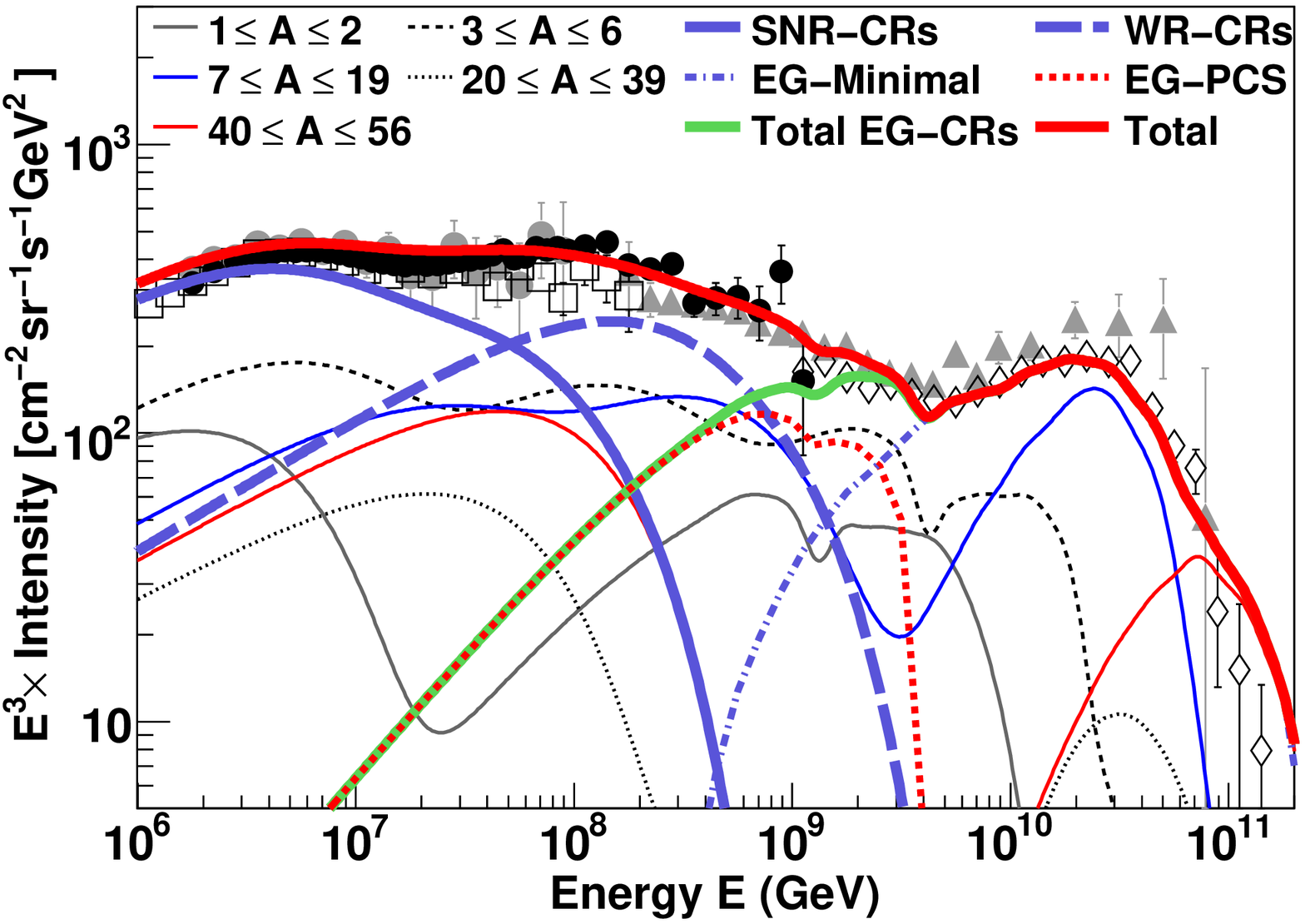}\\
\includegraphics*[width=\columnwidth,angle=0,clip]{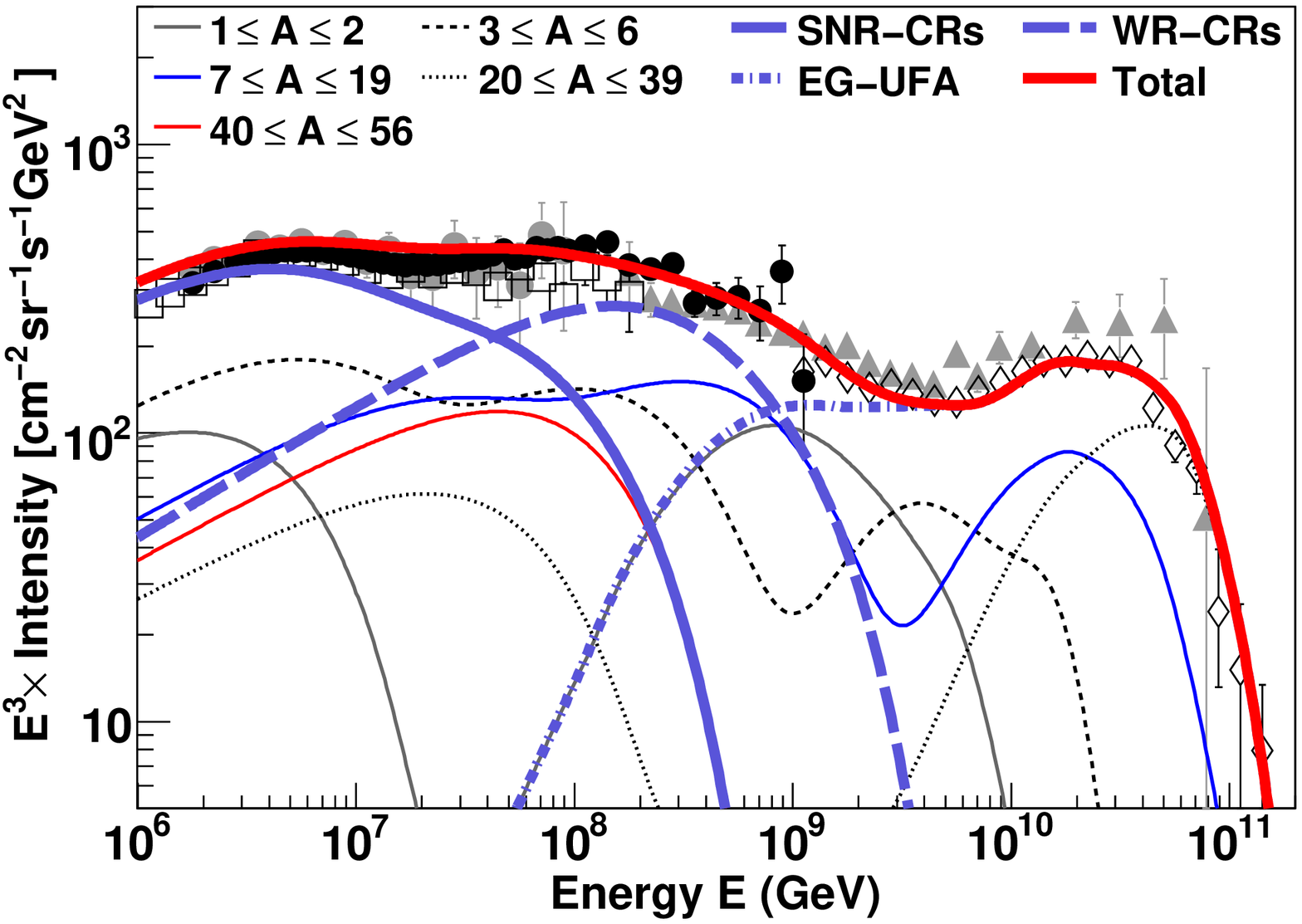}
\caption{\label {fig-spectrum-EG-models-WR2} All-particle spectrum for the three different models of EG-CRs -- Minimal ({\textit {Top}}), PCS ({\textit {middle}}), and UFA ({\textit {bottom}}) -- combined with the WR-CR ($\mathrm{C/He}=0.4$) model for the additional Galactic component. SNR-CR spectra shown are the same as in Figure \ref{fig-total-spectrum-WR} (bottom). Data are the same as in Figure \ref{fig-all-particle-SNR}. For results using WR-CR ($\mathrm{C/He}=0.1$) model, see Appendix \ref{appendix-B}.}
\end{figure}

\begin{figure}
\centering
\includegraphics*[width=\columnwidth,angle=0,clip]{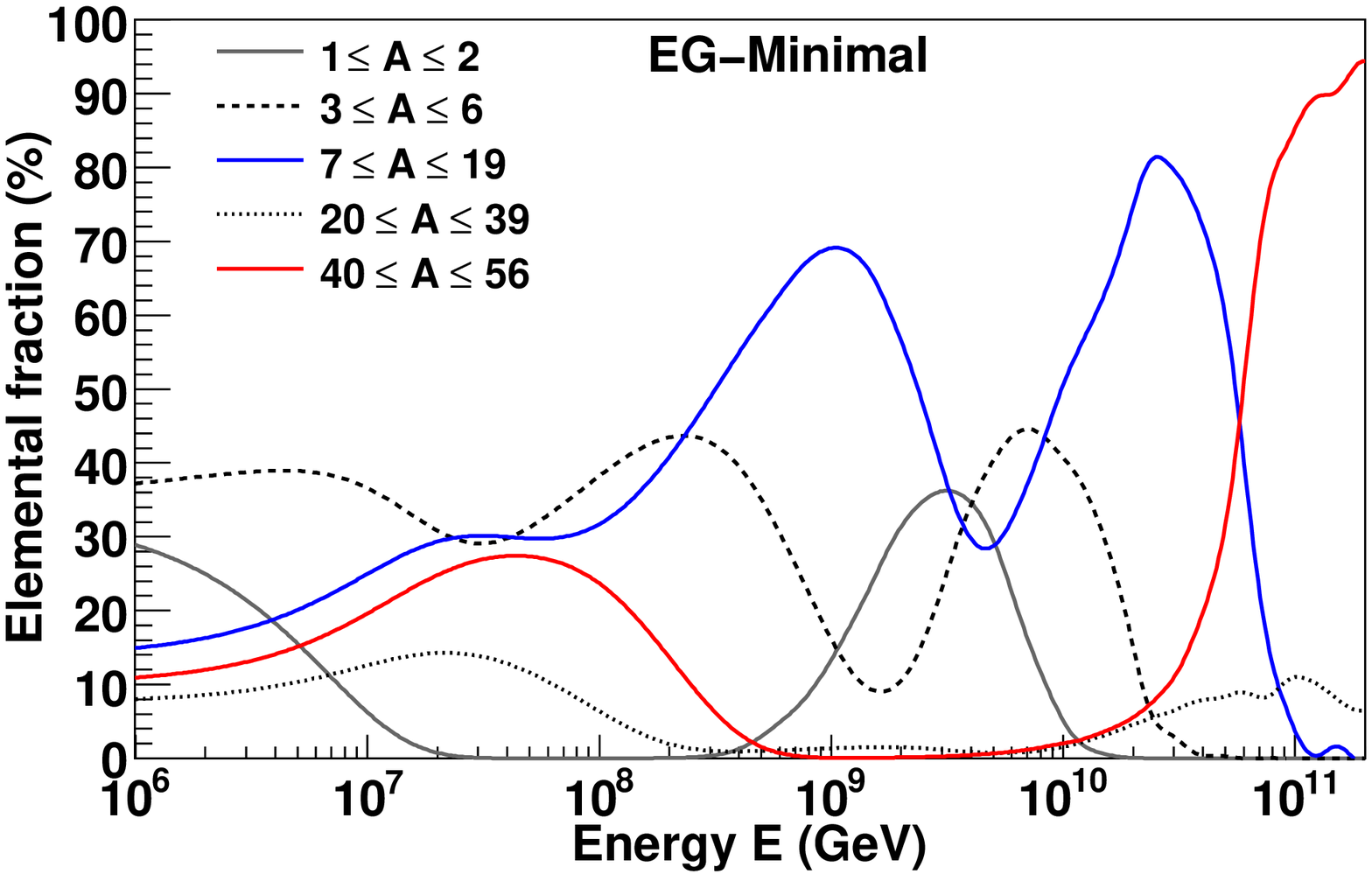}\\
\includegraphics*[width=\columnwidth,angle=0,clip]{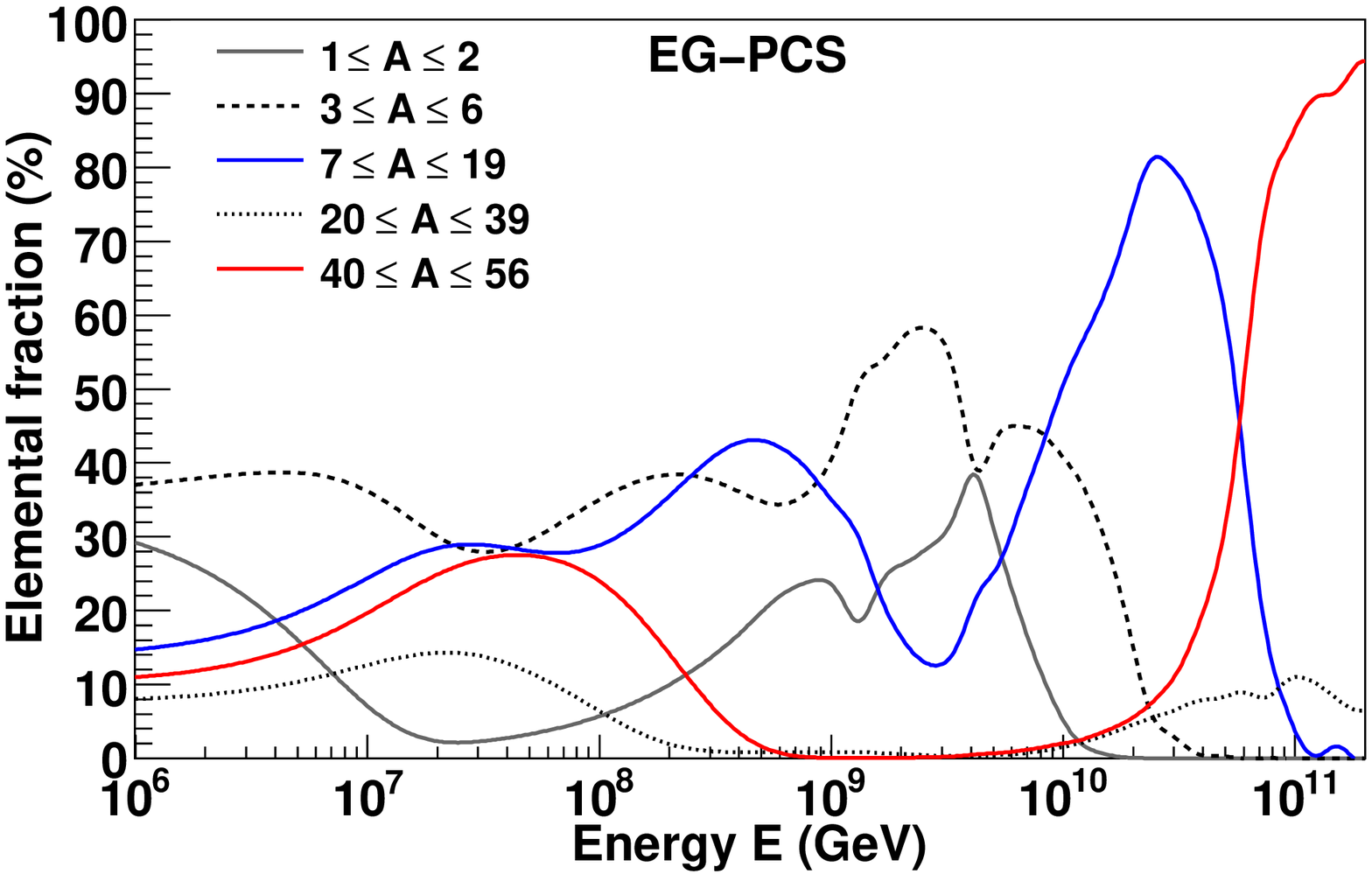}\\
\includegraphics*[width=\columnwidth,angle=0,clip]{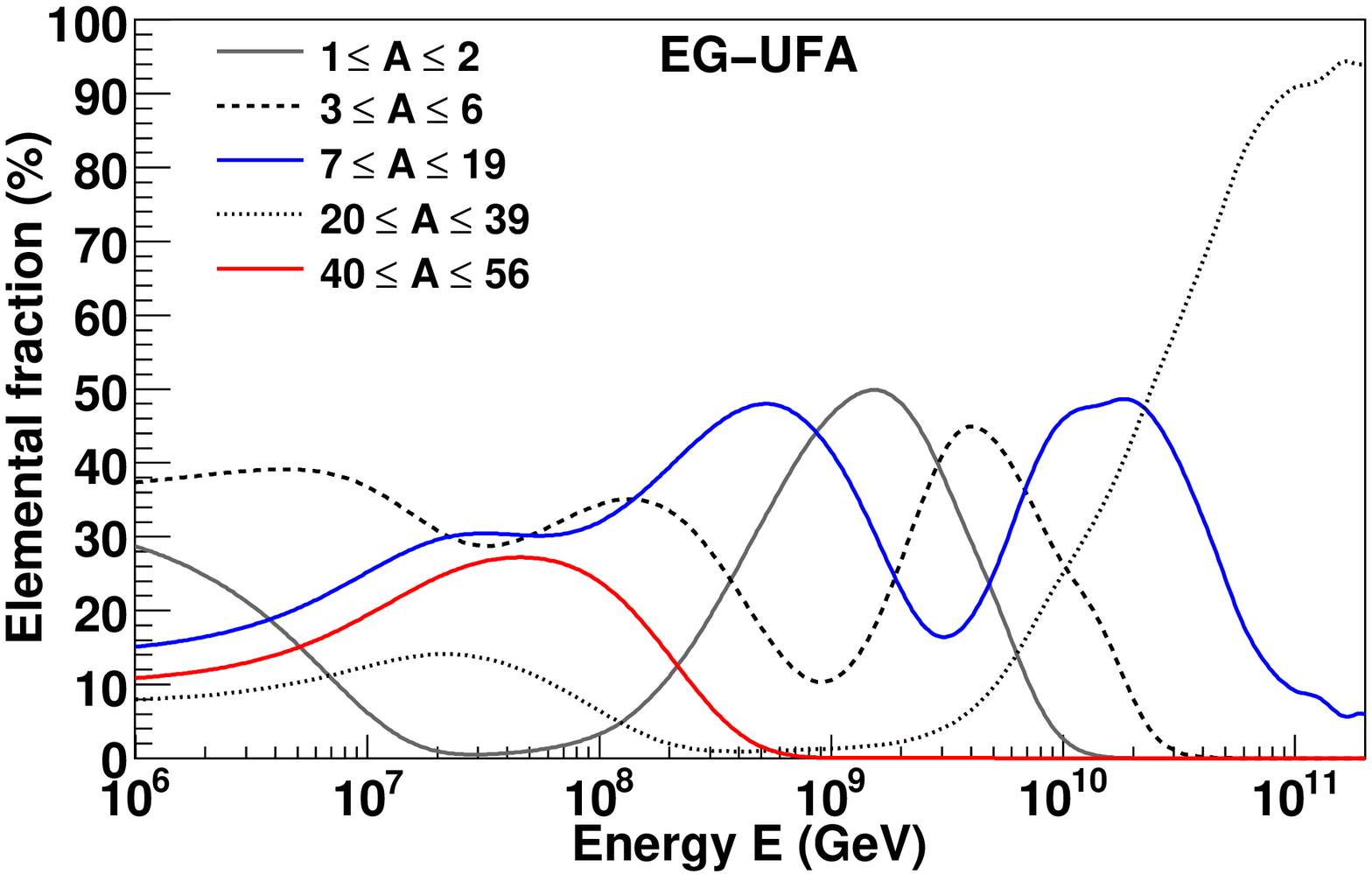}
\caption{\label {fig-fraction-EG-models-WR2} Elemental fraction of the five different mass groups shown in Figure \ref{fig-spectrum-EG-models-WR2} for the three different EG-CR models: minimal ({\textit {top}}), PCS ({\textit {middle}}), and UFA ({\textit {bottom}}), combined with the WR-CRs ($\mathrm{C/He}=0.4$) model for the additional Galactic component. Results obtained using WR-CR ($\mathrm{C/He}=0.1$) model are given in Appendix \ref{appendix-B}.}
\end{figure}

In Figure \ref{fig-spectrum-EG-models-WR2}, we present the all-particle spectrum above $10^6$~GeV obtained using the three different EG-CR models -- minimal model only, UFA and PCS model. The galactic contributions are from SNR-CRs and WR-CRs ($\mathrm{C/He}=0.4$). For the SNR-CRs, all the model parameters are the same as in Figure \ref{fig-total-spectrum-WR} (bottom). For the WR-CRs, the cut-off energy and the normalisation of the source spectrum are re-adjusted in order to produce an overall good fit to the measured spectrum and composition. They are allowed to vary in the three different cases. For the minimal model, the best-fit  proton cut-off energy of the WR-CRs is found to be $1.7\times 10^8$~GeV. This is approximately a factor $1.3$ larger than the value used in Figure \ref{fig-total-spectrum-WR}. For the PCS and the UFA models, the proton cut-off energies are almost the same at $1.1\times 10^8$~GeV, which are about a factor 1.5 less than that of the minimal model. This relaxation in the cut-off energy is due to the strong contribution of EG-CRs below the ankle in the two models. In the minimal model, the transition from Galactic to extra-galactic components occurs around the ankle, while in the PCS and UFA models, it occurs at $\sim 7\times 10^8$~GeV. The variation in the injection energy of WR-CRs remain within $6\%$ between the three models. In Figure \ref{fig-spectrum-EG-models-WR2}, spectra of five different mass groups are also shown. The elemental fraction of these mass groups are shown in Figure \ref{fig-fraction-EG-models-WR2}.

\begin{figure*}
\centering
\includegraphics*[width=0.75\textwidth,angle=0,clip]{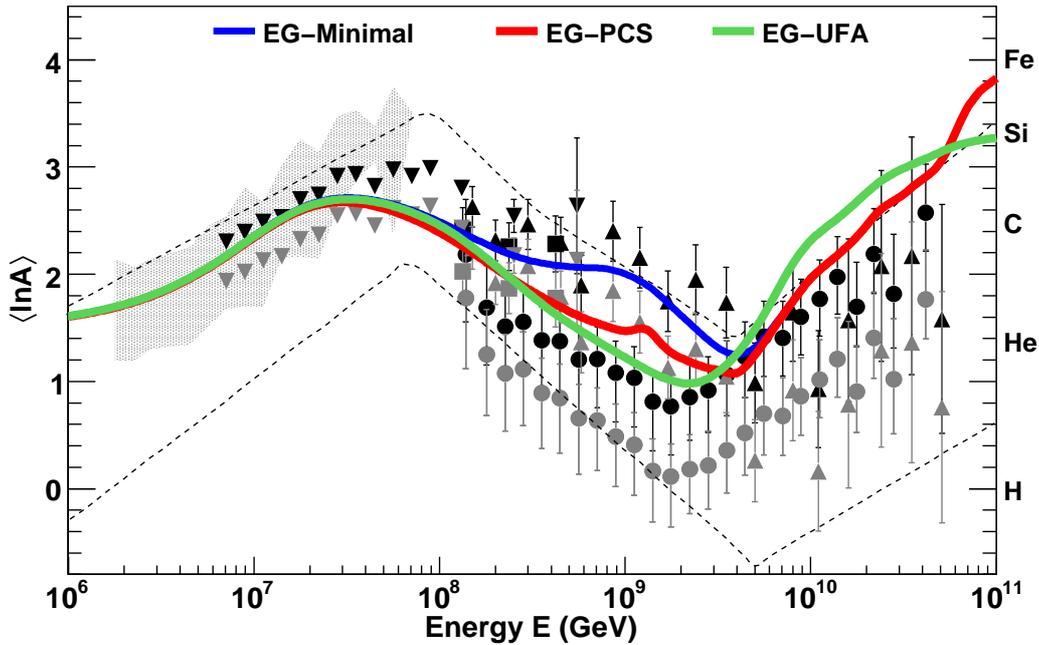}
\caption{\label {fig-lnA-EG-models-WR2} Mean logarithmic mass for the three different EG-CR models combined with the WR-CR ($\mathrm{C/He}=0.4$) model. Data are the same as in Figure \ref{fig-lnA}. Results obtained using WR-CR ($\mathrm{C/He}=0.1$) model are shown in Appendix \ref{appendix-B}.}
\end{figure*}

In Figure \ref{fig-lnA-EG-models-WR2}, we show $\langle\mathrm{lnA}\rangle$ predicted by the three EG-CRs model after adding the Galactic contribution. At energies between ${\sim}\,3\times 10^8$~GeV and $3\times 10^9$~GeV, the minimal model shows a bump that follows the trend of LOFAR and the data from other experiments, but contradicts the composition data from the Pierre Auger Observatory at ${\sim}\,10^9$~GeV. The UFA model over predicts the data above the ankle as the model is also tuned to the variance of $\langle\mathrm{lnA}\rangle$, but it is well within the systematic uncertainties (experimental as well as theoretical) as discussed in \cite{Unger2015}. The sharp feature present just above $10^9$~GeV in the PCS model is due to the dip in the proton spectrum (Figure \ref{fig-spectrum-EG-models-WR2}, middle panel, black-thin-solid line) that results from the intersection of the components from galaxy clusters and the minimal model, and is partially an artefact of the simplified propagation approach applied to this model. We expect it to be much smoother for realistic propagation. At energies below ${\sim}\,10^9$~GeV, both the PCS and the UFA models produce similar results which are in better agreement with the observed trend of the composition, but do not introduce a significant improvement over the canonical extra-galactic component used in Section \ref{sec-total-spectrum}. In all the three cases for the EG-CR model, the CNO group dominates the composition of Galactic cosmic rays at the transition region from Galactic to extra-galactic cosmic rays. A clear distinction between the models would be possible from a detailed measurement of the five major mass groups shown in Figure \ref{fig-fraction-EG-models-WR2}, in which they all have their characteristic `fingerprint': for example, around $10^9\,$GeV the minimal model is dominated by the CNO group, the PCS model by helium, and the UFA model by protons.

Results obtained using the WR-CR ($\mathrm{C/He}=0.1$) scenario are given in Appendix \ref{appendix-B}. The main difference from the results of the $\mathrm{C/He}=0.4$ scenario is the significant dominance of helium up to the transition energy region from Galactic to extra-galactic cosmic rays (see Figures \ref{fig-spectrum-EG-models-WR1} and \ref{fig-fraction-EG-models-WR1}). The main results and the parameter values of the different models discussed in the present work are summarised in Table \ref{table-models}.  

\section{Discussions}
\label{sec-discussion}
Our study has demonstrated that cosmic rays below ${\sim}\,10^9$~GeV can be predominantly of Galactic origin. Above $10^9$~GeV, they are most likely to have an extra-galactic origin. We show that both the observed all-particle spectrum and the composition at high energies can be explained if the Galactic contribution consists of two components: (i) SNR-CRs which dominates the spectrum up to ${\sim}\,10^7$~GeV, and (ii) GW-CRs or preferably WR-CRs which dominates at higher energies up to $\sim 10^9$~GeV. When combined with an extra-galactic component expected from strong radio galaxies or a source population with similar cosmological evolution, the WR-CR scenarios predict a transition from Galactic to extra-galactic cosmic rays at around $(6-8)\times 10^8$~GeV, with a Galactic composition mainly dominated by helium or the CNO group, in contrast to most common assumptions. In the following, we discuss our results for the SNR-CRs, GW-CRs, and WR-CRs in the context of other views on the Galactic cosmic rays below $10^9$~GeV, the implication of our results on the strength of magnetic fields in the Galactic halo and Wolf-Rayet stars, and also the case of a steep extra-galactic component extending below the second knee.

\subsection{SNR-CRs}
The maximum contribution of the SNR-CRs to the all-particle spectrum is obtained at a proton cut-off energy of ${\sim}\,4.5\times 10^6$~GeV (see Figure \ref{fig-all-particle-SNR}). Such a high energy is not readily achievable under the standard model of diffusive shock acceleration theory in supernova remnants for magnetic field values typical of that in the interstellar medium (see e.g. \citealp{Lagage1983}). However, numerical simulations have shown that the magnetic field near supernova shocks can be amplified considerably up to ${\sim}\,10-100$ times the mean interstellar value \citep{Lucek2000, Reville2012}. This is also supported by observations of thin X-ray filaments in supernova remnants which can be explained as due to rapid synchrotron losses of energetic electrons in the presence of strong magnetic fields \citep{Vink2003, Parizot2006}. Such strong fields may lead to proton acceleration up to energies close to the cut-off energy obtain in our study \citep{Bell2004}.

The main composition of cosmic rays predicted by the SNR-CRs {\textit{alone}} looks similar to the prediction of the poly-gonato model \citep{Hoerandel2003a}. Both show a helium dominance over proton around the knee, and iron taking over at higher energies at  ${\sim}\,10^7$~GeV in the SNR-CRs, and at ${\sim}\,6\times 10^6$~GeV in the poly-gonato model.
\begin{table*}
\begin{center}
\caption{\label{table-models}Summary of the different models for cosmic rays, and their results presented in this work. In all the models, the Galactic contribution consists of two components: the first component which is produced by regular supernova explosions in the Galaxy (SNR-CRs), and the second component which is considered to be produced either by cosmic-ray re-acceleration by Galactic wind termination shocks (GW-CRs) or by explosions of Wolf-Rayet stars in the Galaxy (WR-CRs). The source spectral indices for the second Galactic component in all the models are assumed to be the same as for the SNR-CRs (see Table \ref{table-SNR-CRs}). For the extra-galactic component, the different models considered are: (a) sources with strong cosmological evolution like strong radio galaxies (EG-RSB93) (b) extra-galactic contribution mainly above the ankle irrespective of the nature of the sources (EG-Minimal), (c) significant photo-disintigration of cosmic-rays in a source region with high photon density (EG-UFA), and (d) cosmic rays accelerated by accretion shocks in clusters of galaxies (EG-PCS). The all-particle spectra predicted by the different combinations of the Galactic and extra-galactic components are quite similar, and show good agreement with the measured spectrum. On the other hand, although the $\langle\mathrm{lnA}\rangle$ predicted by the different models are almost within the range of the different measurements compiled by \cite{KampertUnger2012}, they show distinctive differences especially in the energy range between the second knee and the ankle. For the model using GW-CRs, the predicted $\langle\mathrm{lnA}\rangle$ also show deviation from the prediction of other models between $\sim\,10^7$ and $10^8$~GeV. The comments on $\langle\mathrm{lnA}\rangle$ given in the table are with respect to the measurements from TUNKA \citep{Berezhnev2013}, LOFAR \citep{Buitink2016}, Yakutsk \citep{Knurenko2010}, and the Pierre Auger Observatory \citep{Porcelli2015} between the second knee and the ankle. QGSJET in the table refers to the QGSJET-II-04 model.}
\vspace{\baselineskip}
\label{fit-parameters}
\resizebox{\textwidth}{!}
{
\begin{tabular}{lllllllll}
\hline
\multicolumn{2}{c}{Model}& \multicolumn{1}{l}{Reference}&\multicolumn{1}{l}{Reference}& \multicolumn{2}{c}{Cut-off rigidities (GV)}&\multicolumn{1}{l}{Composition at:}&\multicolumn{1}{l}{Extra-galactic}&\multicolumn{1}{l}{Predicted $\langle\mathrm{lnA}\rangle$ between the second knee}\\
\cline{1-2}\cline{5-6}
Second	    	& Extra-galactic 	& sections	& figures			& First 			& Second	  		& $10^8$~GeV,	& contribution	at			& and the ankle\\
Galactic 		& component	 	&			&				& Galactic		& Galactic			& $10^9$~GeV	& $(10^8, 10^9)$~GeV	&\\
component	&				&			&				& component		& component			& (p, He, CNO, Fe) &						&\\
\hline
&&&&&&&\\
GW-CRs 	& EG-RSB93 	& \ref{section-galactic-wind} \& \ref{sec-total-spectrum}& \ref{fig-total-spectrum-wind}, \ref{fig-fraction} \& \ref{fig-lnA}  & $3.0\times 10^6$& $9.5\times 10^7$			& $(20\%, 32\%, 12\%, 24\%),$	& $(4\%, 30\%)$	& Good agreement with TUNKA (QGSJET) \\
			&				&											& 													&				& 						& $(32\%, 2\%, 18\%, 30\%)$	&				& and LOFAR/Yakutsk (EPOS-LHC) data,\\
			&				&											&													&				&						&							&				& but strong disagreement with Auger data\\
			&				&											&													&				&						& 							&				&\\
WR-CRs		& EG-RSB93 	& \ref{subsection-WR} \& \ref{sec-total-spectrum}	& \ref{fig-total-spectrum-WR}, \ref{fig-fraction} \& \ref{fig-lnA}	& $4.1\times 10^6$& $1.8\times 10^8$			& $(6\%, 51\%, 14\%, 24\%),$	& $(6\%, 50\%)$	& Moderate agreement with LOFAR and\\
(C/He=0.1)	&				&											&													&				&						& $(48\%, 25\%, 26\%, 0\%)$	&				& Yakutsk (QGSJET) data, and excellent\\
			&				&											&													&				&						&							&				& agreement with Auger (EPOS-LHC) data\\
			&				&											&													&				&						& 							&				&\\
WR-CRs		& EG-RSB93 	& \ref{subsection-WR} \& \ref{sec-total-spectrum}	& \ref{fig-total-spectrum-WR}, \ref{fig-fraction} \& \ref{fig-lnA}	& $4.1\times 10^6$& $1.3\times 10^8$			& $(6\%, 34\%, 30\%, 24\%),$	& $(5\%, 45\%)$	& Good agreement with LOFAR (QGSJET)\\
(C/He=0.4)	&				&											& 													&				&						& $(44\%, 6\%, 49\%, 0\%)$		&				& data, and moderate agreement with Yakutsk\\
			&				&											&													&				&						&							&				& (QGSJET)  and Auger (EPOS-LHC) data\\
			&				&											&													&				&						& 							&				&\\
WR-CRs		& EG-Minimal 	& \ref{section-EG-models} \& \ref{appendix-B}			&\ref{fig-spectrum-EG-models-WR1}, \ref{fig-fraction-EG-models-WR1}& $4.1\times 10^6$& $2.4\times 10^8$	& $(0\%, 57\%, 14\%, 24\%),$	& $(0\%, 16\%)$	& Excellent agreement with LOFAR\\
(C/He=0.1)	&				&											& \& \ref{fig-lnA-EG-models-WR1}							&				& 						&$(15\%, 51\%, 35\%, 0\%)$		&				& (QGSJET) and moderate agreement\\
			&				&											&													&				&						&							&				& with TUNKA/Yakutsk (QGSJET) data,\\
			&				&											&													&				&						&							&				& but strong disagreement with Auger data\\
			&				&											&													&				&						& 							&	  			&\\
WR-CRs		& EG-PCS		& \ref{section-EG-models} \& \ref{appendix-B}		&\ref{fig-spectrum-EG-models-WR1}, \ref{fig-fraction-EG-models-WR1}& $4.1\times 10^6$& $1.5\times 10^8$	& $(6\%, 52\%, 13\%, 24\%),$	& $(10\%, 66\%)$	& Moderate agreement with LOFAR and\\
(C/He=0.1)	&				&											& \& \ref{fig-lnA-EG-models-WR1}							& 				&						& $(25\%, 53\%, 21\%, 0\%)$	&				& Yakutsk (QGSJET) data, and good\\
			&				&											&													&				&						&							&				& agreement with Auger (EPOS-LHC) data\\
			&				&											&													&				&						& 							&				&\\
WR-CRs		& EG-UFA		& \ref{section-EG-models} \& \ref{appendix-B}		&\ref{fig-spectrum-EG-models-WR1}, \ref{fig-fraction-EG-models-WR1}& $4.1\times 10^6$& $1.6\times 10^8$ 	& $(4\%, 52\%, 14\%, 24\%),$	& $(3\%, 58\%)$	& Moderate agreement with LOFAR\\
(C/He=0.1)	&				&											& \& \ref{fig-lnA-EG-models-WR1}							& 				&						& $(49\%, 25\%, 25\%, 0\%)$	&				& (QGSJET) data, and excellent agreement\\
			&				&											&													&				&						&							&				& with Auger (EPOS-LHC) data\\
			&				&											&													&				&						&							&				&\\
WR-CRs		& EG-Minimal 		& \ref{section-EG-models} 						& \ref{fig-spectrum-EG-models-WR2}, \ref{fig-fraction-EG-models-WR2} & $4.1\times 10^6$& $1.7\times 10^8$& $(0\%, 38\%, 32\%, 24\%),$		& $(0\%, 15\%)$	& Good agreement with TUNKA (QGSJET)\\
(C/He=0.4)	&				&											& \& \ref{fig-lnA-EG-models-WR2}							&				&						& $(14\%, 15\%, 69\%, 0\%)$	&				& and LOFAR (EPOS-LHC) data, and\\
			&				&											&													&				&						&							&				& moderate agreement with Yakutsk data,\\
			&				&											&													&				&						&							&				& but strong disagreement with Auger data\\
			&				&											&													&				&						&							&				&\\
WR-CRs		& EG-PCS		& \ref{section-EG-models}						& \ref{fig-spectrum-EG-models-WR2}, \ref{fig-fraction-EG-models-WR2} & $4.1\times 10^6$& $1.1\times 10^8$& $(6\%, 36\%, 29\%, 24\%),$		& $(10\%, 62\%)$	& Moderate agreement with LOFAR/Yakutsk\\
(C/He=0.4)	&				&											& \& \ref{fig-lnA-EG-models-WR2}							&				&						& $(24\%, 42\%, 35\%, 0\%)$	&				& (QGSJET) and Auger (EPOS-LHC) data\\
			&				&											&													&				&						&							&				&\\
WR-CRs		& EG-UFA		& \ref{section-EG-models}						& \ref{fig-spectrum-EG-models-WR2}, \ref{fig-fraction-EG-models-WR2}& $4.1\times 10^6$& $1.1\times 10^8$& $(3\%, 35\%, 32\%, 24\%),$		& $(3\%, 55\%)$	& Moderate agreement with LOFAR/Yakutsk\\
(C/He=0.4)	&				&											&  \& \ref{fig-lnA-EG-models-WR2}						&				&						& $(47\%, 10\%, 41\%, 0\%)$	&				& (QGSJET) data, and good agreement with\\
			&				&											&													&				&						&							&				& Auger (EPOS-LHC) data\\

\hline
\end{tabular}
}
\end{center}
\end{table*}
The helium dominance is more significant in the SNR-CRs than in the poly-gonato model which is due to the flatter spectral index required to reproduce the recent measurements from CREAM and ATIC experiments with the SNR-CRs. The main difference, however, is in the total contribution above ${\sim}\,2\times\,10^7$~GeV. SNR-CRs alone cannot explain the observed all-particle spectrum above ${\sim}\,2\times 10^7$~GeV. They contribute only ${\sim}\,30\%$ of the observed cosmic rays at ${\sim}\,10^8$~GeV. On the other hand, in the poly-gonato model, the total contribution from elements with $1\leq\,Z\leq\,28$ can explain the observed spectrum up to energies close to $10^8$~GeV. This difference is mainly due to the difference in the shapes of the spectral cut-offs of particles between the two models. For the SNR-CRs, we consider a power-law with an exponential cut-off, while the poly-gonato model assumes a broken power-law with a smooth break around the cut-off (break) energy. This leads to a higher flux around the cut-off energy in the poly-gonato model. On adding GW-CRs or WR-CRs as an additional Galactic component, the composition above ${\sim}\,10^7$~GeV in our model has a large fraction of helium or a mixture of helium and CNO group, which is quite different from the prediction of the poly-gonato model where the composition is mainly dominated by iron nuclei. Our prediction (in particular, that of the WR-CR scenario) is more in agreement with the $X_\mathrm{max}$ measurements from fluorescence and Cherenkov light detectors, while the poly-gonato model is in agreement with data from the measurements of air shower particles on the ground.

Recently, \cite{Globus2015} claimed that a single Galactic component with rigidity dependent cut-off is sufficient to explain the observed all-particle spectrum when combined with an extra-galactic component. Their claim that an additional Galactic component is not needed does not contradict our claim of having one. It is simply that they assume the particle spectrum as a {\textit{broken}} power law with an exponential cut-off which leads to an increased flux above the break energy (knee) as in the poly-gonato model. However, we have demonstrated that if one considers a power-law spectrum with an exponential cut-off which is expected for particles produced by diffusive shock acceleration process in supernova remnants \citep{Malkov2001}, a single component cannot explain the observed spectrum beyond the knee, and a second Galactic component is inevitable. Their single component, which they had not assigned to any specific source class, would correspond to the superposition of multiple components similar to the ones proposed in our model. Based on the physical models of the most plausible sources and the propagation of cosmic rays in the Galaxy, we show that two Galactic components are sufficient to explain the measured spectrum, but do not exclude the existence of more than two components.

\subsection{GW-CRs}
Assuming that the maximum energy of particles produced by the Galactic wind termination shock is limited by the condition that the particle diffusion length must be less than the size of the shock, the maximum energy under Bohm diffusion can be written as, $E_\mathrm{m}{\sim}\,3ZeB(V_\mathrm{s}/c)R_\mathrm{s}$, where $B$ is the magnetic field, $V_\mathrm{s}$ is the shock velocity and $R_\mathrm{s}$ is the shock radius. From the GW-CR parameters obtained in our study, we can take  $R_\mathrm{s}=R_\mathrm{sh}=96$~kpc, $V_\mathrm{s}=\tilde{V} R_\mathrm{s}=1443$~km~s$^{-1}$ which is the terminal wind velocity, and $E_\mathrm{m}=9.5\times 10^7$~GeV which is the proton cut-off energy. Using these values, the magnetic field strength in the Galactic halo is estimated to be ${\sim}\,73$~nG. This is approximately a factor $3$ less than the value obtained assuming Parker's magnetic field topology for the solar wind (Equation \ref{eq-magnetic-field}).

An intrinsic issue in the case of re-acceleration by Galactic wind termination shock is the difficulty to observe the re-accelerated particles in the Galactic disk because of advection by the wind flow, except for the highest energy particles, as discussed in Section \ref{section-galactic-wind}. As a consequence, the spectrum in the disk may not show a continuous transition between the SNR-CRs and GW-CRs (see e.g.  \citealp{Zirakashvili2006}). This effect is actually visible in the predicted spectra of the individual elements shown in Figure \ref{fig-total-spectrum-wind}. However, we notice that the superposition of the individual spectra smears out this effect in the all-particle spectrum. Nevertheless, in order to avoid this effect, \cite{Zirakashvili2006} considered termination shocks which are stronger near the Galactic poles and weaker towards the Galactic equator, unlike in our study where the shocks are considered to have equal strengths in all the directions. In their configuration, the maximum energy of particles decreases from the poles towards the equator, and therefore, the superposition of spectra from different colatitudes produces a continuity in the total spectrum. Another consideration is the particle re-acceleration by spiral shocks in the Galactic wind which are formed by the interaction between fast winds originating from the Galactic spiral arms and slow winds from the interarm regions \citep{Voelk2004}. These shocks, which can be formed at distances of ${\sim}\,50-100$~kpc, can accelerate SNR-CRs up to ${\sim}\,Z\times 10^8$~GeV. An alternative possibility is the re-acceleration by multiple shock waves in the Galactic wind generated by time dependent outflows of gas from the Galactic disk \citep{Dorfi2012}. These shocks, which are long-lived like the termination shocks, can accelerate particles up to ${\sim}\,10^8-10^9$~GeV in the lower Galactic halo. An attractive feature of this model is the advection of particles downstream of the shocks towards the Galactic disk, thereby, resolving the difficulty of observing the re-accelerated particles in the disk. Despite having different features, the cosmic-ray composition predicted by all these different models in the energy range of ${\sim}\,10^7-10^9$~GeV are expected to be similar to the result presented here since they consider the same seed particles (cosmic rays from the Galactic disk) for re-acceleration as in our study. Below ${\sim}\,10^7$~GeV where the GW-CRs are significantly suppressed in our case, the other wind models discussed above will give a different result.
\begin{figure}
\centering
\includegraphics*[width=\columnwidth,angle=0,clip]{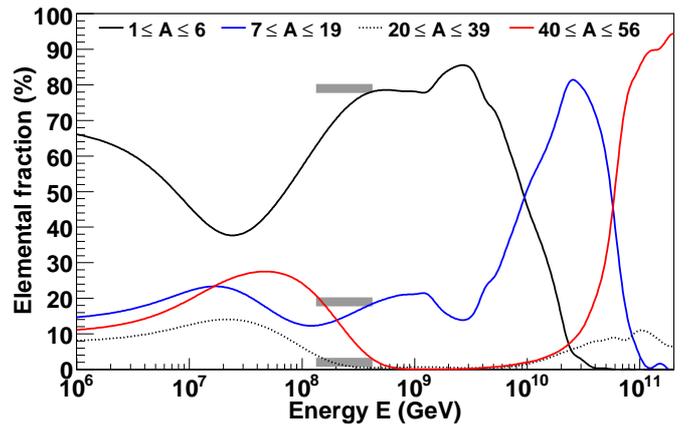}
\caption{\label {fig-fraction-EG-clusterLOFAR-WR1} Elemental fraction for four mass groups obtained using the PCS model of EG-CRs and WR-CRs ($\mathrm{C/He}=0.1$). The proton fraction (not shown in the figure) predicted by the model in the LOFAR energy range is ${\sim}\,10\%$. The grey bands, from top to bottom, represent the best-fit LOFAR measurements of $79\%$ helium, $19\%$ nitrogen and $2\%$ iron nuclei in the energy range of $(1.3-4.2) \times 10^8$~GeV \citep{Buitink2016}. At $99\%$ confidence level, the measured proton plus helium fraction can vary in the range of $(38-98)\%$, and the combined nitrogen and iron fraction within $(2-62)\%$.}
\end{figure}

\subsection{WR-CRs}
The prediction of a large helium fraction and a small iron fraction between around $10^8$ and $10^9$~GeV by the WR-CR ($\mathrm{C/He}=0.1$) model seems to be in agreement with new measurements from the LOFAR radio telescope \citep{Buitink2016}, and the Pierre Auger Observatory \citep{Aab2014}. These measurements have revealed a strong light component, and an almost negligible iron component above ${\sim}\,10^8$~GeV. In Figure \ref{fig-fraction-EG-clusterLOFAR-WR1}, the elemental fraction predicted by the WR-CR ($\mathrm{C/He}=0.1$) model combined with the PCS model for the EG-CRs is compared with the best-fit composition of the LOFAR data for four mass groups: $1\leq \mathrm{A}\leq 6$, $7\leq \mathrm{A}\leq 19$, $20\leq \mathrm{A}\leq 39$, and $40\leq\mathrm{A}\leq 56$. The model predictions are found to show a good agreement with the data.

Using the maximum energy of particles for the WR-CRs, it is possible to estimate the strength of the magnetic field at the surface of Wolf-Rayet stars. Assuming that the magnetic field configuration in the Wolf-Rayet winds follows Parker's model \citep{Parker1958}, the toroidal magnetic field strength near the equatorial plane of the star at the position $R_\mathrm{w}$ from the star follows the relation,
\begin{equation}
\label{eq-magnetic-field}
B=B_0\frac{\omega R_{\star}^2}{V_\mathrm{w} R_\mathrm{w}},
\end{equation}
where $B_0$ is the magnetic field at the surface of the star, $\omega$ is the angular rotation velocity, $R_{\star}$ is the radius of the star, and $V_\mathrm{w}$ is the wind velocity. Using the relation $E_\mathrm{m}{\sim}\,3ZeB(V_\mathrm{s}/c)R_\mathrm{s}$ for the maximum energy as in the case of the GW-CRs, and the proton cut-off energy of $E_\mathrm{m}=1.1\times 10^{8}$~GeV for the WR-CRs ($\mathrm{C/He}=0.4$) obtained using the PCS/UFA model, we get $BR_\mathrm{s}{\sim}\,1.2\times 10^{15}$~G~cm, where we take the shock velocity $V_\mathrm{s}=0.1~c$ \citep{Soderberg2012}. Using this value of $BR_\mathrm{s}$ in Parker's magnetic field configuration (Equation \ref{eq-magnetic-field}) by taking $R_\mathrm{w}=R_\mathrm{s}$ and other Wolf-Rayet star parameters as $R_{\star}=3\times 10^{12}$~cm, $\omega=10^{-6}$~s$^{-1}$, and $V_\mathrm{w}=2000$~km~s$^{-1}$ \citep{Berezhko2000}, we obtain the magnetic field at the surface of the star as $B_0{\sim}\,1.5\times 10^4$~G. Such a strong magnetic field was also predicted in an earlier study by \cite{Biermann1993}, and is found to be in agreement with recent magnetic field measurements from Wolf-Rayet stars. Based on an upper limit of $100$~G in the observable parts of Wolf-Rayet winds, \cite{Chevrotiere2013} estimated an upper limit for the surface magnetic field of ${\sim}\,5400$~G. An even stronger field in the wind, up to ${\sim}\,2000$~G, has been reported \citep{Chevrotiere2014}, which indicates that the surface magnetic field of these stars can go well above the order of $10^4$~G.

From the total energy of $1.4\times 10^{49}$~ergs injected into WR-CRs by a single supernova explosion, and the  explosion rate of Wolf-Rayet stars in the Galaxy of $1/210$~yr$^{-1}$, we estimate the total power injected into WR-CRs as $2.1\times 10^{39}$~ergs s$^{-1}$. This is approximately a factor $40$ less than the power injected into SNR-CRs by supernova explosions in the interstellar medium. The required amount of supernova explosion energy injected into helium nuclei for WR-CRs  is about $1.2-1.6$ times that of the SNR-CRs. This indicates that the average abundance of helium nuclei swept up by supernova shocks in the Wolf-Rayet winds must be higher than the helium abundance present in the interstellar medium if the particle injection fraction and the acceleration efficiency of the shocks are the same for the SNR-CRs and the WR-CRs.

Our results for the WR-CRs are obtained by assuming that the particle injection fraction into the shocks is the same for all the different elements. The injection fraction may depend on the type of the element, and the nature of this dependence is not quite understood. By taking the ratio of the SNR-CRs source spectra (Equation \ref{eq-source}) at a fixed rigidity to the known Solar system elemental abundances \citep{Lodders2009}, we estimate the relative injection fraction of particles for the different elements. Applying these relative injection fractions to the WR-CRs, we find that the composition is significantly dominated by carbon nuclei, in contrast to the results shown in Figure \ref{fig-spectrum-only-WR} where the composition is mainly dominated by helium or a mixture helium and carbon nuclei. Thus, the contribution of WR-CRs in this case is strongly constrained by the measured carbon spectrum at low energies. The all-particle spectrum for this case, after adding the contributions of SNR-CRs and EG-CRs, underpredicts the measured data between the second knee and the ankle. This problem might be resolved if we consider that both GW-CRs and WR-CRs contribute at the same time. In future, we will explore the parameter space of this combined scenario.

\subsection{Comparison with Hillas's `Component B'}
\cite{BellLucek2001} showed that magnetic field upstream of supernova shock fronts can be amplified non-linearly by cosmic rays up to many times the pre-shock magnetic field. They showed that these highly amplified magnetic fields can facilitate cosmic-ray acceleration up to energies $Z\times 10^8$~GeV for supernova shocks expanding in the interstellar medium, even higher by an order of magnitude for shocks expanding into pre-existing stellar winds. Based on the Bell-Lucek's version of diffusive shock acceleration, \cite{Hillas2005} proposed a second Galactic component `Component B', produced by Type~II supernova remnants in the Galaxy expanding into dense slow winds of the preceding red supergiants, to accommodate for the observed cosmic rays above ${\sim}\,10^7$~GeV. In the \cite{Hillas2005} model, a Galactic component `Component A', produced by Type~Ia supernova remnants in the Galaxy, dominates the all-particle energy spectrum below ${\sim}\,10^7$~GeV. The `Component A' has a similar composition to the SNR-CRs in our model, but the `Component B' has a large iron fraction in contrast to the WR-CR component in our model which is dominated mostly by helium or a mixture of helium and CNO group with a small iron fraction. Between ${\sim}\,10^8$ and $10^9$~GeV, the predicted all-particle spectrum in \cite{Hillas2005} consists of a significant iron fraction which may be in agreement with the $\langle\mathrm{lnA}\rangle$ data when mixed with a strong extra-galactic proton component, but is in tension with the small iron fraction (${\sim}\,2-10\%$) preferred by the recent measurements of LOFAR \citep{Buitink2016} and the Pierre Auger Observatory \citep{Aab2014}. These new measurements disfavour the general view that the Galactic component above the second knee is dominated by heavy (iron) nuclei.

\subsection{A steep EG-CR component extending below the second knee}
\begin{figure}
\centering
\includegraphics*[width=\columnwidth,angle=0,clip]{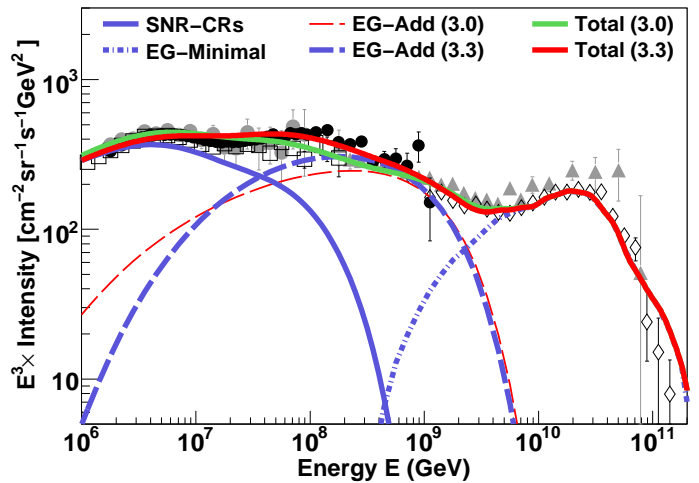}
\caption{\label {fig-spectrum-total-Xgal-add} All-particle energy spectrum for an additional component of EG-CR protons extended down to low energies and modulated by Galactic wind. The numbers within the parentheses denote the injection index for the additional extra-galactic component. For the SNR-CRs, an exponential cut-off energy for protons at $4.1\times 10^6$~GeV is assumed. See text for other details. The EG-Minimal component is the same as in Figure \ref{fig-spectrum-EG-models-WR2} ({\textit{top}}). Data: Same as in Figure \ref{fig-all-particle-SNR}.}
\end{figure}

An alternative model that does not require the introduction of an additional Galactic component is to assume that EG-CRs have a significant contribution down to energies below the second knee. Such a scenario would require a steep spectrum of $\sim E^{-3}$ and a  strong flux suppression below $\sim 10^8$~GeV (see \citealp{Hillas2005} for a brief discussion, and also \citealp{Muraishi2005} in the context of the origin of the knee). To explore this scenario, we inject an additional extra-galactic component of pure protons at the position of the Galactic wind termination shocks, and allow them to propagate diffusively towards the Galactic disk in the presence of the Galactic wind outflow. The injection spectrum is assumed to follow $E^{-\gamma}\exp(-E/E_\mathrm{c})$. The propagation is treated exactly the same as the propagation of GW-CRs from the termination shock towards the Galactic disk. All propagation parameters are kept the same, except for the wind velocity constant $\tilde{V}$ which is treated as a free parameter. The best-fit all-particle spectrum obtained after adding the contribution of SNR-CRs and EG-CRs from the minimal model is shown in Figure \ref{fig-spectrum-total-Xgal-add}. The best-fit parameters are $\gamma=3.3$, $E_\mathrm{c}=4.1\times 10^6$~GeV for the SNR-CRs protons, $E_\mathrm{c}=1.5\times10^9$~GeV for the additional EG-CRs, and $\tilde{V}=200.5$~$\mathrm{km/s/kpc}$. This value of $\tilde{V}$ gives a wind velocity which is about a factor $13$ larger than the wind velocity used in the study of GW-CRs. Such a fast wind is required in order to generate a strong modulation for particles below the second knee so that the predicted flux does not exceed the observed data at low energies. For $\gamma<3.3$, the required wind velocity is lower, but the model prediction does not fit  the observed data very well (see e.g. the case of $\gamma=3$ in Figure \ref{fig-spectrum-total-Xgal-add}). Replacing the additional extra-galactic protons with heavier elements only slightly reduces the required wind velocity. Having a strong Galactic wind can have serious effects on the spectrum and distribution of low-energy cosmic rays in the Galaxy (see e.g. \citealp{Bloemen1993}). In the presence of a strong wind, cosmic-ray transport will be dominated by advection rather than diffusion, and will produce a cosmic-ray distribution that resembles the distribution of the sources. But, the cosmic-ray distribution inferred from the observations of diffuse gamma-ray emission from the Galaxy indicates a radial gradient weaker than the distribution of supernova remnants or pulsars in the Galaxy. These observations suggest that if supernova remnants are the main sources of cosmic rays in the Galaxy, the propagation of cosmic rays should be dominated by diffusion, not by advection. In addition, if the transport is dominated by advection, the cosmic-ray spectrum is expected to exhibit a break (steepening) at an energy where the advection boundary, $z_\mathrm{c}\propto {[D(E)/\tilde{V}]^{1/2}}$, equals the halo boundary $L$. Such a break is not observed below the knee, except at $\sim 10$~GeV which is due to Solar modulation. Attributing the knee to such a break raises issues regarding the cosmic-ray injection index. Below the break, cosmic-ray transport is advection dominated and the spectrum is expected to follow $E^{-(\gamma+a/2)}$, where $\gamma$ is the source index and $a$ is the diffusion index. For the observed spectral index of $\sim 2.7$ and $a=0.33$ used in our study, we get $\gamma=2.53$. This is incompatible with the prediction of diffusive shock acceleration theory which predicts an index close to $2$ for the strong shocks present in supernova remnants \citep{Ptuskin2010, Caprioli2011}. Choosing $a=0.6$, as in pure diffusion propagation models, gives $\gamma=2.4$. This relaxes the tension a bit, but such a high value of $a$ is not favoured by the observed small level of cosmic-ray anisotropy. Another strong constraint on the Galactic wind velocity is provided by the abundance ratio of radioactive secondary to stable secondary. Measurement of $^{10}$Be/$^9$Be ratio puts a constraint at $\tilde{V}\leq 45$~$\mathrm{km/s/kpc}$ \citep{Bloemen1993}. All these arguments pose a serious problem to the alternative scenario of a strong EG-CR component with a steep spectrum extending below the second knee, and modulating by Galactic wind. One possibility, but rather unrealistic, for this scenario to work is if the additional EG-CR component has a spectrum and composition almost similar to that of the GW-CRs produced at the Galactic wind termination shocks.

An alternative to the modulation of EG-CRs by the Galactic wind is the `magnetic horizon effect' \citep{StanevEtAl2000, Lemoine2005, AloisioBerezinsky2005}, which leads to a flattening of the extra-galactic spectrum below an energy where the diffusive propagation distance in a partly turbulent extra-galactic magnetic field, over the time scale set by energy losses of the cosmic rays through interactions with ambient photon backgrounds, gets below the average distance of cosmic ray sources. Assuming a relatively strong ($\gtrsim 1\,$~nG) extra-galactic field with a constant coherence length extending over the entire universe, this effect could set in at around $10^9$~GeV, effectively cutting off the extra-galactic component at lower energies slightly below the ankle \citep{Aloisio2012}, or even above \citep{MollerachRoulet2013}. However, more detailed treatments in the context of large scale structure formation \citep{KoteraLemoine2008}, have indicated that this effect is much less efficient due to the large voids in the universe which are essentially free of magnetic fields. As shown recently in detailed simulations, the magnetic horizon effect should play virtually no role above the second knee for any type of nuclei, and for protons in some extra-galactic magnetic field scenarios, not even above the knee \citep{BatistaSigl2014}.

We point out that neither the Galactic wind nor the magnetic horizon effects discussed above prevent a hard extra-galactic component,  like the light component with $\gamma=2.7$ as indicated by the KASCADE-Grande measurements above ${\sim}\,10^8$~GeV \citep{Apel2013}, from contributing around the second knee as such a hard component will be already consistent with the measured data at low energies. Even if such a hard extra-galactic component is present, an additional Galactic component will still be required as the extra-galactic component will remain subdominant in the all-particle spectrum below $10^8$~GeV. 

An additional problem for EG-CRs with an overall spectrum steeper than $E^{-2.7}$ is that, if one assumes that they fill the extra-galactic space homogeneously with energies from ${\sim}\,1$~GeV to $10^9$~GeV, it contains more energy than the gravitational binding energy released in the universe during structure formation  \citep{Rachen2016}. Using realistically low efficiencies for this energy -- which is, besides the lower overall nuclear binding energy released in fusion by all primordial baryonic matter going into stars, the only fundamental energy budget present in the late universe -- to be converted into cosmic rays, one can conclude that spectral indices as discussed here for a dominant extra-galactic component below the second knee cannot easily be reconciled with this energy budget, no matter which kind of sources one proposes. Mainly on the basis of this argument, together with the difficulties of a sufficient spectral modification at low energies discussed above, we consider a dominantly extra-galactic explanation of cosmic rays below $10^8$~GeV as implausible.

\section{Conclusions}
\label{sec-conclusion}
We have demonstrated that a single Galactic component with progressive energy cut-offs in the individual spectra of different elements, and describing the low-energy measurements below ${\sim}\,10^6$~GeV from balloon and satellite-borne experiments, cannot explain simultaneously the knee and the second knee observed in the all-particle spectrum. We show that a two-component Galactic model, the first component dominating up to ${\sim}\,5\times10^7$~GeV and the second component dominating in the range of ${\sim}\,5\times10^7-10^9$~GeV, can explain almost all observed features in the all-particle spectrum and composition when combined with an extra-galactic component dominating above ${\sim}\,10^9$~GeV. Discussing two different scenarios for the second Galactic component, we find that a contribution of Wolf-Rayet supernovae explain best both the measured energy spectrum and composition. Our main result is that this component predicts a Galactic contribution at and above the second knee which is mainly dominated by helium or a mixture of helium and CNO nuclei, and is consistent with a `regular' extra-galactic contribution from sources with a flat spectral index and a cosmological evolution typical for AGNs or star formation. Using re-acceleration at the  Galactic wind termination shock as a second Galactic component also allows to fit the all-particle energy spectrum, but not the observed composition very well. Tests of the two-component Galactic model using different hypotheses for a significant extra-galactic cosmic-ray component below the ankle, do neither significantly improve nor deteriorate this result, mostly because \textit{both} the Galactic and extra-galactic components have a rather light composition, and contain little or no heavy nuclei like iron, in contrast to common assumptions. In all cases, the transition from Galactic to extra-galactic cosmic rays occurs between the second knee and the ankle, and we see neither the need nor a theoretical case for an extra-galactic component significantly contributing at or below $10^8$~GeV.  Our findings are in agreement with recent measurements from LOFAR and the Pierre Auger Observatory, which have revealed a strong light component and a rather low iron fraction between ${\sim}\,10^8$ and $10^9$~GeV. A clear distinction of the various discussed Galactic and extra-galactic scenarios would be possible if we could separately measure the spectra of at least four major mass groups, that is protons, helium, CNO, and heavier, at energies between the second knee and the ankle.

\section*{Acknowledgements}
We wish to thank Michael Unger, Glennys Farrar and Luis Anchordoqui, for their comments on the manuscript, and for providing us results for the UFA model. We also thank Peter Biermann for his insightful comments, and Torsten En{\ss}lin and Christoph Pfrommer for helping to improve the manuscript during a discussion at the meeting of `International Team 323' at the International Space Science Institute in Bern. ST wishes to thank Onno Pols for discussions on Wolf-Rayet stars. AvV acknowledges financial support from the NWO Astroparticle physics grant WARP. We furthermore acknowledge financial support from an Advanced Grant of the European Research Council (grant agreement no. 227610), European Union's Horizon 2020 research and innovation programme (grant agreement no. 640130), the NWO TOP grant (grant agreement no. 614.001.454), and the Crafoord Foundation (grant no. 20140718).

\appendix
\section{Derivation of Equation \ref{eq-solution-wind}}
\label{appendix-A}
The Green's function, $G(\textbf{r},\textbf{r}^\prime,p,p^\prime)$, of Equation \ref{eq-transport-wind} satisfies,
\begin{equation}
\nabla.(D_\mathrm{w}\nabla G-\textbf{V}G)+\frac{\partial}{\partial p}\left\lbrace\frac{\nabla.\textbf{V}}{3}pG\right\rbrace=-\delta(\textbf{r}-\textbf{r}^\prime)\delta(p-p^\prime).
\end{equation}
In rectangular coordinates, the above equation can be written as,
\begin{align}
\label{eq-diffusion}
&D_\mathrm{w}\frac{\partial^2 G}{\partial x^2}+D_\mathrm{w}\frac{\partial^2 G}{\partial y^2}+D_\mathrm{w}\frac{\partial^2 G}{\partial z^2}-\tilde{V}\frac{\partial}{\partial x}(xG)-\tilde{V}\frac{\partial}{\partial y}(yG)\nonumber\\
&-\tilde{V}\frac{\partial}{\partial z}(zG)+ \frac{\partial}{\partial p}(\tilde{V} p N)=-\delta(x-x^\prime)\delta(y-y^\prime)\delta(z)\delta(p-p^\prime),
\end{align}
where we have written $\textbf{V}=\tilde{V}(x\hat{i}+y\hat{j}+z\hat{k})$ with $\hat{i}$, $\hat{j}$ and $\hat{k}$ representing the unit vectors along the $x$, $y$ and $z$ directions. Following a similar procedure adopted in \cite{Lerche1982b}, we express,
\begin{align}
\label{green-function}
&G(x,x^\prime,y,y^\prime,z,z^\prime,p,p^\prime)=\nonumber\\
&\int_{-\infty}^\infty dk_x\int_{-\infty}^\infty dk_y\int_{-\infty}^\infty dk_z\;\bar{G}(k_x,x^\prime,k_y,y^\prime,k_z,z^\prime,p,p^\prime)\nonumber\\
&\times e^{ik_x\left(x-x^\prime\right)}e^{ik_y\left(y-y^\prime\right)}e^{ik_z\left(z-z^\prime\right)},
\end{align}
and,
\begin{align}
\label{delta-function}
&\delta(x-x^\prime)=\frac{1}{2\pi}\int_{-\infty}^\infty dk_x\;e^{ik_x\left(x-x^\prime\right)},\nonumber\\
&\delta(y-y^\prime)=\frac{1}{2\pi}\int_{-\infty}^\infty dk_y\;e^{ik_x\left(y-y^\prime\right)},\nonumber\\
&\delta(z-z^\prime)=\frac{1}{2\pi}\int_{-\infty}^\infty dk_z\;e^{ik_x\left(z-z^\prime\right)}.
\end{align}
Inserting Equations \ref{green-function} and \ref{delta-function} into Equation \ref{eq-diffusion}, we get,
\begin{align}
\label{eq-diffusion2}
&-D_\mathrm{w}\left(k^2_x+k^2_y+k^2_z\right) G-i\tilde{V}\left(k_x x^\prime+k_y y^\prime+k_z z^\prime\right)\bar{G}\nonumber\\
&+\tilde{V}\left(k_x\frac{\partial\bar{G}}{\partial k_x}+k_y\frac{\partial\bar{G}}{\partial k_y}+k_z\frac{\partial\bar{G}}{\partial k_z}\right)+\tilde{V}p\frac{\partial \bar{G}}{\partial p}+\tilde{V}\bar{G}\nonumber\\
&=-\frac{1}{8\pi^3}\delta(p-p^\prime).
\end{align}
We now introduce variables $\psi_x$, $\psi_y$ and $\psi_z$ such that $k_x=\psi_x F(p)$, $k_y=\psi_y F(p)$ and $k_z=\psi_z F(p)$, where
\begin{equation}
\label{eq-fp}
F(p)=\exp\left(\tilde{V}\int^p du\;\frac{1}{\tilde{V} u}\right).
\end{equation}
This reduces Equation \ref{eq-diffusion2} to
\begin{align}
\label{eq-diffusion3}
\tilde{V}p\frac{\partial \bar{G}}{\partial p}+B(p)\bar{G}=-\frac{1}{8\pi^3}\delta(p-p^\prime),
\end{align}
where,
\begin{align}
B(p)=&-D_\mathrm{w}(p)\left(\psi^2_x+\psi^2_y+\psi^2_z\right)F^2(p)\nonumber\\
&-i\left(\psi_x x^\prime+\psi_y y^\prime+\psi_z z^\prime\right)F\tilde{V}+\tilde{V}.
\end{align}
The solution of Equation \ref{eq-diffusion3} is given by,
\begin{align}
\label{eq-diffusion4}
\bar{G}(k_x,x^\prime,k_y,y^\prime,k_z,z^\prime,p,p^\prime)&=\frac{1-\mathrm{H}\left[p-p^\prime\right]}{8\pi^3\tilde{V}p^\prime}\nonumber\\
&\times\exp\left[\int^E_{E^\prime}du\;\frac{B(u)}{\tilde{V}u}\right],
\end{align}
where the Heaviside step function $\mathrm{H}\left[p-p^\prime\right]=1 (0)$ for $p>p^\prime (<p^\prime)$. Taking inverse Fourier transform of $\bar{G}$, we obtain the required Green's function as,
\begin{align}
G(x,x^\prime,y,y^\prime,z,z^\prime,p,&p^\prime)=\frac{1-\mathrm{H}\left[p-p^\prime\right]}{8\pi^3\tilde{V}p}\left(\frac{\pi}{I_{p,p^\prime}}\right)^{3/2}\nonumber\\
&\times\exp\left[-\frac{\left(C^2_{x,x^\prime}+C^2_{y,y^\prime}+C^2_{z,z^\prime}\right)}{4I_{p,p^\prime}}\right]
\end{align}
where,
\begin{align}
&C_{x,x^\prime}=\tilde{V}x^\prime\int^p_{p^\prime}du\;\frac{1}{\tilde{V}u}\frac{F(u)}{F(p)}-x^\prime+x,\nonumber\\
&C_{y,y^\prime}=\tilde{V}y^\prime\int^p_{p^\prime}du\;\frac{1}{\tilde{V}u}\frac{F(u)}{F(p)}-y^\prime+y,\nonumber\\
&C_{x,x^\prime}=\tilde{V}z^\prime\int^p_{p^\prime}du\;\frac{1}{\tilde{V}u}\frac{F(u)}{F(p)}-z^\prime+z,
\end{align}
and,
\begin{equation}
I_{p,p^\prime}=\int^{p^\prime}_p du\;\frac{D_\mathrm{w}(u)}{\tilde{V}u}\left(\frac{F(u)}{F(p)}\right)^2.
\end{equation}
Then, for a given cosmic-ray source characterised by $q(x^\prime,y^\prime,z^\prime,p^\prime)$, the differential number density of particles with momentum $p$ at a distance $(x,y,z)$ is given by,
\begin{align}
\label{eq-sol-appen}
N(x,y,z,p)&=\int^{-\infty}_\infty dx^\prime\int^{-\infty}_\infty dy^\prime \int^{-\infty}_\infty dz^\prime\int^{-\infty}_\infty dp^\prime\nonumber\\
&\times G(x,x^\prime,y,y^\prime,z,z^\prime,p,p^\prime) q(x^\prime,y^\prime,z^\prime,p^\prime).
\end{align}
For any point source located at $(0,0,0)$ and emitting $q(p)$ spectrum of particles, that is $q(x^\prime,y^\prime,z^\prime,p^\prime)=\delta(x^\prime)\delta(y^\prime)\delta(z^\prime)q(p^\prime)$, the solution becomes,
\begin{align}
\label{eq-solu1}
N(x,y,z,p)&=\frac{1}{8\pi^3 \tilde{V}p}\int^\infty_p dp^\prime q(p^\prime) \left(\frac{\pi}{I_{p,p^\prime}}\right)^{3/2}\nonumber\\
&\times\exp\left[-\frac{\left(x^2+y^2+z^2\right)}{4I_{p,p^\prime}}\right].
\end{align}
From Equation \ref{eq-fp}, since $F(p)$ reduces to $p$, and so also $F(u)$ to $u$, by writing $(x^2+y^2+z^2)=r^2$ in spherical coordinates and replacing $q(p)$ by $Q_\mathrm{esc}(p)$ as given by Equation \ref{eq-qesc}, Equation \ref{eq-solu1} can be reduced in the form of Equation \ref{eq-solution-wind}:
\begin{align}
N(r,p)=\frac{\sqrt{\tilde{V}}p^2}{8\pi^{3/2}}\int^\infty_0 &dp^\prime \frac{Q_\mathrm{esc}(p^\prime)}{\left[{\int^{p^\prime}_p u D_\mathrm{w}(u)du}\right]^{3/2}}\nonumber\\
&\times \exp\left({-\frac{r^2\tilde{V} p^2}{4\int^{p^\prime}_p u D_\mathrm{w}(u)du}}\right).
\end{align}

\section{All-particle spectrum and composition of cosmic rays obtained using different EG-CR models and WR-CRs ($\mathrm{C/He}=0.1$)}
\label{appendix-B}
\begin{figure}[H]
\centering
\includegraphics*[width=\columnwidth,angle=0,clip]{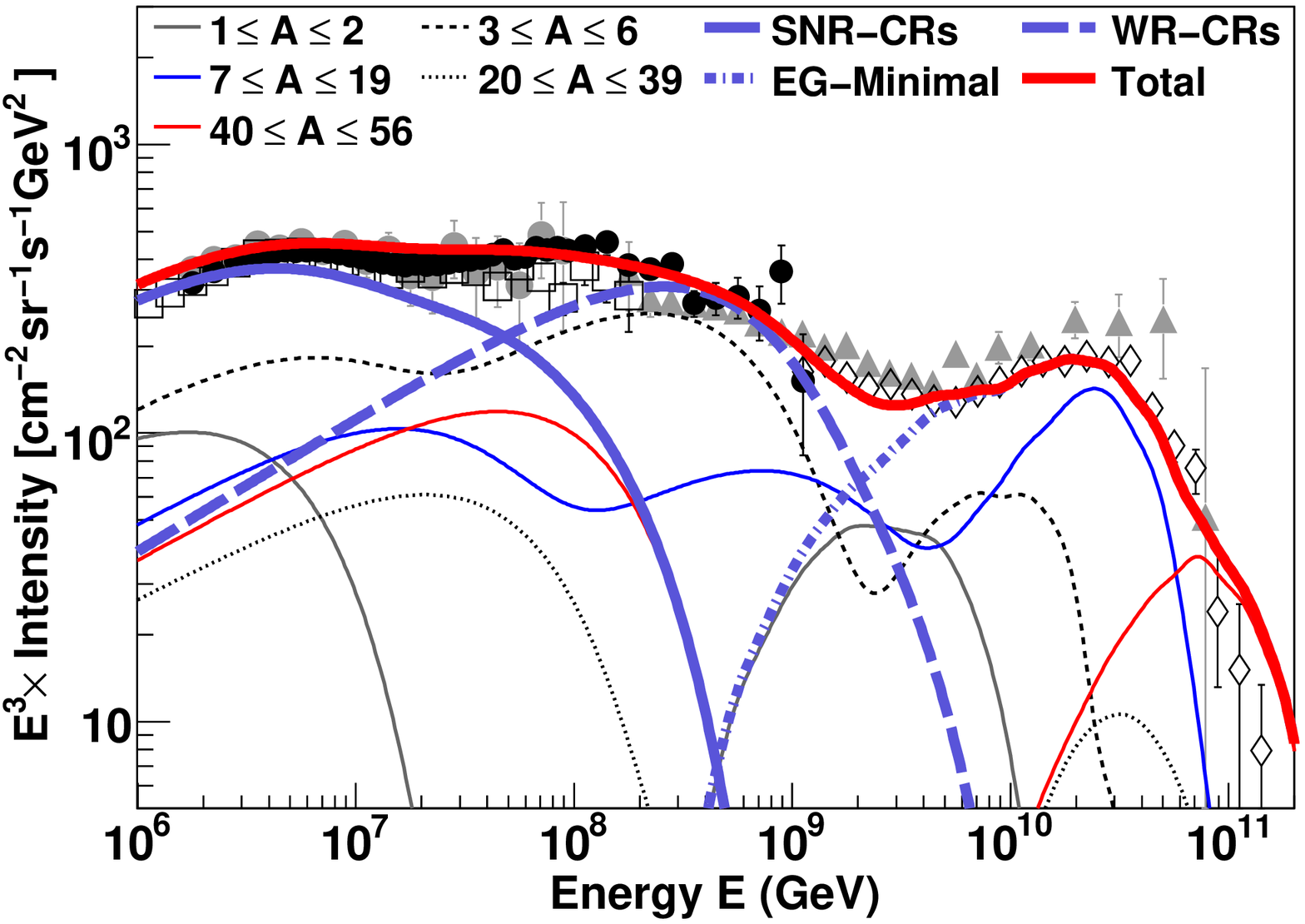}\\
\includegraphics*[width=\columnwidth,angle=0,clip]{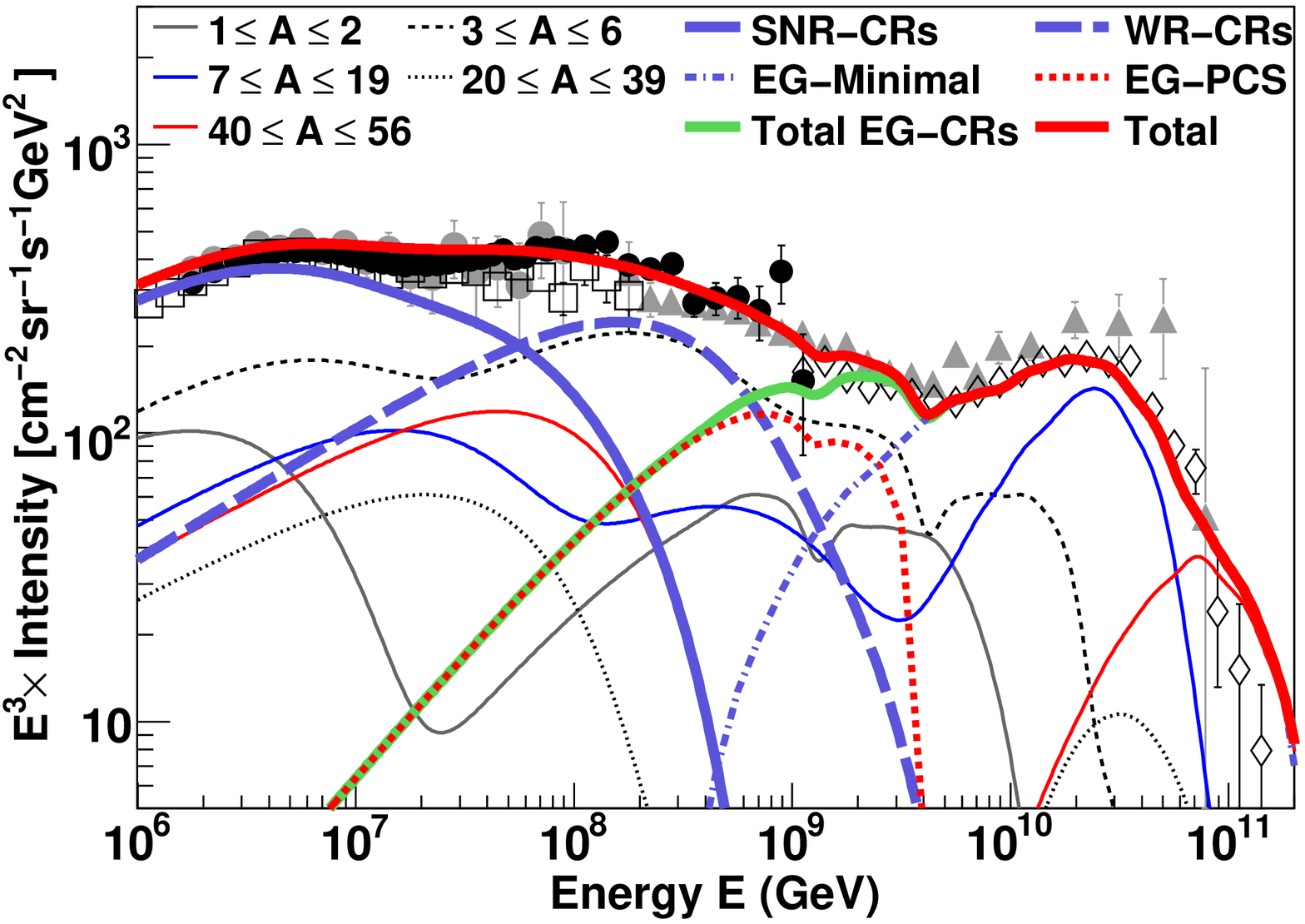}\\
\includegraphics*[width=\columnwidth,angle=0,clip]{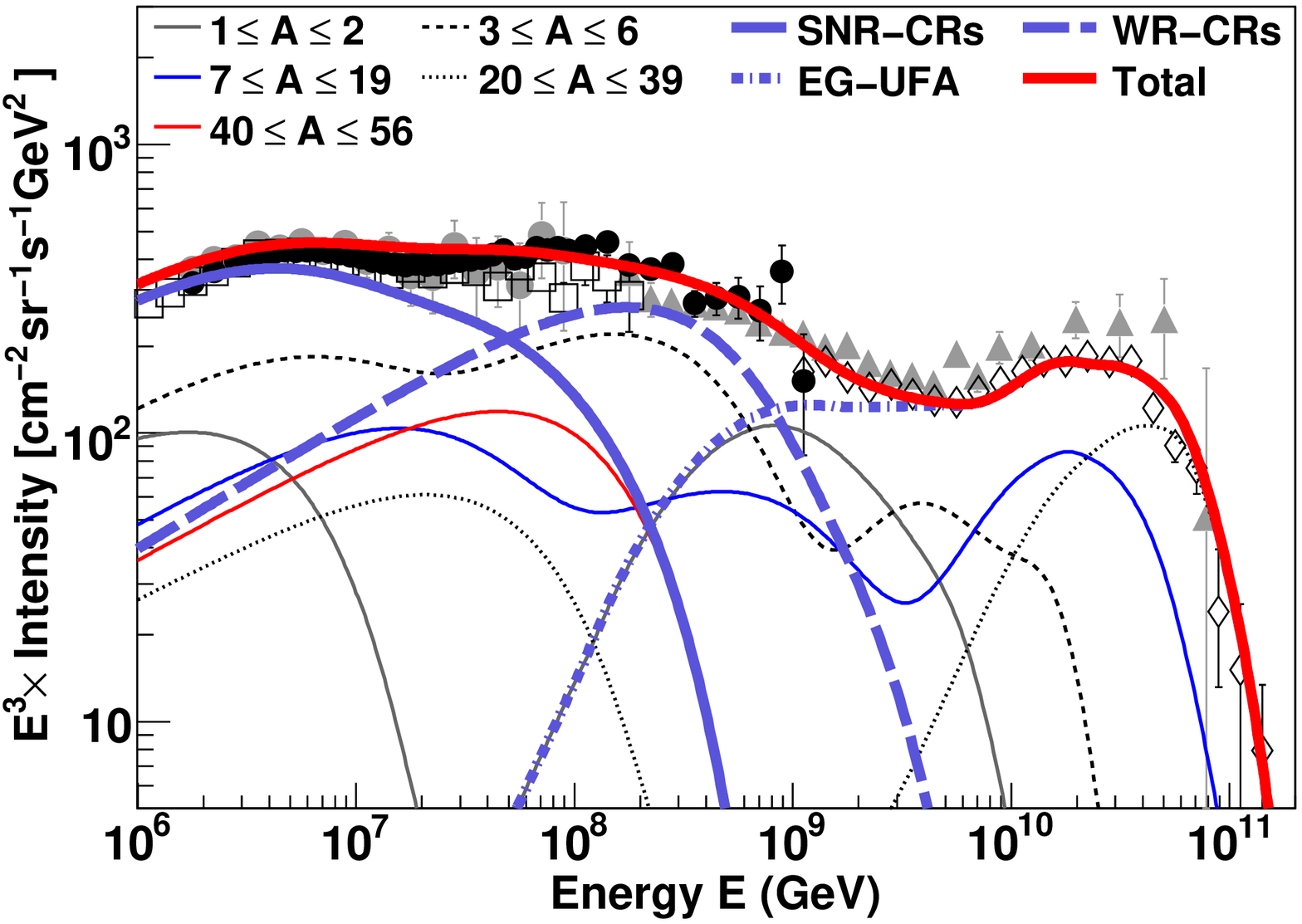}
\caption{\label {fig-spectrum-EG-models-WR1} All-particle spectrum for the three different EG-CR models: minimal ({\textit {top}}), PCS ({\textit {middle}}), and UFA ({\textit {bottom}}), obtained using WR-CRs ($\mathrm{C/He}=0.1$) as the additional Galactic component.   The proton cut-off energies for the WR-CRs used in the calculation are $2.4\times10^8$~GeV for the minimal model, $1.5\times10^8$~GeV for the PCS model, and $1.6\times10^8$~GeV for the UFA model. The injection energy of the WR-CRs varies within $6\%$ between the three models. SNR-CR spectra are the same as in Figure \ref{fig-total-spectrum-WR} ({\textit {top}}). Data are the same as in Figure \ref{fig-all-particle-SNR}.}
\end{figure}

\begin{figure}
\centering
\includegraphics*[width=\columnwidth,angle=0,clip]{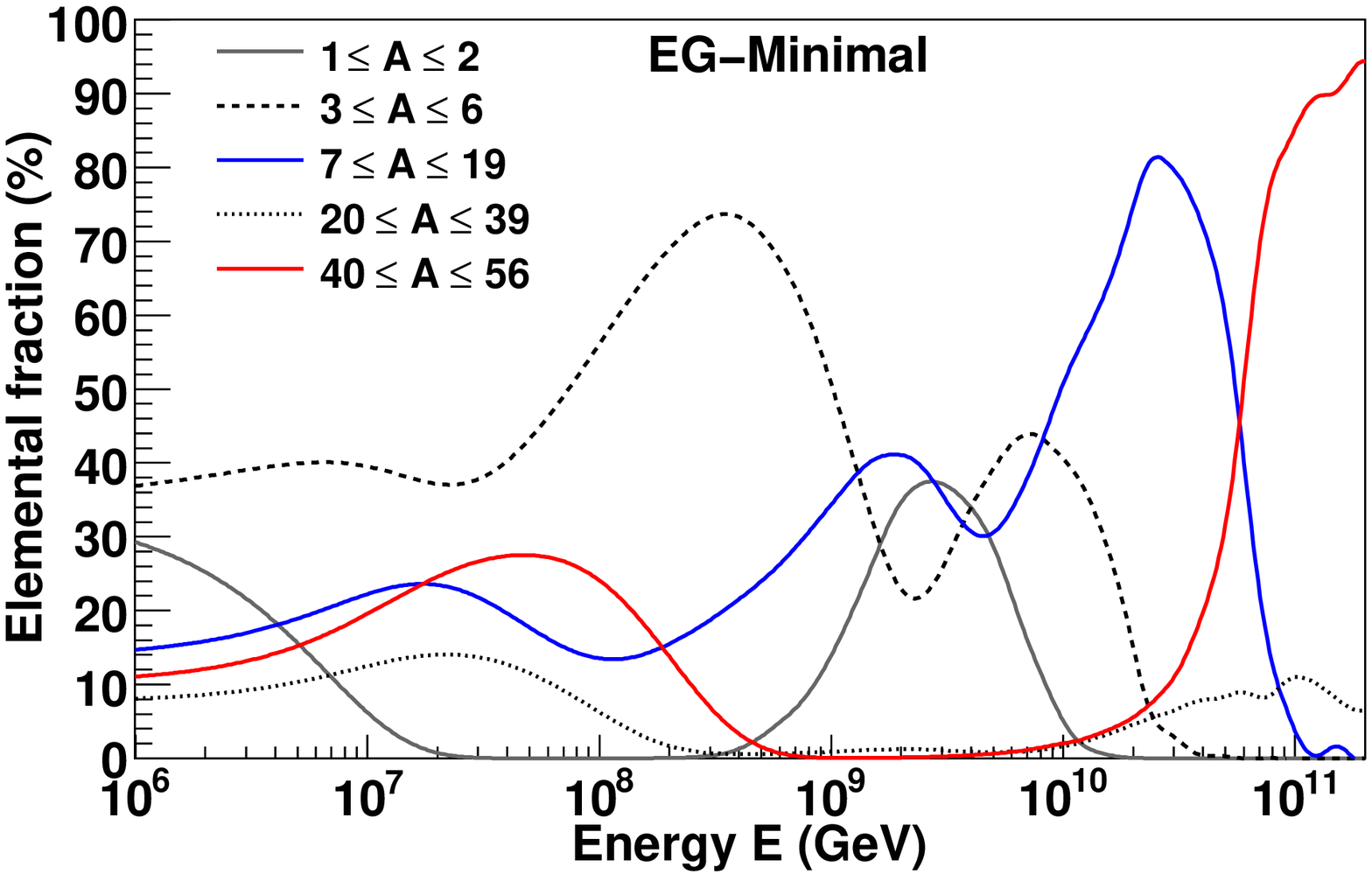}\\
\includegraphics*[width=\columnwidth,angle=0,clip]{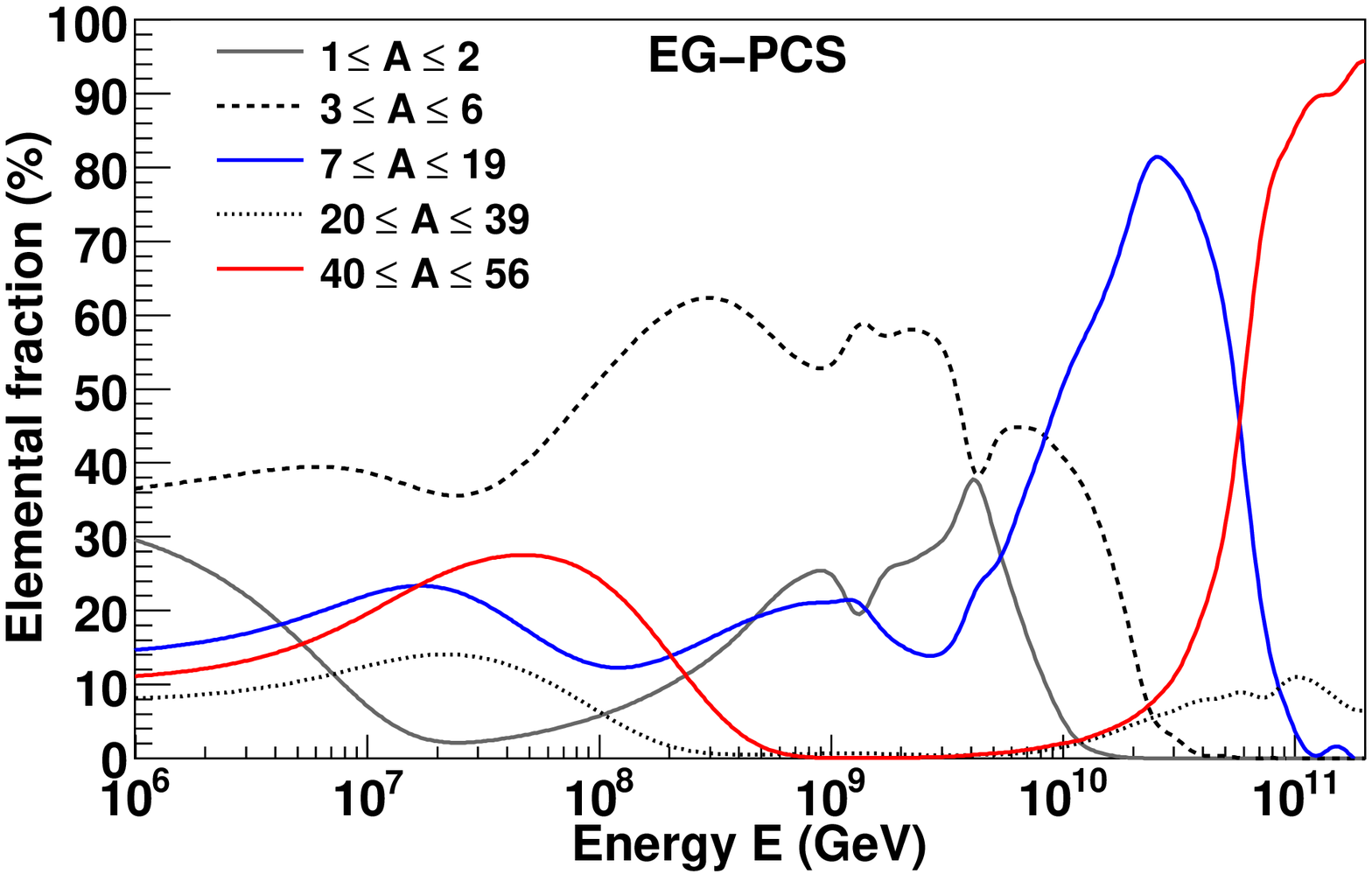}\\
\includegraphics*[width=\columnwidth,angle=0,clip]{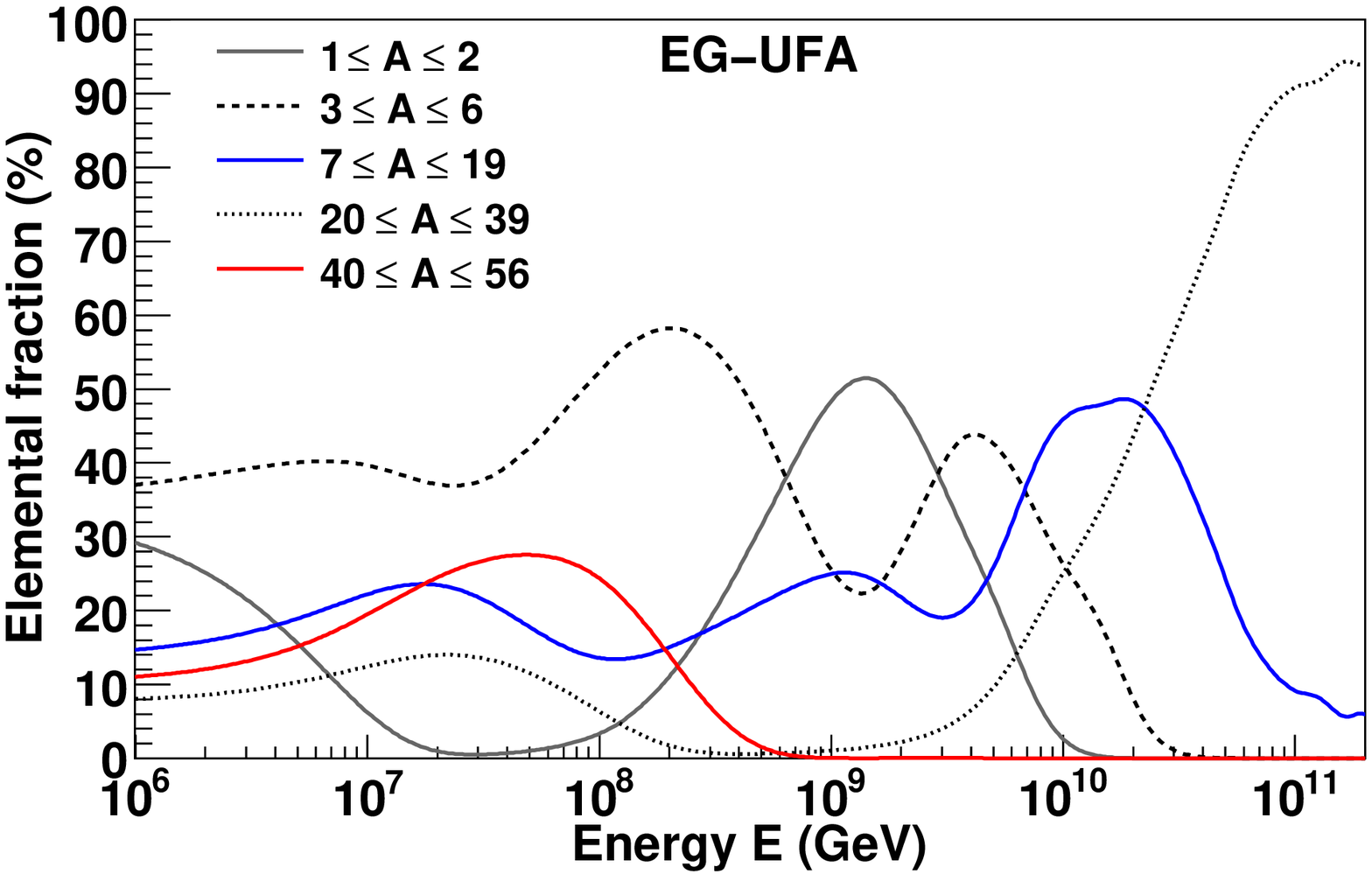}
\caption{\label {fig-fraction-EG-models-WR1} Elemental fraction of the five different mass groups shown in Figure \ref{fig-spectrum-EG-models-WR1} for the three different EG-CR models: minimal ({\textit {top}}), PCS ({\textit {middle}}), and UFA ({\textit {bottom}}), obtained using WR-CRs ($\mathrm{C/He}=0.1$) as the additional Galactic component.}
\end{figure}

\begin{figure*}
\centering
\includegraphics*[width=0.7\textwidth,angle=0,clip]{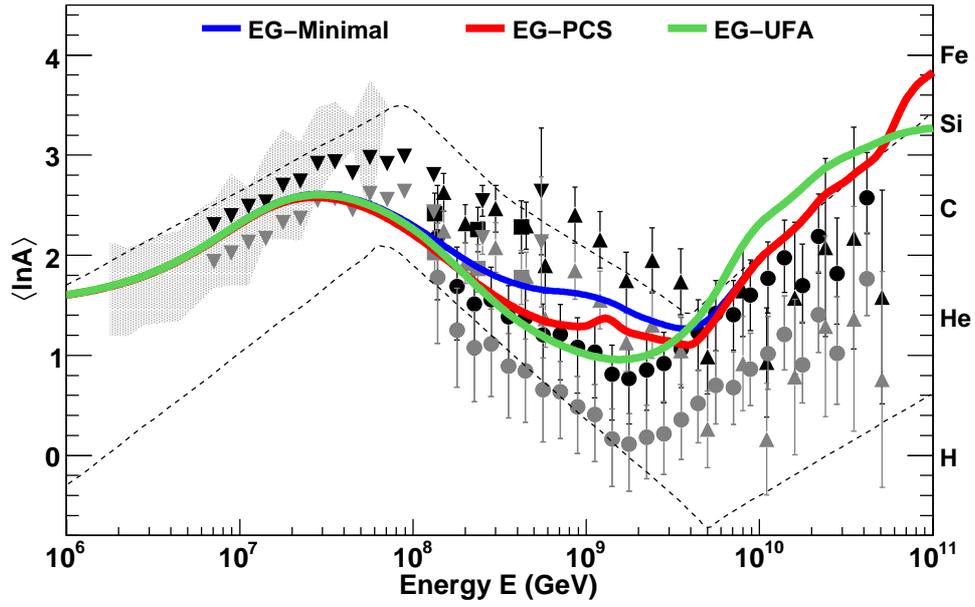}
\caption{\label {fig-lnA-EG-models-WR1} Mean logarithmic mass of cosmic rays for the minimal, PCS and UFA models of EG-CRs obtained using WR-CRs ($\mathrm{C/He}=0.1$) as the additional Galactic component. Data are the same as in Figure \ref{fig-lnA}.}
\end{figure*}

The predicted all-particle spectrum, elemental fraction and $\langle\mathrm{lnA}\rangle$ obtained for the three different models of EG-CRs (the minimal, PCS and UFA), combined with the WR-CR ($\mathrm{C/He}=0.1$) scenario for the additional Galactic component, are shown in Figures \ref{fig-spectrum-EG-models-WR1}, \ref{fig-fraction-EG-models-WR1}, and \ref{fig-lnA-EG-models-WR1}, respectively. The proton cut-off energies for the WR-CRs required to produce a good-fit to the measured spectrum are $2.4\times10^8$~GeV for the minimal model, $1.5\times10^8$~GeV for the PCS model, and $1.6\times10^8$~GeV for the UFA model.These values are about a factor $1.4$ larger than the cut-off energies obtained in the case of $\mathrm{C/He}=0.4$. The variation in the injection energy of WR-CRs between the three cases remain within $6\%$ as in the $\mathrm{C/He}=0.4$ scenario.

The predicted composition is dominated by helium nuclei up to around the second knee for the minimal and the UFA models, while for the PCS model, helium dominates up to around $10^{10}$~GeV. The Galactic component at the transition energy region from Galactic  to extra-galactic cosmic rays is dominated by helium, unlike in the case of $\mathrm{C/He}=0.4$, where it is dominated by a mixture of helium and CNO group.
The $\langle\mathrm{lnA}\rangle$ predicted by the minimal model shows some deviation from the general trend of the measurements between ${\sim}\,10^8$ and $5\times10^9$~GeV, although the discrepancy is less than that observed in the $\mathrm{C/He}=0.4$ scenario. The predictions of both the PCS and the UFA models show better agreement with the data below ${\sim}\,10^9$~GeV. Between around $10^7$ and $10^9$~GeV, they predict a mean mass lighter than the prediction of the $\mathrm{C/He}=0.4$ case, and show a better agreement with the data (EPOS-LHC) from the Pierre Auger Observatory in the $10^8-10^9$~GeV energy range, but slightly under predict the available measurements at around ${\sim}\,10^8$~GeV. The two WR-CR scenarios should be possible to differentiate by accurate measurements of the elemental composition between $10^7$ and $10^9$~GeV.

\end{document}